\documentclass[12pt]{article}
\usepackage{amsmath, amssymb, amsfonts}
\usepackage{color, multirow, multicol, array}
\usepackage{graphicx}
\usepackage{epsfig}
\usepackage[english]{babel}

\begin{document}

\title{Using the PPML approach for constructing a low-dissipation, operator-splitting scheme 
for numerical simulations of hydrodynamic flows}

\author{Kulikov Igor, Vorobyov Eduard}
\maketitle

\begin{abstract}
An approach for constructing a low-dissipation numerical method is described. The method is based 
on a combination of the operator-splitting method, Godunov method, and piecewise-parabolic method 
on the local stencil. Numerical method was tested on a standard suite of hydrodynamic test problems. In addition, the performance of the method is demonstrated on a global test problem showing the development of a spiral structure in a gravitationally unstable gaseous galactic disk.
\end{abstract}

\section{INTRODUCTION}
During the last two decades, two main approaches have been used for numeral hydrodynamics 
simulations of astrophysical flows: the Lagrangian smooth particle hydrodynamics methods 
(hereafter, the SPH method) and Eulerian mesh-based methods. Numerous comparisons between the SPH 
and mesh-based methods have been performed \cite{Agertz_2007,Tasker_2008}. The main 
disadvantages of most SPH methods are the inaccurate computation of large gradients and 
discontinuities \cite{Vshivkov_2009}, suppression of physical instabilities \cite{Agertz_2007}, 
difficulty with choosing the proper smoothing kernel \cite{Attwood_2007} and the use of artificial 
viscosity \cite{Sijacki_2006}. The main disadvantages of most mesh-based methods are their 
Galilean non-invariance on the mesh \cite{Tasker_2008,Wadsley_2008}, difficulty with coding and 
implementation, and difficulty with treating multi-component systems, such as 
stars and gas \cite{Mitchell_2013}. 
During the last decade, the combined Eulerian-Lagrangian approach has 
been developed and actively used for numerical simulations of astrophysical hydrodynamics flows 
\cite{Murphy_2008,Springel_2010}. These methods unite the advantages of both approaches, while
attempting to reduce the disadvantages.

Recently, another Eulerian-Lagrangian numerical scheme for astrophysical flows has been presented, 
which employs the operator-splitting methodology and Godunov-type methods 
\cite{Vshivkov_2009,Kulikov_2013,Vshivkov_2007,Vshivkov_2011_a,Vshivkov_2011_b}. This numerical 
method is based on the solution of hydrodynamic equations in two stages. At the first Eulerian step,
the hydrodynamic equations are solved excluding the advective terms. At the second Lagrangian step,
equations for the advective transport are solved. This separation into two stages can effectively 
solve the problem of Galilean-invariance and the use of Godunov methods for the Eulerian step 
enables correct modelling of discontinuous solutions. This numerical method has been successfully 
employed for modeling collisions between galaxies \cite{Vshivkov_2011_a,Tutukov_2011}. 
 {A significant disadvantage of the method lays} in the fact that it was first-order accurate. The main shortcoming of first-order schemes is dissipation of the numerical solution on 
discontinuities. It is therefore our main motivation to extend the original method to a higher 
order of accuracy. 

There are several well-known high-order numerical hydrodynamics methods such as the MUSCL (Monotonic Upstream-Centered Scheme for Conservation Laws)  method \cite{KurganovTadmor_2000,VanLeer_1979}, total variation diminishing (TVD) method \cite{Jin_1995,Jiang_1996,Balsara_2000,Balsara_2009,Henrick_2005}, and piecewise parabolic method (PPM) \cite{Collela_1984}. The general idea of their approach is the construction of a piecewise-polynomial function on  each cell of the numerical mesh. It may be a piecewise-linear reconstruction (in the case of the MUSCL scheme) or piecewise-parabolic reconstruction (in the case of the PPM method). For the construction of a monotonized numerical solution (needed to avoid the growth
of spurious extrema) the limiters are usually employed in TVD methods \cite{Watersona_2007}. 
The problem of selection of limiters is analogous to the choice of artificial viscosity in SPH methods: the wrong choice of artificial viscosity for the SPH method (or limiters for the TVD method) can  cause a substantial distortion of the numerical solution. The PPM method does not have 
the monotonicity problem, because piecewise-polynomial solutions on each cell are constructed without extrema. The main disadvantage of the PPM method is the use of a non-local stencil for computing 
hydrodynamic quantities on the next time step. The non-local stencil has
problems with the proper choice of boundary conditions, domain decomposition, and dissipation 
of numerical solution. To resolve these problems, a modification of the PPM method was suggested --
the so-called piecewise-parabolic method on the local stencil (PPML) \cite{Ustyugov_2007,Ustyugov_2008}. The main idea of the PPML method is the use of a piecewise parabolic numerical solution on  the previous time step for computing the Riemann problem. 

Numerical simulations of galaxies are an indispensable tool for studying 
their properties and evolution, taking into account the complexity of physical processes 
involved.  The main galactic components such as gas and dust can be modelled 
as fluids, making numerical hydrodynamics simulations the main numerical approach.
The main focus of this paper is concentrated on the extension of the original method 
\cite{Vshivkov_2009,Kulikov_2013,Vshivkov_2007,Vshivkov_2011_a,Vshivkov_2011_b}
based on the PPML approach to a higher order of accuracy. 
 {The methods presented here are specific to astrophysical flows where the
extraordinarily high flow speeds, as well as the effect of self-gravity, impose
further restrictions on the kinds of flow solvers that are designed. The operator
splitting between the force terms and advected terms is certainly
non-standard and would not be appropriate for a traditional flow solver.}

In the first section, the details of the original method and the construction of a high order extension are described. The second section contains a verification of the high order nature of the modified method. In the third section, we provide computational experiments dealing with gravitational instability in disk galaxies and the development of a spiral structure.

\section{NUMERICAL SCHEME}
We aim at solving numerically the following system of hydrodynamic equations describing the
dynamics of gas in galaxies:
\begin{equation}
\label{dens}
\frac{\partial \rho}{\partial t} + \nabla \cdot (\rho \vec{v}) = 0,
\end{equation}
\begin{equation}
\frac{\partial \rho \vec{v}}{\partial t} + \nabla \cdot (\vec{v} \rho \vec{v}) = - \nabla p - \rho \nabla \Phi,
\end{equation}
\begin{equation}
\frac{\partial \rho E}{\partial t} + \nabla \cdot (\rho E \vec{v}) = - (\nabla \cdot p\vec{v}) - \rho \nabla \Phi \cdot \vec{v},
\end{equation}
\begin{equation}
\frac{\partial \rho \epsilon}{\partial t} + \nabla \cdot (\rho \epsilon \vec{v}) = - (\gamma - 1)\rho \epsilon \nabla \cdot \vec{v},
\label{inten}
\end{equation}
where $p$ is the gas pressure, $\rho$ is the gas volume density, $\vec{v}$ is the gas velocity, 
$E$ is the total energy per unit mass, $\Phi$ is the gravitational potential computed 
using the Poisson equation $\Delta \Phi = 4\pi G \rho$. The gas pressure is related to 
the internal energy per unit mass using the following equation of state $p = (\gamma - 1)\rho \epsilon$, where $\gamma$ is the ratio of specific heats. The reason for using equations for both  $\epsilon$ 
and $E$ will be described later.

 {The entropy equation is sometimes used instead of the non-conservative
equation for internal energy equation \cite{Balsara_1999,Ryu_1993,Springel_2002,GodunovKulikov_2014}. This approach has a clear advantage in case of high Mach numbers \cite{Balsara_1999}. 
We decided to use the equation for internal energy, because this form can be easier expanded to include the effects of radiative cooling and heating, and also chemistry, which are important extensions of the classic hydrodynamics equations in astrophysics.}

The solver for system (\ref{dens}-\ref{inten}) is based on a combination of the operator-splitting method and Godunov method \cite{Vshivkov_2011_a}. 
The method consists of two steps: the Eulerian one and Lagrangian one. During the first (Eulerian) 
step, system (\ref{dens}-\ref{inten})  is solved excluding the advective terms (second terms on the
left-hand-side). During the second (Lagrangian) step, the advective transport of hydrodynamic variables $\rho$, $\vec{v}$, $E$, and $\epsilon$ are computed. 

Let us consider for simplicity a one-dimensional analogue to system ~(\ref{dens})-~(\ref{inten}) written in the following non-conservative form:
\begin{equation}
\frac{\partial}{\partial t} \left( 
\begin{array}{c}
\rho \\ 
v \\
p
\end{array}  \right) + 
\left(
\begin{array}{ccc}
v & \rho & 0 \\ 
0 & v & \rho^{-1} \\ 
0 & \gamma p & v
\end{array} 
\right)
\frac{\partial}{\partial x} \left( 
\begin{array}{c}
\rho \\ 
v \\
p
\end{array}  \right) = 0.
\label{noncons1dfull} 
\end{equation}
Here, we made use of the adopted equation of state to substitute the internal energy $\epsilon$
with the gas pressure $p$. This set of one-dimensional equations will be used for constructing the Riemann problem at the Eulerian step and advective transport at the Lagrangian step.

\subsection{The hydrodynamic solver at the Eulerian step}
Using the adopted equation of state, system~(\ref{noncons1dfull}) can be reduced to the following non-conservative matrix form:
\begin{equation}
\label{matrix1D}
\frac{\partial}{\partial t} \left( 
\begin{array}{c}
v \\
p
\end{array}  \right) + 
\left(
\begin{array}{cc}
0 & \rho^{-1} \\ 
\gamma p & 0
\end{array} 
\right)
\frac{\partial}{\partial x} \left( 
\begin{array}{c}
v \\
p
\end{array}  \right) = 0.
\end{equation}
We note that the continuity equation~(\ref{dens}) for the density $\rho$ contains only  Lagrangian terms.  {The traditionally known operator splitting methods do not justify going from 
equation~(\ref{noncons1dfull}) to equation~(\ref{matrix1D}), especially since 
equation~(\ref{matrix1D}) is in non-conservative form. However, its
usage here is, unfortunately, a compromise needed to remove the advective terms from 
equation~(\ref{noncons1dfull}). } 

We can solve the Riemann problem for this system using a Godunov-type method as described below.
Let us rewrite system~(\ref{matrix1D}) in the following form:
\begin{equation}
\label{hyperbolic}
\frac{\partial u}{\partial t} + B \frac{\partial u}{\partial x} = 0,
\end{equation}
where the vector  $u =\left( 
\begin{array}{c}
v \\
p
\end{array}  \right)$ and matrix $B$ is written as:
\begin{equation}
B = \left(
\begin{array}{cc}
0 & \widehat{\rho}^{-1} \\ 
\gamma \widehat{p}  & 0
\end{array} 
\right),
\end{equation}
where $\widehat{\rho}$ and $\widehat{p}$ are the space-averaged density and pressure on the cell interfaces. The method for computing these quantities is described later in this section.

Unfortunately, the matrix equation~(\ref{hyperbolic})  
{is not easy to solve analytically, because the matrix $B$ has a peculiar non-diagonal form}. 
We note, however, that system~(\ref{hyperbolic}) is hyperbolic and hence matrix $B$ 
can be represented as $B = R \times \Lambda \times L$, where $R$ and $L$ are the right-hand-side and left-hand-side eigenvector matrices,
respectively, and $\Lambda$ is a diagonal matrix of the eigenvalues of matrix $B$:
\begin{equation}
R = 
\left(
\begin{array}{cc}
\frac{1}{\sqrt{1 + \gamma \widehat{p} \widehat{\rho}}} & \frac{1}{\sqrt{1 + \gamma \widehat{p} \widehat{\rho}}} \\ 
\frac{\widehat{\rho} \sqrt{\gamma \widehat{p} / \widehat{\rho}}}{\sqrt{1 + \gamma \widehat{p} \widehat{\rho}}} & -\frac{\widehat{\rho} \sqrt{\gamma \widehat{p} / \widehat{\rho}}}{\sqrt{1 + \gamma \widehat{p} \widehat{\rho}}}
\end{array} 
\right),
\end{equation}
\begin{equation}
L = 
\left(
\begin{array}{cc}
\frac{ \widehat{\rho} \left( \widehat{\rho}^{-1} + \gamma \widehat{p} \right) }{ 2  \sqrt{1 + \gamma \widehat{p} \widehat{\rho}}} & \frac{ \left( \widehat{\rho}^{-1} + \gamma \widehat{p} \right) }{ 2  \sqrt{1 + \gamma \widehat{p} \widehat{\rho}} \sqrt{\gamma \widehat{p} / \widehat{\rho}} } \\
\frac{ \widehat{\rho} \left( \widehat{\rho}^{-1} + \gamma \widehat{p} \right) }{ 2  \sqrt{1 + \gamma \widehat{p} \widehat{\rho}}} & -\frac{ \left( \widehat{\rho}^{-1} + \gamma \widehat{p} \right) }{ 2  \sqrt{1 + \gamma \widehat{p} \widehat{\rho}} \sqrt{\gamma \widehat{p} / \widehat{\rho}} }
\end{array} 
\right),
\end{equation}
\begin{equation}
\Lambda = 
\left(
\begin{array}{cc}
\sqrt{\gamma \widehat{p} / \widehat{\rho}} &  0 \\ 
0 & -\sqrt{\gamma \widehat{p} / \widehat{\rho}}
\end{array} 
\right).
\end{equation}
Let us multiply system~(\ref{hyperbolic}) with matrix $L$:
\begin{equation}
L  \frac{\partial u}{\partial t} + L \times  R \times \Lambda \times L \frac{\partial u}{\partial x} = 0.
\end{equation}
Using the identity for eigenvectors $L\times R = R \times L = I$, where $I$ is the unit matrix, 
and making the substitution $w = Lu$, the last system can be written as:
\begin{equation}
\frac{\partial w}{\partial t} + \Lambda \frac{\partial w}{\partial x} = 0,
\label{hyperbolic_new}
\end{equation}
where $\Lambda$ is a diagonal, sign-definite matrix with eigenvalues $\lambda_{j}$ ($j$=1,2).
\begin{equation}
\label{lambdaeulerian}
\lambda_{1} = \sqrt{\gamma \widehat{p} / \widehat{\rho}} \qquad \lambda_{2} = -\sqrt{\gamma \widehat{p} / \widehat{\rho}}.
\end{equation}

The matrix equation~(\ref{hyperbolic_new})  {with the 
diagonal matrix $\Lambda$ can now be easily solved  analytically}. 
To do this, we need to define the values of $w$ on the left and right cell 
interfaces. In the case of piecewise-constant functions, the initial conditions for 
system~(\ref{hyperbolic_new}) can be written as:
\begin{equation}
w(x,0) = w^{0}(x) = \left\lbrace \begin{array}{c}
w^{L}, x < 0 \\ 
w^{R}, x > 0
\end{array} \right. = 
\left\lbrace \begin{array}{c}
L u^{L}, x < 0 \\ 
L u^{R}, x > 0
\end{array} \right. ,
\end{equation}
where $x$ is the x-coordinate of the cell interface, $u^{L}$ and $u^{R}$ are piecewise-constant initial conditions at the l.h.s. and r.h.s. of the cell interface and $w^{L} = L u^{L}$ and $w^{R} = L u^{R}$.

The last system has an analytic solution $w_{j}(x,t) = w^{0}_{j}(x - \lambda_{j} t)$, 
where $j=1,2$. Once $w_1(x,t)$ and $w_2(x,t)$ have been found, we proceed with determining the vector $u$  (i.e., the velocity and pressure) using the inverse substitution 
\begin{equation}
u(x,t) = R w(x,t),
\label{solution}
\end{equation}
where $w(x,t) =\left( 
\begin{array}{c}
w_{1}(x,t) \\
w_{2}(x,t)
\end{array}  \right)$. We note that $u(x,t) =\left( 
\begin{array}{c}
u_{1}(x,t) \\
u_{2}(x,t)
\end{array}  \right)$ can be used to find the solution of the Riemann problem.

Now, we proceed with defining the space-averaged elements of matrices $\Lambda$, $R$, and $L$ 
at the cell interfaces. To do this, we will use a modification of the Roe approach \cite{Roe_1997} and the average density and pressure on the cell interfaces can be written as:
\begin{equation}
\widehat{\rho} = \frac{\rho^{3/2}_{L} + \rho^{3/2}_{R}}{\sqrt{\rho_{L}} + \sqrt{\rho_{R}}} ,
\end{equation}
\begin{equation}
\widehat{p} = \frac{ p_{L}\sqrt{\rho_{L}} + p_{R}\sqrt{\rho_{R}} }{ \sqrt{\rho_{L}} + \sqrt{\rho_{R}} } .
\end{equation}
The reason for this modification of the Roe approach is that it allows for an accurate calculation 
of the boundary between the high- and low-density gas in isothermal flows. 
 {Let us consider the interface between the high- and low-density regions 
with density and pressure of gas equal to 1.0 and $10^{-4k}, k > 0$, respectively. Using the 
Roe approach \cite{Roe_1997}, the speed of sound at the interface is $\approx10^k$, while 
on the left and right hand side from the interface the sound speed $\approx 1.0$. In the case of 
our modification, the sound speed is $\approx 1.0$ everywhere, including at the interface. 
This modification allows us to manage higher Courant-Friedrichs-Lewy numbers. Additionally, 
with this modification we can better solve the problem of gas expanding into vacuum, as 
demonstrated in Section~4. }
Finally, the values of $\widehat{\rho}$ and $\widehat{p}$ are used to
find the vector $u(x,t)$ using equation~(\ref{solution}).

Now, we can proceed with solving one-dimensional 
equations of hydrodynamics (along the $x$-coordinate) at the Eulerian step 
in conservative variables
\begin{equation}
\label{densEulerian}
\frac{\partial \rho v}{\partial t} = - \frac{\partial p}{\partial x},
\end{equation}
\begin{equation}
\frac{\partial \rho E}{\partial t} = - \frac{\partial p v}{\partial x},
\end{equation}
\begin{equation}
\frac{\partial \rho \epsilon}{\partial t} = - (\gamma - 1)\rho \epsilon \frac{\partial v}{\partial x}.
\label{intenEulerian}
\end{equation}
The numerical scheme for the one-dimensional system (\ref{densEulerian})-(\ref{intenEulerian}) can be written in the following form:
\begin{equation}
\label{SCHdensEulerian}
\frac{ \rho^{n+1/2}_{i} - \rho^{n}_{i}}{\tau} = 0, 
\end{equation}
\begin{equation}
\frac{ \rho v^{n+1/2}_{i} - \rho v^{n}_{i}}{\tau} = -\frac{P_{i+1/2} - P_{i-1/2}}{h},
\end{equation}
\begin{equation}
\frac{ \rho E^{n+1/2}_{i} -\rho E^{n}_{i}}{\tau} = -\frac{P_{i+1/2} V_{i+1/2} - P_{i-1/2} V_{i-1/2}}{h},
\end{equation}
\begin{equation}
\frac{ \rho \epsilon^{n+1/2}_{i} - \rho \epsilon^{n}_{i}} {\tau} = - (\gamma - 1) \rho \epsilon^{n}_{i} \frac{V_{i+1/2} - V_{i-1/2}}{h},
\label{SCHintenEulerian}
\end{equation}
where the values with indices $i \pm 1/2$ are the solution of the Riemann problem at the cell interfaces, while the values with indices $i$ are hydrodynamic variables defined at the cell centers. The index $n$ defines the values at the current time step, while the index $n+1/2$ corresponds to the values updated during the Eulerian step.

The solution of the Riemann problem for the normal velocity and pressure at the cell interfaces can be written as:
\begin{equation}
V = u_1(0,\tau) = \frac{v_{L} + v_{R}}{2} + \frac{p_{L} - p_{R}}{2} \sqrt{\frac{ (\sqrt{\rho_{L}} + \sqrt{\rho_{R}})^2 }{ \gamma (\rho^{3/2}_{L} + \rho^{3/2}_{R}) (p_{L}\sqrt{\rho_{L}} + p_{R}\sqrt{\rho_{R}}) }} ,
\end{equation}
\begin{equation}
P = u_2(0,\tau)= \frac{p_{L} + p_{R}}{2} + \frac{v_{L} - v_{R}}{2} \sqrt{\frac{ \gamma (\rho^{3/2}_{L} + \rho^{3/2}_{R}) (p_{L}\sqrt{\rho_{L}} + p_{R}\sqrt{\rho_{R}}) }{ (\sqrt{\rho_{L}} + \sqrt{\rho_{R}})^2 } } .
\end{equation}

The time step $\tau$ is chosen using the Courant--Friedrichs--Lewy condition:
\begin{equation}
\frac{\tau \times \left( \vert v_{max} \vert + \vert c_{max} \vert \right)}{h} = CFL < 1 ,
\end{equation}
where $v_{max}$ is the maximal gas velocity, $c_{max}$ is the maximal sound speed, $CFL$ is the  Courant number. In our simulation the Courant number was chosen to be $CFL = 0.2$. 

\subsection{Low-dissipation hydrodynamic solver at the Eulerian step} 
In the previous section, we used a piecewise-constant function on 
every cell to define the initial conditions for system~(\ref{hyperbolic_new}):
\begin{equation}
w(x,0) = w^{0}(x) = \left\lbrace \begin{array}{c}
w^{L}, x < 0 \\ 
w^{R}, x > 0
\end{array} \right. = 
\left\lbrace \begin{array}{c}
L u^{L}, x < 0 \\ 
L u^{R}, x > 0
\end{array} \right. .
\end{equation}
This scheme may be rather dissipative, because it assumes a piecewise-constant function in every numerical
cell. We can improve the accuracy of our scheme by using a local three-point stencil (the current cell and one adjacent cell on each side) to construct a piecewise-parabolic function. Then the Riemann problem for the Eulerian stage can be formulated for piecewise-polynomial initial conditions.   In this case, the initial conditions can be written as:
\begin{equation}
w(x,0) = w^{0}(x) = \left\lbrace \begin{array}{c}
w^{L}(x), x < 0 \\ 
w^{R}(x), x > 0
\end{array} \right. = 
\left\lbrace \begin{array}{c}
L u^{L}(x), x < 0 \\ 
L u^{R}(x), x > 0
\end{array} \right. ,
\end{equation}
where $u^{L}(x)$ and $u^{R}(x)$ are piecewise-parabolic initial condition and $w^{L}(x) = L u^{L}(x)$ and $w^{R}(x) = L u^{R}(x)$.
The last system has an analytical solution $w_{j}(x,t) = w^{0}_{j}(x - \lambda_{j} t), j=1,2$,
where  the values of $\lambda_{j}$ are defined similar to Equation~(\ref{lambdaeulerian})
\begin{equation}
\lambda_{1,2} =  \pm \sqrt{\frac{ \gamma (p_{L}\sqrt{\rho_{L}} + p_{R}\sqrt{\rho_{R}}) }{\rho^{3/2}_{L} + \rho^{3/2}_{R}} }.
\end{equation}
We note that the analytical solution $w_{j}(x,t) = w^{0}_{j}(x - \lambda_{j} t), j=1,2$ is 
a piecewise-parabolic function rather than a piecewise-constant function considered in the previous
section. This approach is called the  piecewise-parabolic method on local stencil (PPML) and it was first applied in \cite{Ustyugov_2007,Ustyugov_2008}.
We make an inverse substitution and the solution of the Riemann problem for $u$ can be written as 
\begin{equation}
u(x,t) = R w(x,t),
\end{equation}
where $w(x,t) = \left(w_{1}(x,t), w_{2}(x,t) \right)$. Finally, the solution for the normal velocity and pressure at the cell interfaces can be written as:
\begin{equation}
\label{lowriemannvelocity}
V = \frac{v_{L}(-\lambda t) + v_{R}(\lambda t)}{2} + \frac{p_{L}(-\lambda t) - p_{R}(\lambda t)}{2} 
\sqrt{\frac{ (\sqrt{\rho_{L}} + \sqrt{\rho_{R}})^2 }{ \gamma (\rho^{3/2}_{L} + \rho^{3/2}_{R}) (p_{L}\sqrt{\rho_{L}} + p_{R}\sqrt{\rho_{R}}) }},
\end{equation}
\begin{equation}
\label{lowriemannpressure}
P = \frac{p_{L}(-\lambda t) + p_{R}(\lambda t)}{2} + \frac{v_{L}(-\lambda t) - v_{R}(\lambda t)}{2} 
\sqrt{\frac{ \gamma (\rho^{3/2}_{L} + \rho^{3/2}_{R}) (p_{L}\sqrt{\rho_{L}} + p_{R}\sqrt{\rho_{R}}) }{ (\sqrt{\rho_{L}} + \sqrt{\rho_{R}})^2 } },
\end{equation}
where the expressions for $p_{L}(-\lambda t)$, $p_{R}(\lambda t)$, $v_{L}(-\lambda t)$, $v_{R}(\lambda t)$ can be found in the Appendix. 

\subsection{Hydrodynamic solver at the Lagrangian step} 
At the Lagrangian step, hydrodynamic equations are written in the following form:
\begin{equation}
\frac{\partial \rho}{\partial t} + \nabla \cdot (\rho \vec{v}) = 0,
\end{equation}
\begin{equation}
\frac{\partial \rho \vec{v}}{\partial t} + \nabla \cdot (\vec{v} \rho \vec{v}) = 0,
\end{equation}
\begin{equation}
\frac{\partial \rho E}{\partial t} + \nabla \cdot (\rho E \vec{v}) = 0,
\end{equation}
\begin{equation}
\frac{\partial \rho \epsilon}{\partial t} + \nabla \cdot (\rho \epsilon \vec{v}) = 0.
\end{equation}
This system can be recast in the following general form:
\begin{equation}
\frac{\partial f}{\partial t} + \nabla \cdot (f \vec{v}) = 0,
\label{genlagrange}
\end{equation}
where $f$ can be the density $\rho$, momentum density $\rho \vec{v}$, total energy density $E$ or internal energy density $\epsilon$. The Lagrangian step describes advective transport 
of all hydrodynamics variables. 
To solve the system of equations~(\ref{genlagrange})  we will use a Godunov-type method. For the calculation of the fluxes $F = f \vec{v}$ at the cell interfaces we use a one-dimensional 
linearized analogue of system~(\ref{genlagrange})
\begin{equation}
\frac{\partial f}{\partial t} + \widehat{v_x} \frac{\partial f}{\partial x} = 0.
\end{equation}
This system has a trivial eigenvalue decomposition: the eigenvalue is equal $\lambda= \widehat{v_x}$. 
In this case, the eigenvalue is not a sign-definite variable, and therefore the solution can be written in the following form:
\begin{equation}
F = \widehat{v_x} \times \left\lbrace 
\begin{array}{c}
f_{L}(- \vert \lambda \vert t), \widehat{v_x} \geq 0 \\ 
f_{R}( \vert \lambda \vert t), \widehat{v_x} < 0
\end{array} 
\right. ,
\label{fluxonlagrangestage}
\end{equation}
where $f_{L}(-\vert \lambda \vert t)$ and $f_{R}(\vert \lambda \vert t)$ are piecewise-parabolic function and the space-averaged velocity at the cell interfaces can be written in the following form:
\begin{equation}
\label{spaceaveragevelocitylagrange}
\widehat{v_x} = \frac{ v_{L}\sqrt{\rho_{L}} + v_{R}\sqrt{\rho_{R}} }{ \sqrt{\rho_{L}} + \sqrt{\rho_{R}} } .
\end{equation}
We use space averaging similar to that applied to the density and pressure at the Eulerian stage.

The numerical scheme for a one-dimensional system ~(\ref{genlagrange}) can be written in the following form:
\begin{equation}
\frac{ f^{n+1}_{i} - f^{n+1/2}_{i}}{\tau} = -\frac{F_{i+1/2} - F_{i-1/2}}{h},
\end{equation}
where the values with indices $i \pm 1/2$ are the fluxes ~(\ref{fluxonlagrangestage}) through the corresponding cell interfaces, while the values with indices $i$ are the hydrodynamic variables defined at the cell centers. The indices $n+1/2$ denote the values taken from the Eulerian step and indices $n+1$ correspond to the values updated during the Lagrangian step. 

\begin{figure}
\centering
\begin{minipage}[h]{0.32\linewidth}
\center{ \includegraphics[bb = 0 0 477 477, width=1\linewidth]{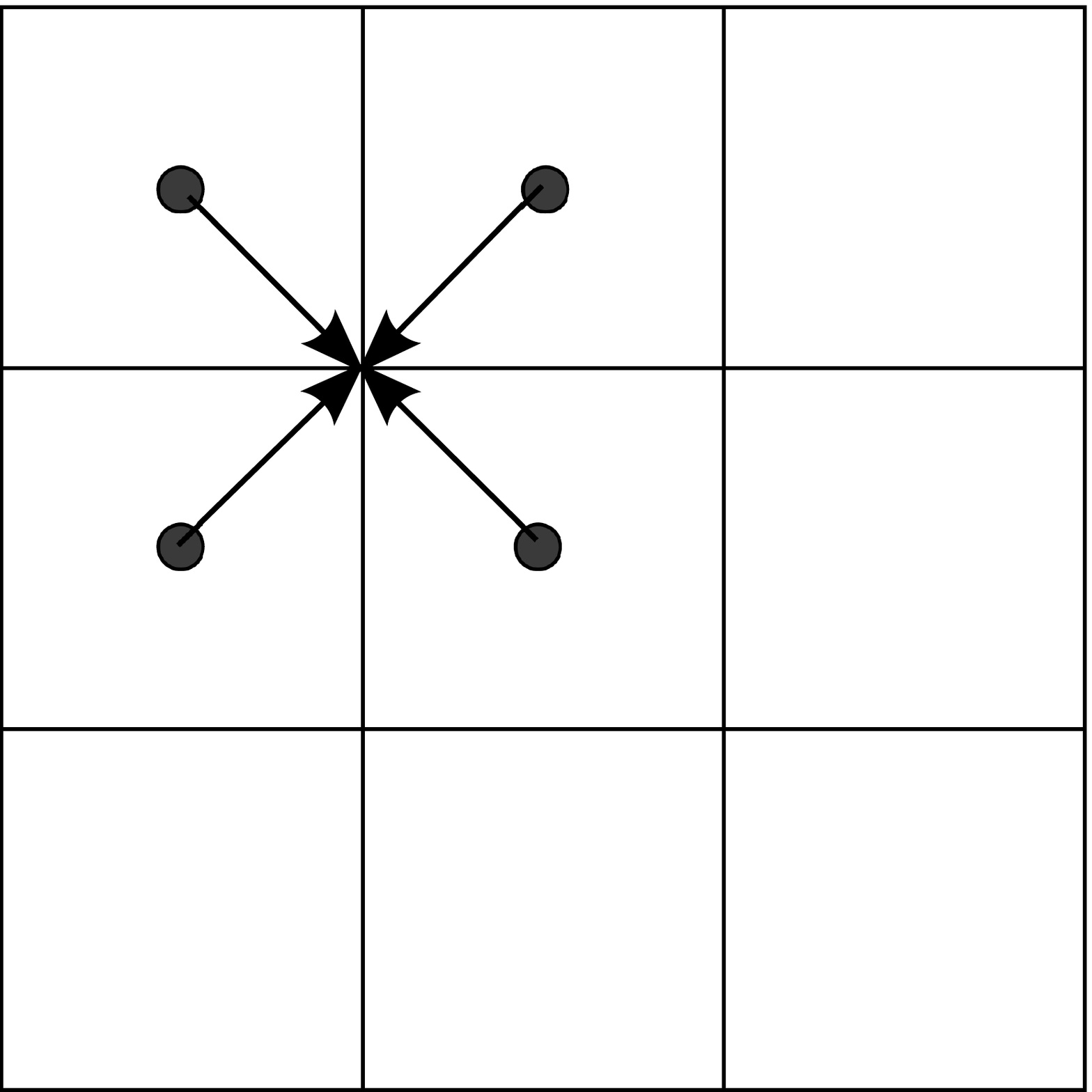} \\ (a)}
\end{minipage}
\hfill
\begin{minipage}[h]{0.32\linewidth}
\center{ \includegraphics[bb = 0 0 477 477, width=1\linewidth]{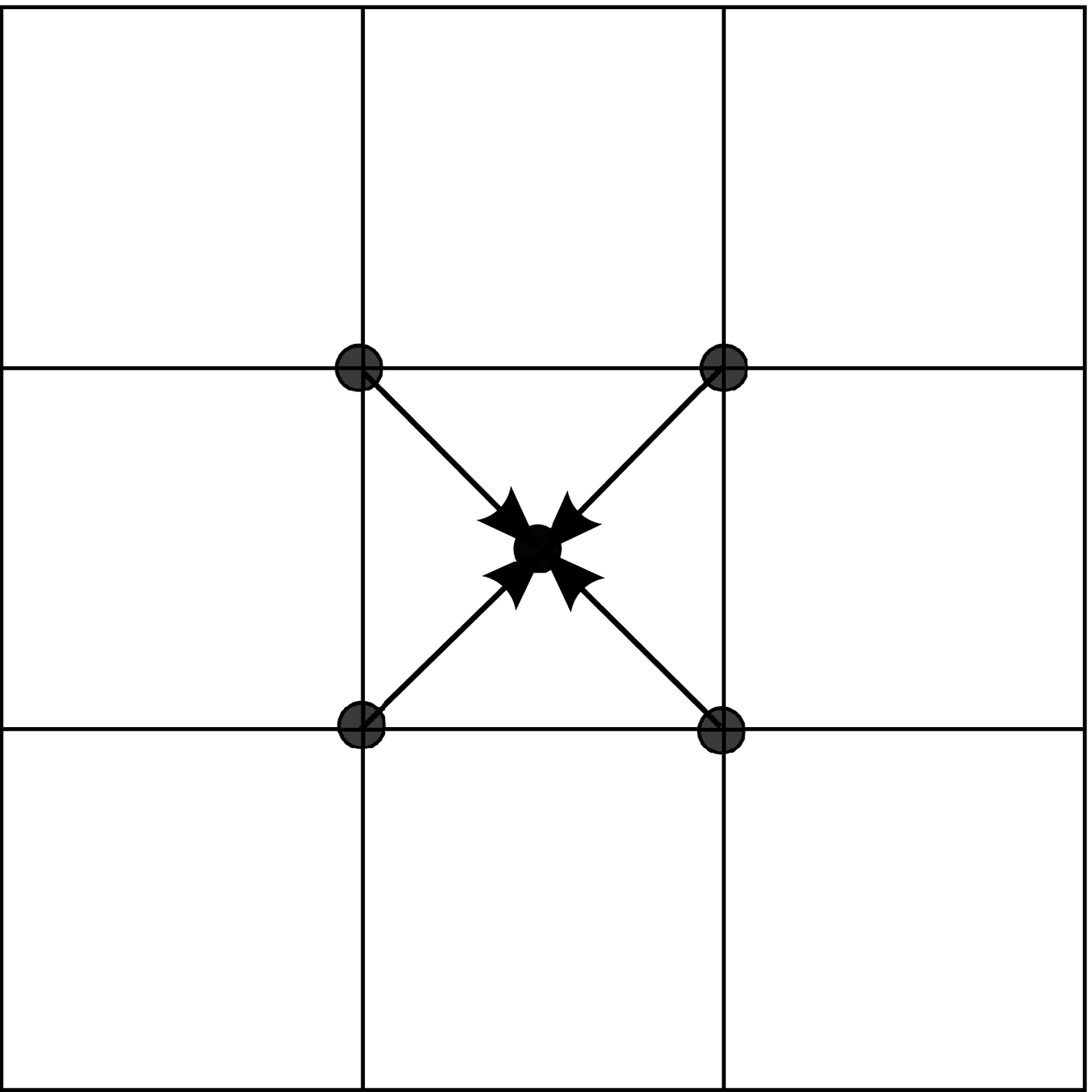} \\ (b)}
\end{minipage}
\hfill
\begin{minipage}[h]{0.32\linewidth}
\center{ \includegraphics[bb = 0 0 477 477, width=1\linewidth]{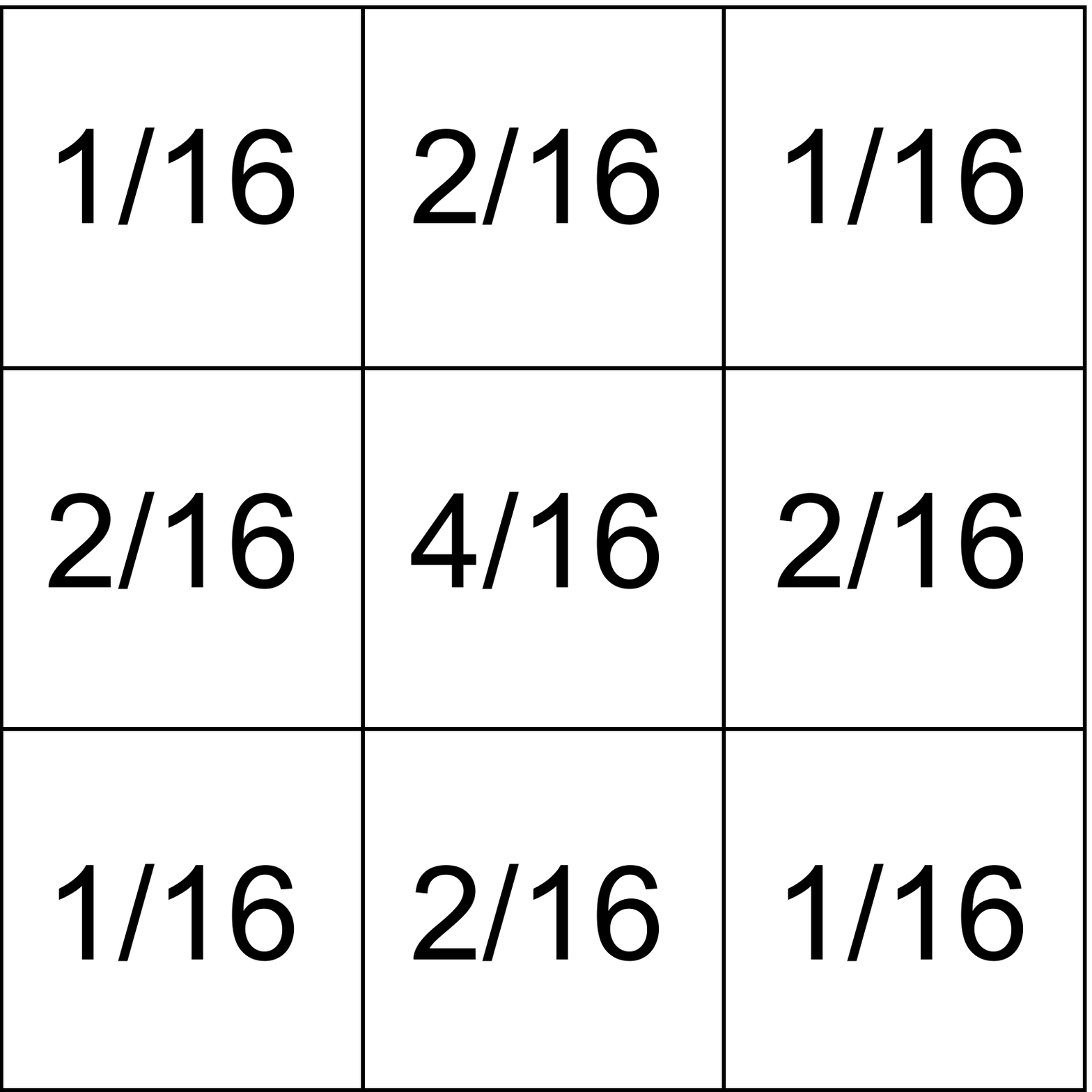} \\ (c)}
\end{minipage}
\caption{Scheme for calculating space-averaged velocities. Project of the normal velocities from the cell centers to the corner of cells (a), from the cells to the center (b) by means weight template (c).}
\label{LagrangeScheme}
\end{figure}

A significant disadvantage of this approach is that the values of 
$\widehat{v_x}$ are defined using only 
the information from  the cells in the $x$-direction, which is strictly valid only in the one-dimensional
case. In multi-dimensions, we use the following equation to calculate the space-averaged velocity 
$\widehat{v_x}$ 
\begin{equation}
\label{modspaceaveragevelocitylagrange}
\widehat{v_x} = \frac{ \widetilde{v_{L}}\sqrt{\rho_{L}} + \widetilde{v_{R}}\sqrt{\rho_{R}} }{ \sqrt{\rho_{L}} + \sqrt{\rho_{R}} },
\end{equation} 
where $\widetilde{v_{L}}$ and $\widetilde{v_{R}}$ are defined using a two-level scheme. At the first level, we project the normal velocities from the cell centers to the corner of the neighboring cells (see fig. \ref{LagrangeScheme}a), thus taking into account the information from the cells in the direction perpendicular to the $x$-direction.
At the second level, we project the normal velocities from the cell corners to the center of the given cell (see fig. \ref{LagrangeScheme}b). This  value will be used to define $\widetilde{v_{L}}$ and $\widetilde{v_{R}}$. The weight template is shown in Figure (\ref{LagrangeScheme}c). 
For convenience, we give an explicit formula to compute velocity $\widetilde{v}$ in cell $(i,k,l)$
\begin{equation}
\label{veltemplate}
\widetilde{v}_{(i,k,l)} = \frac{1}{16} \left( v_{i,k+1,l+1}+v_{i,k+1,l-1}+v_{i,k-1,l+1}+v_{i,k-1,l-1} \right) + 
\end{equation}
$$
\frac{1}{8} \left( v_{i,k+1,l}+v_{i,k-1,l}+v_{i,k,l+1}+v_{i,k,l-1} \right) + \frac{1}{4}v_{i,k,l}.
$$
 {
This approach is based on the geometric transformation described in \cite{Vshivkov_2007,Godunov_2013},
which takes into account fluxes from all possible directions. 
Of course, multidimensional Riemann solvers can also be used \cite{Balsara_2010,Balsara_2012b,Balsara_2014, Balsara_2014b,Balsara_2015,Balsara_2015b,Balsara_2015c,Boscheri_2014,Boscheri_2014b}, which 
may be important when solving the multidimensional magnetohydrodynamics equations.
}

 {
Now, we present the detailed algorithm for computing the Lagrangian step for hydrodynamic 
variables $f$ (density, momentum, internal and total energies) in each cell $(i,k,l)$ of the computational domain. Equation~(\ref{genlagrange}) in the Cartesian coordinates has the following form:
$$
\frac{ f^{n+1}_{i,k,l} - f^{n+1/2}_{i,k,l}}{\tau} = -\frac{F^x_{i+1/2,k,l} - F^x_{i-1/2,k,l}}{h_x}
-\frac{F^y_{i,k+1/2,l} - F^y_{i,k-1/2,l}}{h_y} -\frac{F^z_{i,k,l+1/2} - F^z_{i,k,l-1/2}}{h_z}.
$$
We consider the flux of $f$ along the $x$-direction ($i$-index), the other two fluxes can be 
computed by analogy. First, we calculate the values of $\widetilde{v}_{x}$ in cells $i-1$, $i$, 
and $i+1$ using the following equation:
$$
\widetilde{v}_{x,([i\pm1,i],k,l)} = \frac{1}{16} \left( v_{x,([i\pm1,i],k+1,l+1)}+v_{x,([i\pm1,i],k+1,l-1)}+v_{x,([i\pm1,i],k-1,l+1)}+v_{x,([i\pm1,i],k-1,l-1)} \right) + 
$$
$$
\frac{1}{8} \left( v_{x,([i\pm1,i],k+1,l)}+v_{x,([i\pm1,i],k-1,l)}+v_{x,([i\pm1,i],k,l+1)}+v_{x,([i\pm1,i],k,l-1)} \right) + \frac{1}{4}v_{x,([i\pm1,i],k,l)}.
$$ 
Second, we compute $\widehat{v}_x$ using Equation~(\ref{modspaceaveragevelocitylagrange}):
$$
\widehat{v}_{x,i+1/2,k,l} = \frac{ \widetilde{v}_{x,(i+1,k,l)} \sqrt{\rho_{i+1,k,l}} + \widetilde{v}_{x,(i,k,l)} \sqrt{\rho_{i,k,l}} }{ \sqrt{\rho_{i+1,k,l}} +  \sqrt{\rho_{i,k,l}} } ,
$$
$$
\widehat{v}_{x,i-1/2,k,l} = \frac{ \widetilde{v}_{x,(i-1,k,l)} \sqrt{\rho_{i-1,k,l}} + \widetilde{v}_{x,(i,k,l)} \sqrt{\rho_{i,k,l}} }{ \sqrt{\rho_{i-1,k,l}} +  \sqrt{\rho_{i,k,l}} } .
$$
Finally, the fluxes of hydrodynamic variables can be computed based on Equation~(\ref{fluxonlagrangestage})
as follows: 
$$
F^x_{i+1/2,k,l} = \widehat{v}_{x,i+1/2,k,l} \times \left\lbrace 
\begin{array}{c}
f_{i,k,l}(- \vert \widehat{v}_{x,i+1/2,k,l} \vert t), \widehat{v}_{x,i+1/2,k,l} \geq 0 \\ 
f_{i+1,k,l}( \vert \widehat{v}_{x,i+1/2,k,l} \vert t), \widehat{v}_{x,i+1/2,k,l} < 0
\end{array} 
\right. ,
$$
$$
F^x_{i-1/2,k,l} = \widehat{v}_{x,i-1/2,k,l} \times \left\lbrace 
\begin{array}{c}
f_{i-1,k,l}(- \vert \widehat{v}_{x,i-1/2,k,l} \vert t), \widehat{v}_{x,i-1/2,k,l} \geq 0 \\ 
f_{i,k,l}( \vert \widehat{v}_{x,i-1/2,k,l} \vert t), \widehat{v}_{x,i-1/2,k,l} < 0
\end{array} 
\right. .
$$
}

\begin{figure}
\centering
\begin{minipage}[h]{0.48\linewidth}
\center{ \includegraphics[bb = 20 20 310 215, clip, width=1\linewidth, height=0.74\textwidth]{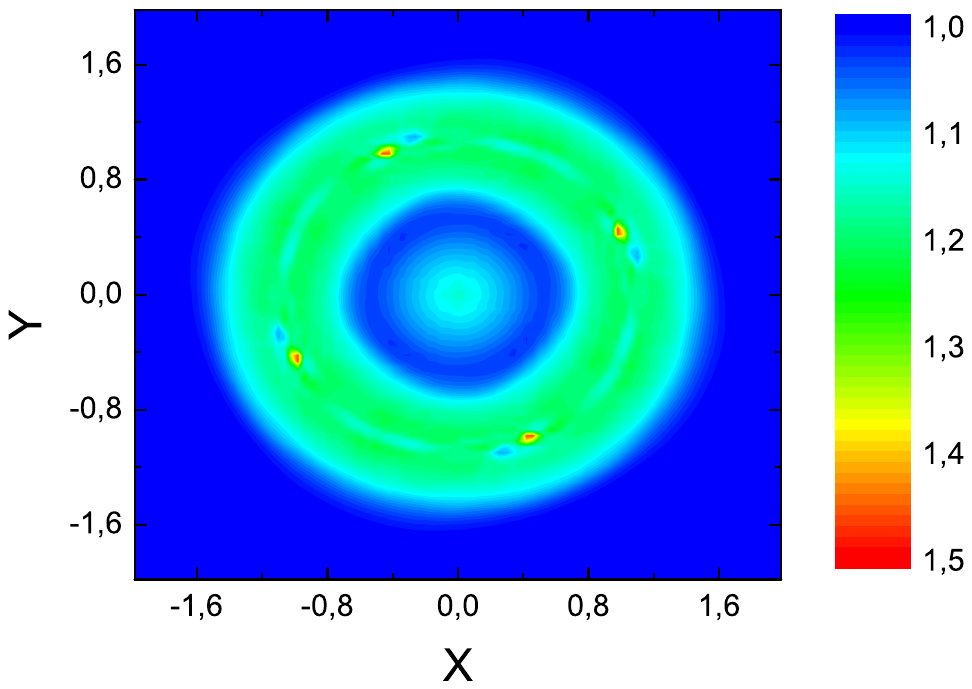} }
\end{minipage}
\begin{minipage}[h]{0.48\linewidth}
\center{ \includegraphics[bb = 20 20 310 215, clip, width=1\linewidth, height=0.74\textwidth]{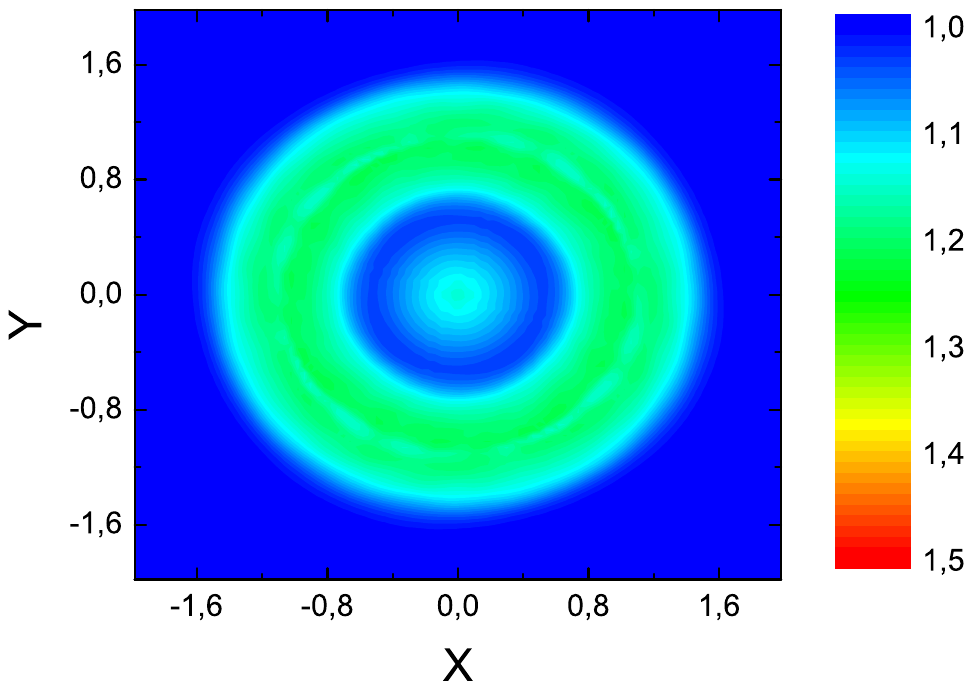} }
\end{minipage}
\caption{Rotation of a gaseous disk. {\bf Left}:  velocity is defined using 
Equation~(\ref{spaceaveragevelocitylagrange}). {\bf Right}: velocity is defined using
Equation~(\ref{modspaceaveragevelocitylagrange}).}
\label{LagrangeCompare}
\end{figure}

We compare the 1-point stencil and 9-point stencil approaches for calculating the velocities $u_{\rm
L}$ and $u_{\rm R}$ on the problem of rotating gaseous disk.  For this test problem, we define
the gas density and pressure as:
$$
p(r) = \rho(r) = 
\left\lbrace 
\begin{array}{c}
2 - r, r < 1 \\ 
1, r \geq 1
\end{array} 
\right. ,
$$
$$
\omega(r) = \left\lbrace 
\begin{array}{c}
1, r < 1 \\ 
0, r \geq 1
\end{array} 
\right. .
$$
Figure (\ref{LagrangeCompare}) shows the results of using of the 1-point stencil (left) and 9-point stencil (right) for the definition of velocity at the Lagrangian step. The red points visible in the left panel are the geometrical artifacts and they disappear when applying the 9-point stencil.

\subsection{The conservation of total energy}
Our system of hydrodynamics equations is overdefined. We use two approaches to conserve the total energy:
\begin{enumerate}
 \item Renormalization of the absolute value of the velocity, with each component of the velocity remaining unchanged (on boundary gas-vacuum) \cite{Vshivkov_2011_b}: $$\vert \vec{v} \vert = \sqrt{2(E-\epsilon)}.$$
 \item The internal energy (or entropy) correction \cite{GodunovKulikov_2014}: $$\vert \rho \epsilon \vert = \left( \rho E - \frac{\rho \vec{v}^{2}}{2} \right).$$
\end{enumerate}
In the very low density regions (more than five orders of magnitude lower than the mean density), 
we use the first approach, while in the other regions the second approach is applies.  Such a modification of the method keeps the detailed energy balance and guaranteed non-decrease of entropy. Similar approaches were used in \cite{Bryan_2014,Clarke_2010}.  {Other authors using approach for retaining positivity of pressure and conservation \cite{Balsara_2012,Zhang_2010}, which are particularly important at high Mach number.}

\subsection{The Poisson solver}
Our hydrodynamics equations include the effect of self-gravity. It is therefore necessary to compute
the gravitational potential using the gas current density distribution. In our method, we use
a finite-difference representation of the Poisson equation on the 27-point stencil:
\begin{equation}
-\frac{38}{9}\Phi_{i,k,l} +  
\label{stencil27point}
\end{equation}
$$
\frac{4}{9} \left( \Phi_{i-1,k,l} + \Phi_{i+1,k,l} + \Phi_{i,k-1,l} + \Phi_{i,k+1,l} + \Phi_{i,k,l-1} + \Phi_{i,k,l+1} \right) + 
$$
$$
\frac{1}{9} \left( \Phi_{i-1,k-1,l} +  \Phi_{i+1,k-1,l} +  \Phi_{i-1,k+1,l} +  \Phi_{i+1,k+1,l} + \Phi_{i-1,k,l-1} + \Phi_{i+1,k,l-1} \right. + 
$$
$$
\left. \Phi_{i-1,k,l+1} + \Phi_{i+1,k,l+1} + \Phi_{i,k-1,l-1} +  \Phi_{i,k+1,l-1} + \Phi_{i,k-1,l+1} + \Phi_{i,k+1,l+1} \right) + 
$$
$$
\frac{1}{36} \left( \Phi_{i-1,k-1,l-1} + \Phi_{i-1,k+1,l-1} + \Phi_{i-1,k-1,l+1} + \Phi_{i-1,k+1,l+1}  \right. +
$$
$$
\left. \Phi_{i+1,k-1,l-1} + \Phi_{i+1,k+1,l-1} + \Phi_{i+1,k-1,l+1} +  \Phi_{i+1,k+1,l+1} \right) = 4 \pi G h^2 \rho_{i,k,l} .
$$
The main motivation for the 27-point stencil is the Galilean invariance of the numerical solution. 
The algorithm for solving Equation~\ref{stencil27point} consists of the following several stages.

 {Stage 1. Formulation of boundary conditions for Poisson equation.} 
To find the gravitational potential at the boundaries, we use the first two terms of the multipole 
expansion: 
\begin{equation}
\Phi(x,y,z) \vert_{D} = -\frac{M}{r} - \frac{1}{r^3} \left( I_x + I_y + I_z - 3 I_0 \right) ,
\end{equation}
where
\begin{equation}
I_0 = \frac{\left( x^2 I_x + y^2 I_y + z^2 I_z \right) -2 \left( xy I_{xy} + xz I_{xz} + yz I_{yz} \right) }{r^2},
\end{equation}
\begin{equation}
I_x = \sum_{j} \left( z^2_j + y^2_j \right) m_j \qquad I_y = \sum_{j} \left( x^2_j + z^2_j \right) m_j \qquad I_z = \sum_{j} \left( x^2_j + y^2_j \right) m_j,
\end{equation}
\begin{equation}
I_{xy} = \sum_{j} x_j y_j m_j \qquad I_{xz} = \sum_{j} x_j z_j m_j \qquad I_{yz} = \sum_{j} y_j z_j m_j,
\end{equation}
where $x, y, z$ are the coordinates of the boundary cells, $r = \sqrt{x^2 + y^2 + z^2}$ the distance
from the boundary cells to the coordinate center, $x_j, y_j, z_j$ the coordinates of computational
cells, $m_j$ the mass in each cell, $M$ the total mass  in the computational domain, and the summation
is performed over all grid cells. Whenever boundary values of the potential appear on the left-hand
side of Equation~(\ref{stencil27point}), they are taken to the right-hand side and added to the 
density term.

 {Stage 2. The Fourier transformation of density to the harmonic space.}
In the second step, we take a direct Fourier transform of the density to obtain their 
harmonic amplitudes 
\begin{equation}
\sigma_{j,m,n} = \sum_{i,k,l} \rho_{i,k,l} exp \left( - \frac{\vec{i} 2\pi ij}{I} - \frac{\vec{i} 2\pi km}{K} - \frac{\vec{i} 2 \pi ln}{L} \right),
\end{equation}
where $I,K,L$ is number of cells along coordinates, $\vec{i}$ is the imaginary unit. 

 {Stage 3. The solution of Poisson equation in the harmonic space.}
We assume that the gravity is represented as a superposition of eigenfunctions of the Laplace operator:
\begin{equation}
\Phi_{i,k,l} = \sum_{j,m,n} \phi_{j,m,n} exp \left( \frac{\vec{i} 2\pi ij}{I} + \frac{\vec{i} 2\pi km}{K} + \frac{\vec{i} 2\pi ln}{L} \right).
\end{equation}
After substituting this expansion into equation~(\ref{stencil27point}), we obtain a simple formula 
to compute the harmonic amplitudes of the gravitational potential:
\begin{equation}
\phi_{j,m,n} = \frac{ \frac{2}{3} \pi h^2 \sigma_{j,m,n}}
{1- \left(1-\frac{2sin^{2}\frac{\pi j}{I}}{3}\right) \left(1-\frac{2sin^{2}\frac{\pi m}{K}}{3}\right) \left(1-\frac{2sin^{2}\frac{\pi n}{L}}{3} \right)}.
\end{equation}
To transform the gravitational potential from the harmonic space into the physical space, the inverse Fourier transform is used. For the numerical implementation of the Fast Fourier Transform, we use the FFTW library \cite{Frigo_1998}. 

The verification of the Poisson equation solver was done on the following density profile
$$
\rho(r) = \left\{
\begin{array}{@{\,}r@{\quad}l@{}}
2r^{3}-2r^{2}+1, & r \le 1  \\ 0, & r > 1 
\end{array}\right. ,
$$
which has the following analytical solution
$$
\Phi(r)= \left\{
\begin{array}{@{\,}r@{\quad}l@{}}
\frac{4\pi}{15}r^{5}-\frac{3\pi}{5}r^{4}+\frac{2\pi}{3}r^{2}-\frac{3\pi}{5}, & r \le 1 \\
 -\frac{4\pi}{15r}, & r > 1 
\end{array}\right. .
$$
 
Table~(\ref{PoissonTestSolver}) shows the behavior of the numerical solution on a sequence of meshes with increasing numerical resolution. Our method is fourth-order accurate.

\section{DISCUSSION}
{\bf The limiter problem.} The main trend in constructing accurate numerical methods (e.g. PPM, TVD and WENO) has been the use
of high-order polynomials in the interpolation schemes. However, this practice often produces 
unphysical oscillations around discontinuous solutions. To resolve this problem, the limiters are introduced to create a monotonic interpolation scheme and remove nonphysical extrema \cite{Goloviznin_1998,Shu_1988,ShuOsher_1988}. 
However, the use of different limiters results often in the incorrect calculation of the shock wave velocity. The operator splitting approach allows us to use  {limiters only 
when calculating piecewise parabolic functions in, e.g., equations~(33), (34), and (41)}. We do not provide  a rigorous proof here, but we simply state that this property of our scheme has been 
experimentally confirmed. 

{\bf The Roe average.} In our numerical scheme, we provide a modification of the classic 
Roe space-averaging of hydrodynamic variables, which behaves better on modelling the gas-vacuum 
boundary. In fact, the use of the classic Roe scheme for density at the gas-vacuum boundary 
(and, in general, at any high density and pressure jumps) yields too large sound speeds.
In general, the gas-vacuum boundary should be modelled by means a gas kinetic approach because
of low collision frequency between gas particles, but that it is technically not feasible today. 

{\bf Advantages and disadvantages of our numerical method.}
In our understanding, the advantages of our numerical scheme are:
1) accuracy on sufficiently smooth solutions and low dissipation  
on discontinuous solutions, 2) limiter-free and artificial-viscosity-free implementation,
3) Galilean invariance, 4) guaranteed non-decrease of the entropy \cite{GodunovKulikov_2014},
5) extensibility on other hyperbolic models, 6) simplicity of program implementation,
and 7) high scalability.
We note that the use of a computational mesh results in distortion of the numerical solution. 
In the case of complex nonlinear hydrodynamic flows, such distortions can lead to the artificial 
alignment of the numerical solution with the cell boundaries. The use of non-regular or adaptive meshes alleviates but does not resolve this problem. Therefore, it is desirable to construct Galilean 
invariant numerical schemes, which produce numerical solutions that does not depend on 
the orientation of the computational mesh. 
We think that our numerical method can be effectively implemented on GPUs, as the GPUPEGAS code \cite{Kulikov_2014}, and on the Intel Phi architecture, as the AstroPhi code \cite{Kulikov_2015}, because it uses the same algorithms.

Our  numerical method has disadvantages, the three most significant of which we discuss below:
\begin{enumerate}
\item A rather simple finite-difference scheme for the time derivative. By example, the WENO code uses
the Runge-Kutta fourth-order-accurate scheme for the time derivative. The advantage of such multilayer schemes is the expansion of  the computational stencil (on regular 3D mesh to $9^3$ cells). It is 
a very strong solution for the Galilean invariance problem. 
\item The construction of the interpolation parabola for problems with complex geometries. When using non-regular meshes (e.g. triangle cells), the construction of piecewise parabola 
is a non-trivial problem, which requires the use of a special spline technique adapted to cells 
with an arbitrary simplex of cells.
\item At the Eulerian step we solve the Riemann problem, which in turn relies on the analytic solution of the spectral problem (because the numerical solution of the eigenvalue/eigenvector problem is ill-posed). Finding the analytic solution for any hyperbolic equations is difficult. 
A possible solution is to use ''the potentials technique'' \cite{Godunov_2013}, but this technique requires the singular value decomposition of matrices, which is very expensive.
\end{enumerate}

{\bf The future work.}
In the future, we plan to expand our numerical method to ''collisionless'' hydrodynamics, which solves
for the first three moments of the collisionless Boltzmann equation \cite{Mitchell_2013,Vorobyov_2006}.
The main features of this expansion will be the formulation of the equation of state for 
collisionless component and the development of thermodynamically consistent star formation process 
with guaranteed non-decrease of the entropy.  

\section{THE VERIFICATION OF NUMERICAL METHOD}
In this section, we test the performance of our numerical code using a comprehensive set of test problems.  {For all numerical simulations, a CFL number equal to 0.2 was used.}

\subsection{The Sod shock-tube problem}
\begin{figure}
\centering
\includegraphics[bb = 0 0 300 230, width=0.5\linewidth]{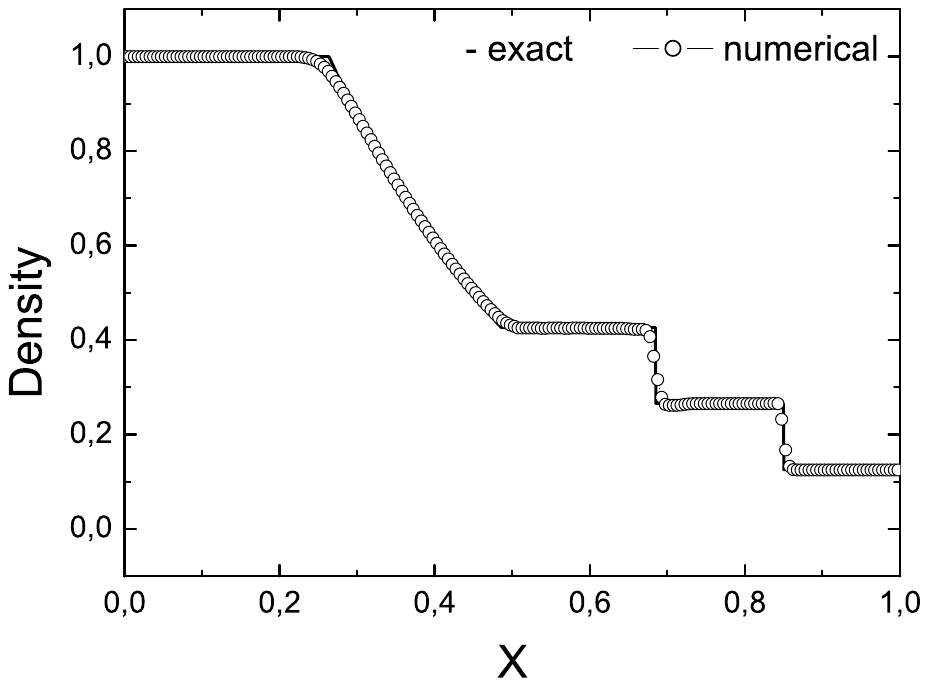}
\includegraphics[bb = 0 0 300 230, width=0.5\linewidth]{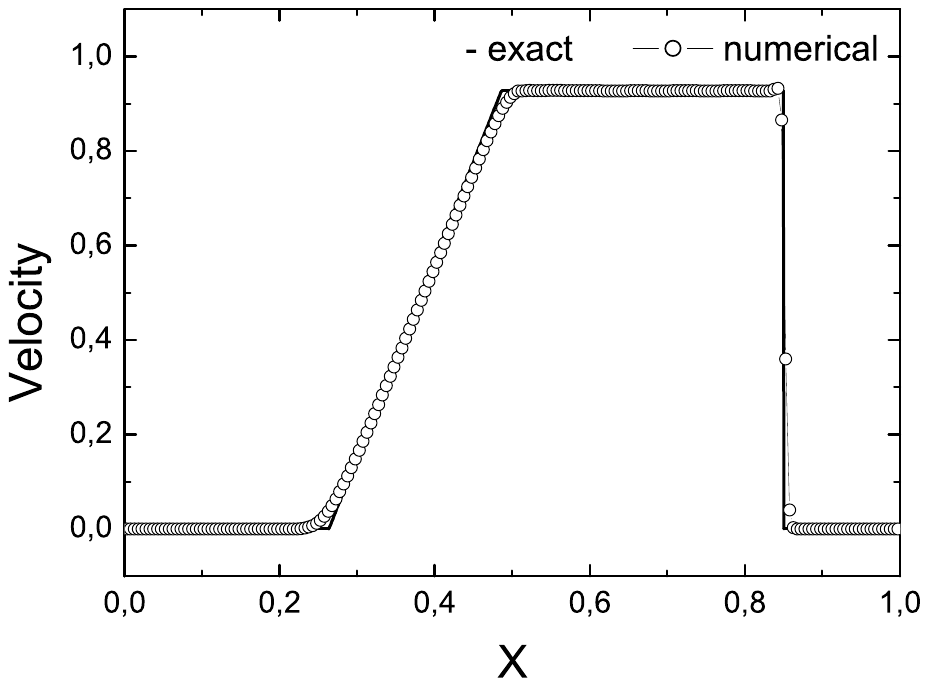} 
\includegraphics[bb = 0 0 300 230, width=0.5\linewidth]{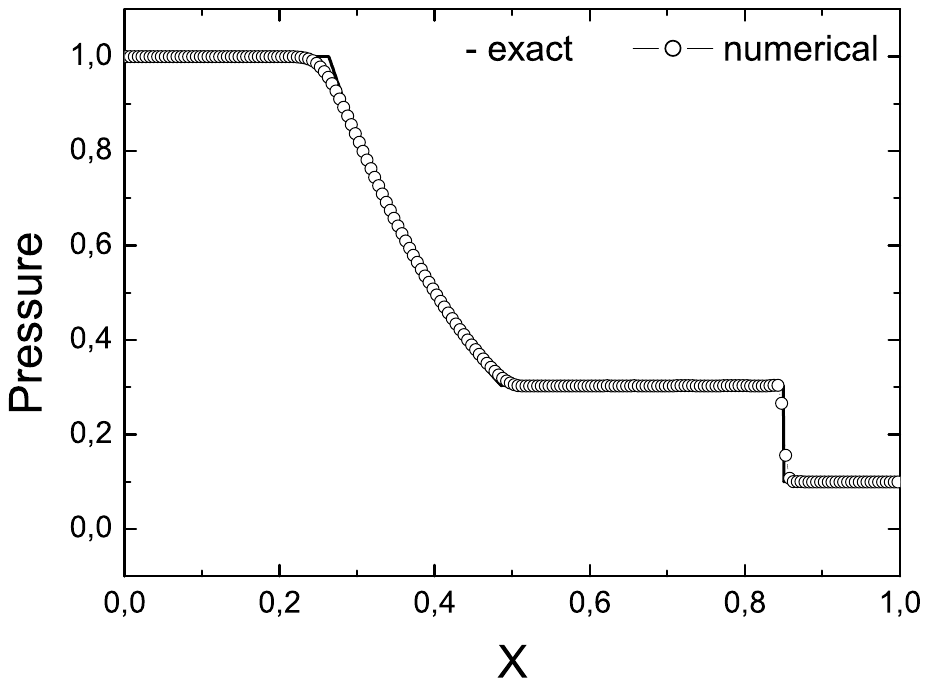} 
\caption{Results of the first shock tube problem.}
\label{ShockTubeSimulation1}
\end{figure}

\begin{figure}
\centering
\includegraphics[bb = 0 0 300 230, width=0.5\linewidth]{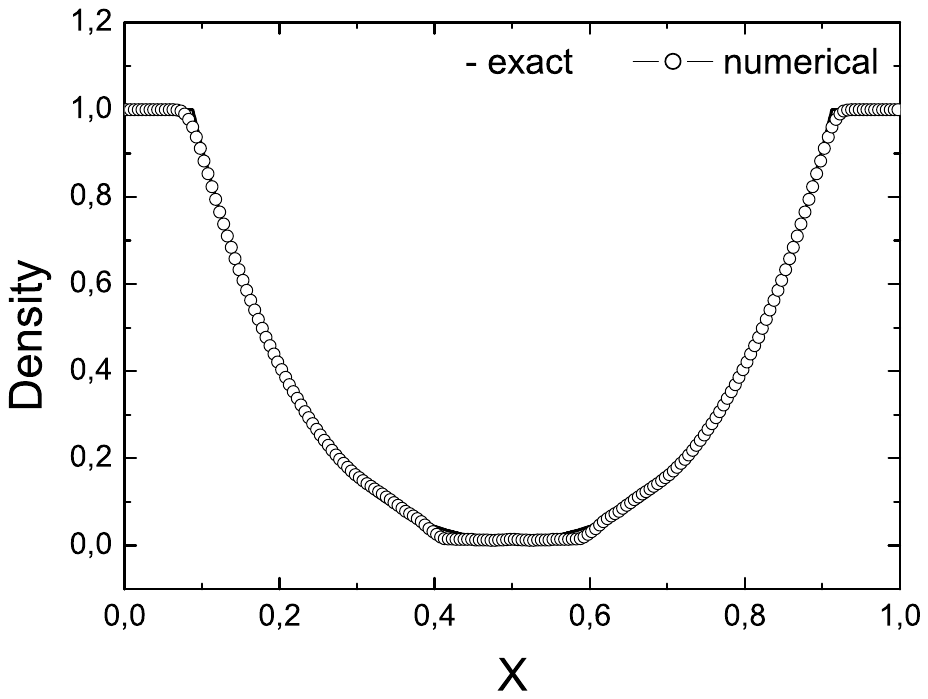}
\includegraphics[bb = 0 0 300 230, width=0.5\linewidth]{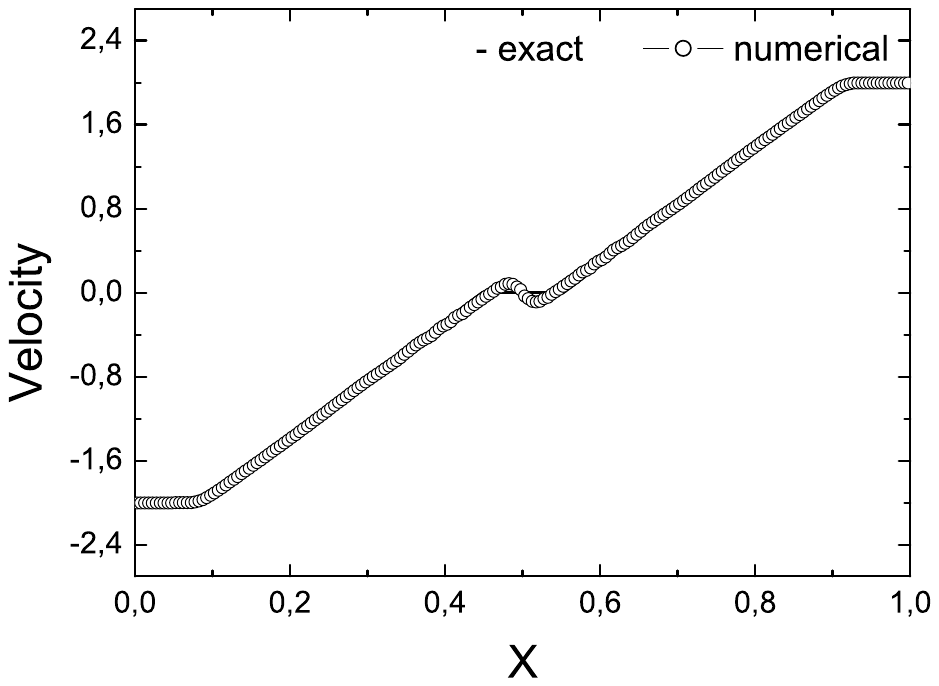}
\includegraphics[bb = 0 0 300 230, width=0.5\linewidth]{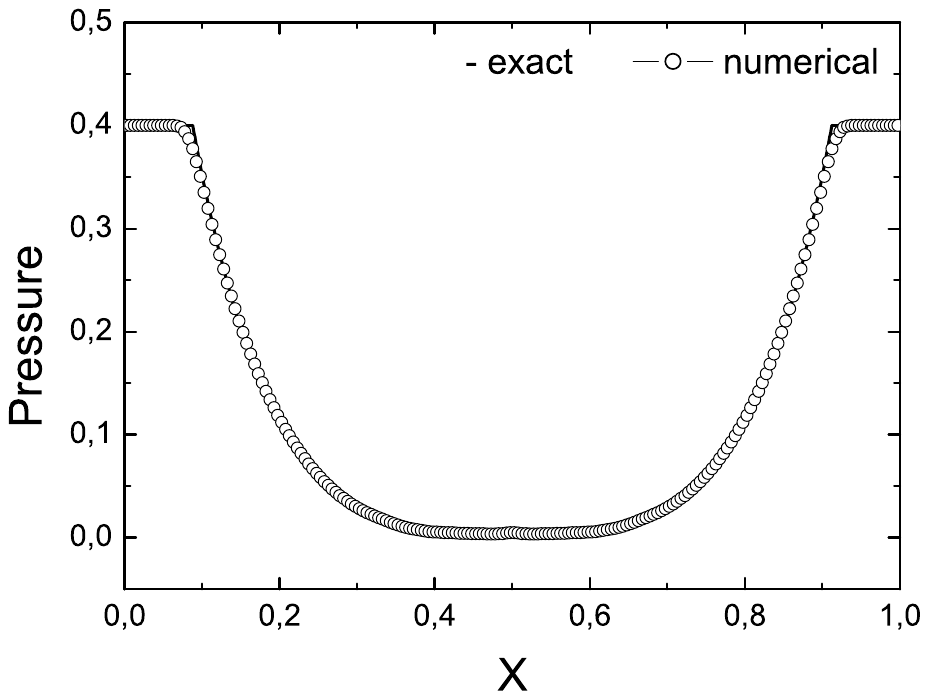} 
\caption{Results of the second shock tube problem.}
\label{ShockTubeSimulation2}
\end{figure}

\begin{figure}
\centering
\includegraphics[bb = 0 0 300 230, width=0.5\linewidth]{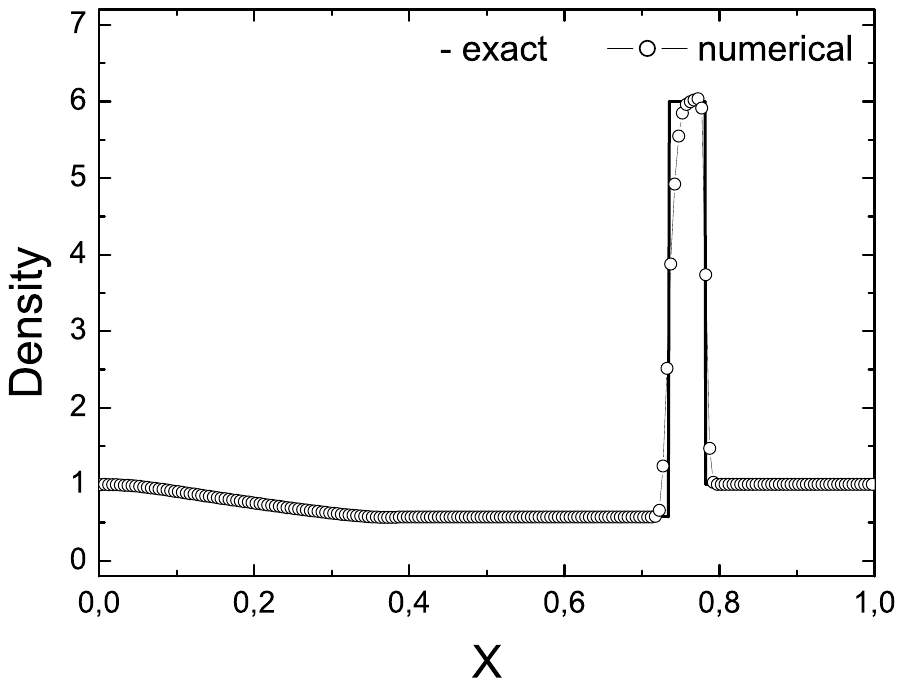}
\includegraphics[bb = 0 0 300 230, width=0.5\linewidth]{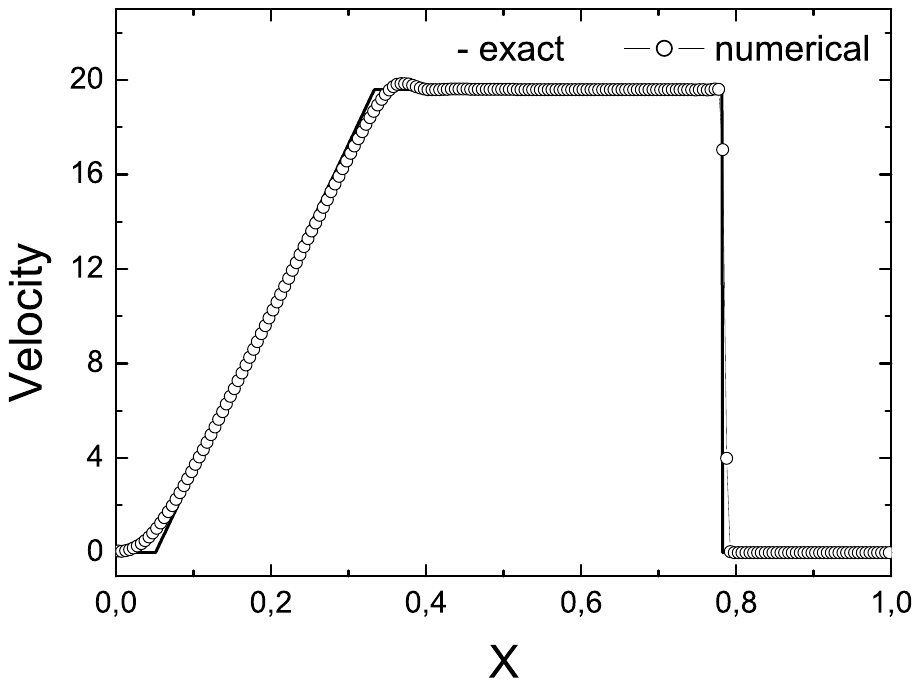} 
\includegraphics[bb = 0 0 300 230, width=0.5\linewidth]{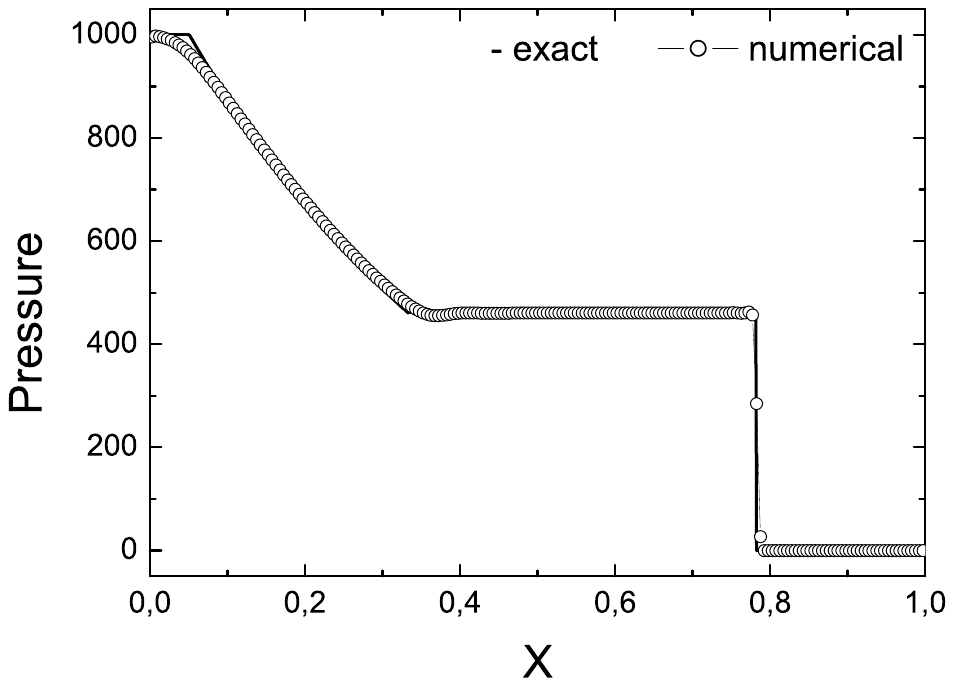}
\caption{Results of the third shock tube problem.}
\label{ShockTubeSimulation3}
\end{figure}

\begin{figure}
\centering
\includegraphics[bb = 0 0 300 230, width=0.48\linewidth]{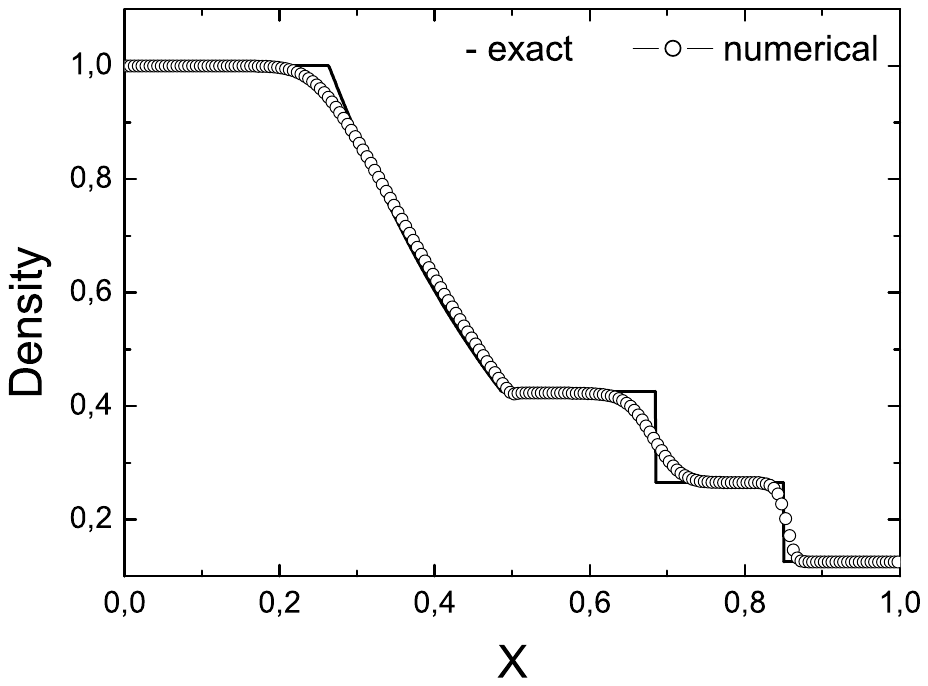}
\includegraphics[bb = 0 0 300 230, width=0.48\linewidth]{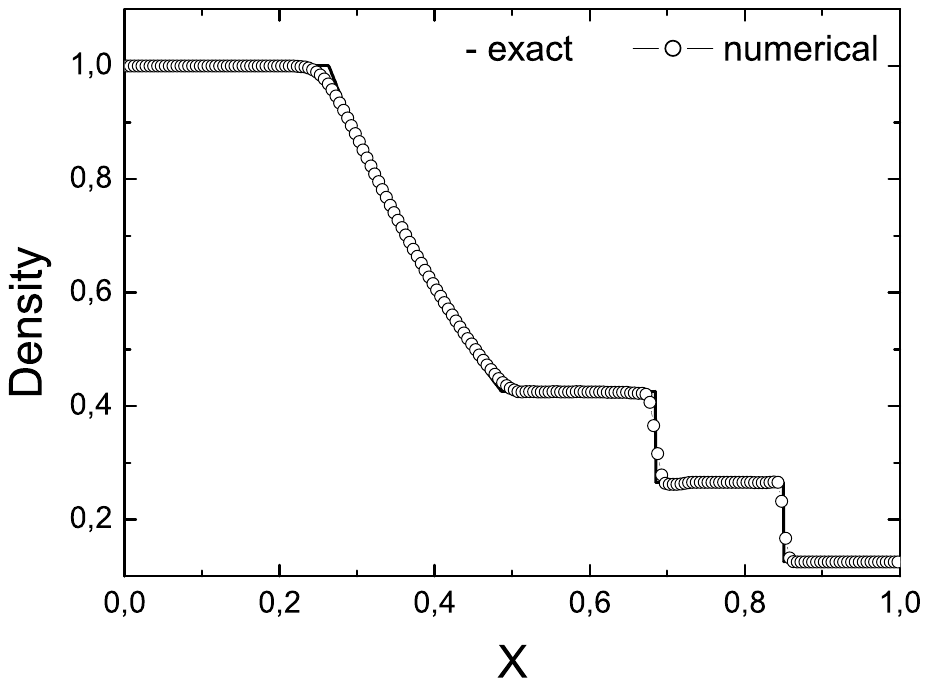} 
\caption{Comparison of the first-order (left) and higher-order (right) methods using the first
Sod shock tube problem.}
\label{ShockTubeCompare}
\end{figure}

We choose the shock-tube problem to demonstrate the advantages of using our high-order method for 
simulating discontinuities with low dissipation. This test can also highlight problems with 
simulating the rarefaction waves, since many methods are known to produce a non-physical increase 
of the internal energy in the rarefaction region. Moreover, a large initial drop of pressure 
(five orders of magnitude) is the standard robustness test showing the capability of numerical methods to simulate the strong perturbations with quickly spreading shock waves. The initial configurations for three different tests are shown in Table (\ref{ShockTubeProblem}), where $x_0$ is the position of the interface between two initial states and indexes L and R represent the l.h.s. and r.h.s. states, respectively. For this test, we use hundred grid cells.
The results of simulations are shown in Figures (\ref{ShockTubeSimulation1}), (\ref{ShockTubeSimulation2}),
and  (\ref{ShockTubeSimulation3}), and the comparison of the high-order and first-order methods 
is presented in Figure (\ref{ShockTubeCompare}). 

The main purpose of the first test shown in Figure (\ref{ShockTubeSimulation1}) is the correct simulation of the shock wave region. Our numerical method performs well on this problem, showing low dissipation and no unphysical oscillations. When compared with the first-order method, the number of cells over which the shock is spread reduces from 12 to just 
two  (see Figure \ref{ShockTubeCompare}). The main purpose of the second test shown in Figure 
(\ref{ShockTubeSimulation2}) is to test the code's ability to simulate correctly the rarefaction region. The main purpose of the third test shown in Figure (\ref{ShockTubeSimulation3}) is to test the code's ability to simulate correctly the strong shock wave. Our numerical method correctly reproduces all types of solutions.

\subsection{Simulation of the Kelvin-Helmholtz and Releigh-Taylor instabilities.}
It is important to show that our numerical method does not suppress physical instabilities that may develop during numerical simulations. To check this, we investigated the growth of the Rayleigh-Taylor and Kelvin-Helmholtz instabilities. The Rayleigh-Taylor instability arises at the interface of two fluids with different densities when the lighter fluid pushes against the heavier one due to 
gravitational acceleration and the Kelvin-Helmholtz instability arises at the interface of two fluids moving with respect to each other. Both instabilities give rise to nonlinear hydrodynamic turbulence.

\begin{figure}
\centering
\begin{minipage}[h]{0.48\linewidth}
\center{ \includegraphics[bb = 20 20 310 220, clip, width=1\linewidth, height=0.74\textwidth]{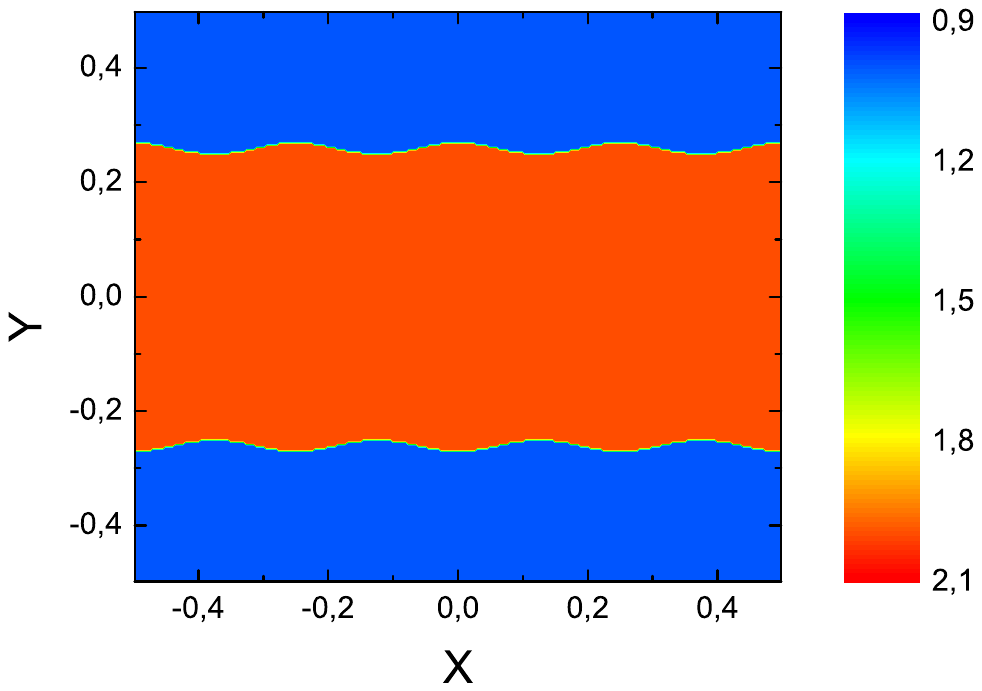} }
\end{minipage}
\begin{minipage}[h]{0.48\linewidth}
\center{ \includegraphics[bb = 20 20 310 220, clip, width=1\linewidth, height=0.74\textwidth]{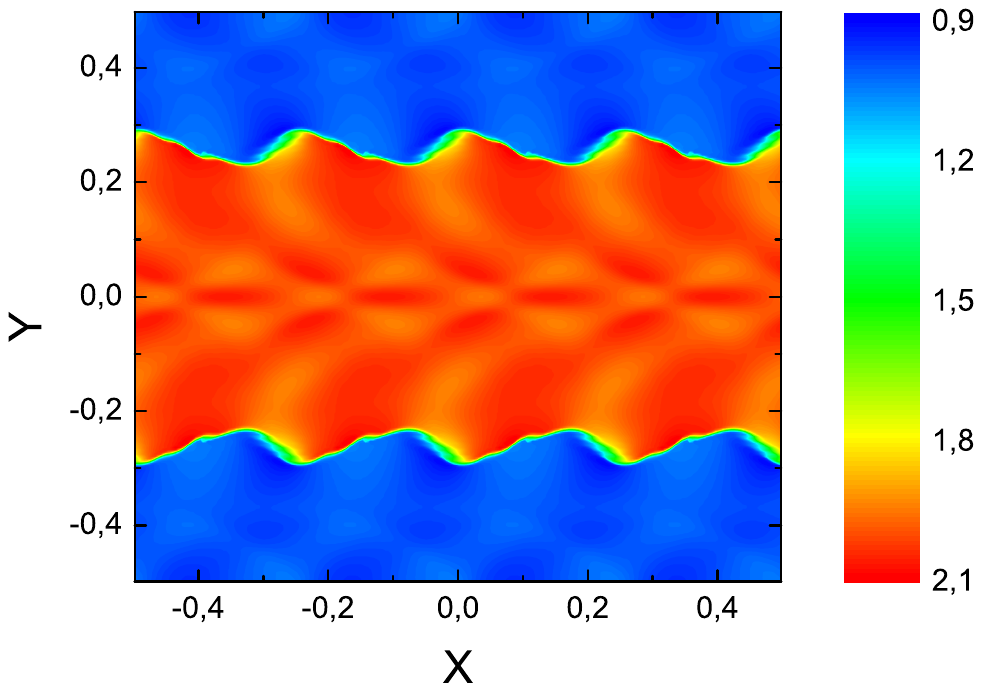}  }
\end{minipage}
\begin{minipage}[h]{0.48\linewidth}
\center{ \includegraphics[bb = 20 20 310 220, clip, width=1\linewidth, height=0.74\textwidth]{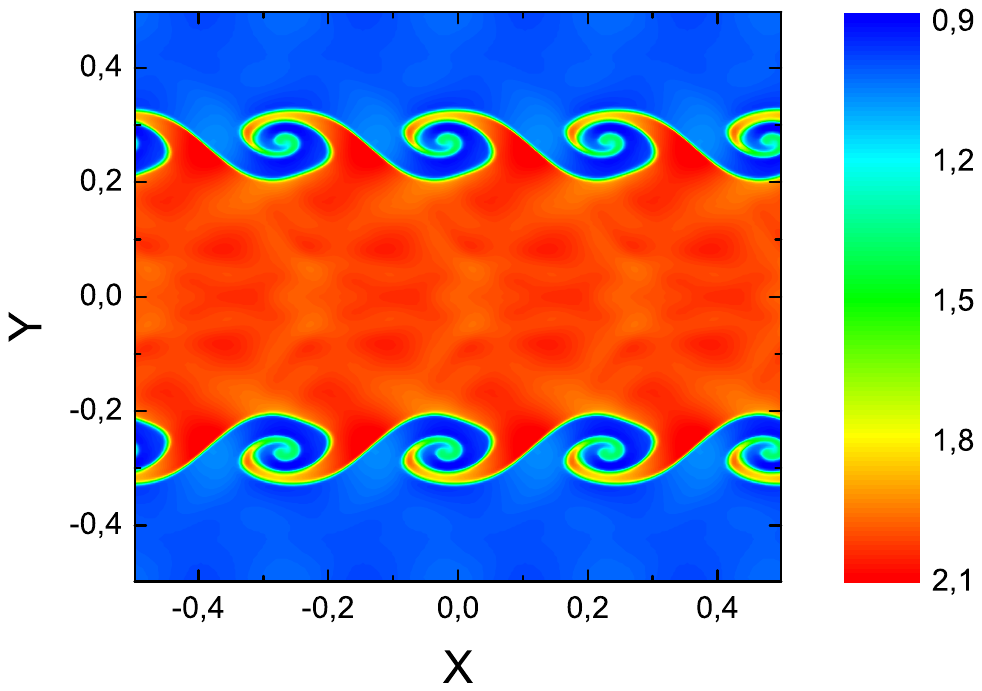} }
\end{minipage}
\begin{minipage}[h]{0.48\linewidth}
\center{ \includegraphics[bb = 20 20 310 220, clip, width=1\linewidth, height=0.74\textwidth]{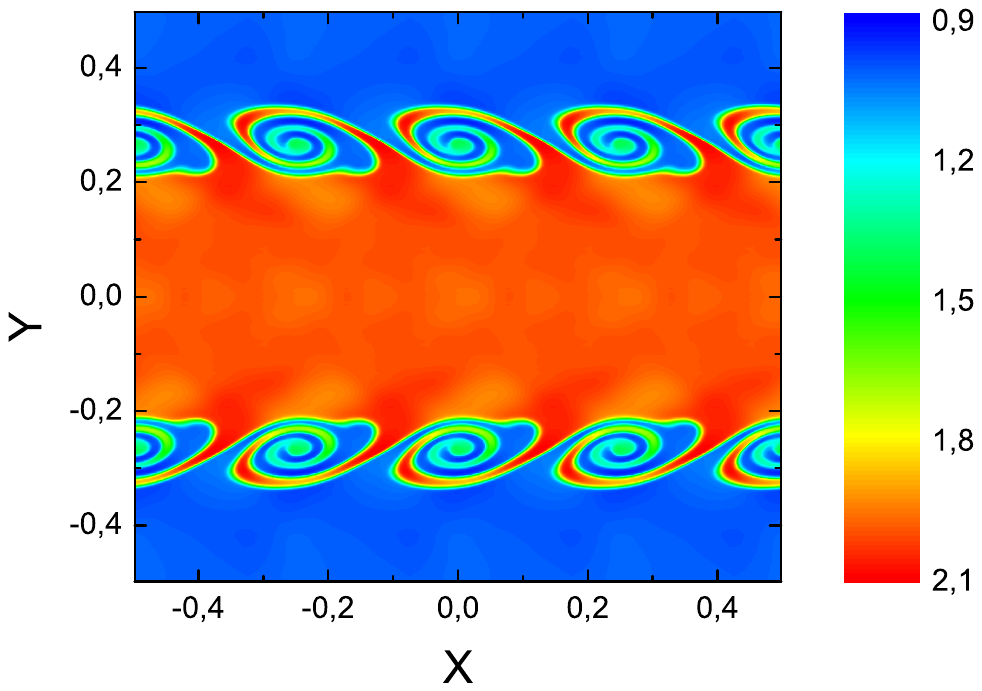} }
\end{minipage}
\caption{Development of the Kelvin-Helmholtz instability. The evolution times are 0.0 
(top-left), 0.2 (top-right), 0.5 (bottom-left) and 0.8 (bottom-right).}
\label{KHInstability}
\end{figure}

To simulate the development of the Kelvin-Helmholtz instability, we set up a square region with size of $[-0.5:0.5]^2$. The initial density and the x-component of velocity are chosen as: 
$$
\rho_{0}(x,y) = \left\{
\begin{array}{c}
1, \vert y \vert > 0.25 + 0.01 (1 + cos(8 \pi x)) \\ 
2, \vert y \vert \leq 0.25 + 0.01 (1 + cos(8 \pi x))
\end{array} 
\right. ,
$$
$$
v_{x,0}(x,y) = \left\{
\begin{array}{c}
-0.5, \vert y \vert > 0.25 + 0.01 (1 + cos(8 \pi x)) \\ 
0.5, \vert y \vert \leq 0.25 + 0.01 (1 + cos(8 \pi x))
\end{array} 
\right. .
$$
 {Here, the perturbation to the interface between two fluids is set by a cosine wave with
amplitude $A=0.01$ and wave number $k=4$.}
The gas pressure is set to $p = 2.5$ and the adiabatic index is set to $\gamma = 1.4$. A small 
sinusoidal perturbation is given to both the gas density and velocity in order to initiate the growth of the instability.  {For numerical simulations, a computational mesh of $256^2$ grid zones was used.} The development of the Kelvin-Helmholtz instability  is shown in Figure (\ref{KHInstability}). To calculate the characteristic growth time of the Kelvin-Helmholtz 
instability we use the following formula:
\begin{equation}
\tau_{KH} = \frac{\lambda \left( \rho_{in} + \rho_{out} \right)}{v_{rel} \sqrt{\rho_{in} \rho_{out}}},
\end{equation} 
where $\lambda = 1/4$ is the inverse frequency of the sinusoidal perturbation, $v_{rel} = 1$ the 
relativity velocity between both fluids, $\rho_{in} = 2$ the density in the inner region, and $\rho_{out} = 1$ the density in the outer region. For these parameters, $\tau_{KH}\approx 0.53$, which is in 
good agreement with our test simulations shown in Figure~(\ref{KHInstability}). Indeed, fully developed
turbulent eddies at the interface between two fluids occur at $t>0.5$. 

To simulate the Rayleigh-Taylor instability, we set the $[-0.25;0.25] \times [-1.5;1.5]$ box with the
following parameters:
$$
\rho_{0}(x,y) = \left\{
\begin{array}{c}
1, \vert y \vert > 0.75 \\ 
2, \vert y \vert \leq 0.75
\end{array} 
\right. ,
$$
$$
p = 0.15 - \rho \cdot g \cdot \vert y \vert,
$$ 
where $p$ is the pressure in hydrostatic equilibrium, $g=0.1$ the acceleration of free fall, 
$y$ the vertical coordinate, and  $\gamma = 1.4$. 

\begin{figure}
\centering
\includegraphics[bb = 0 0 165 230, clip, width=0.4\linewidth]{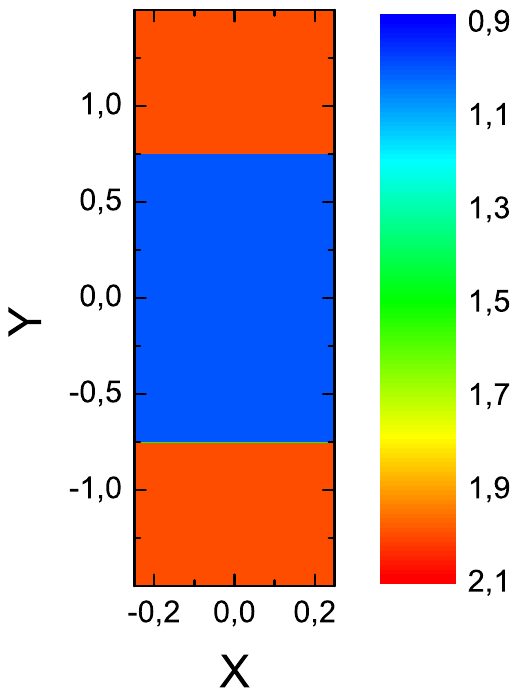}
\includegraphics[bb = 0 0 165 230, clip, width=0.4\linewidth]{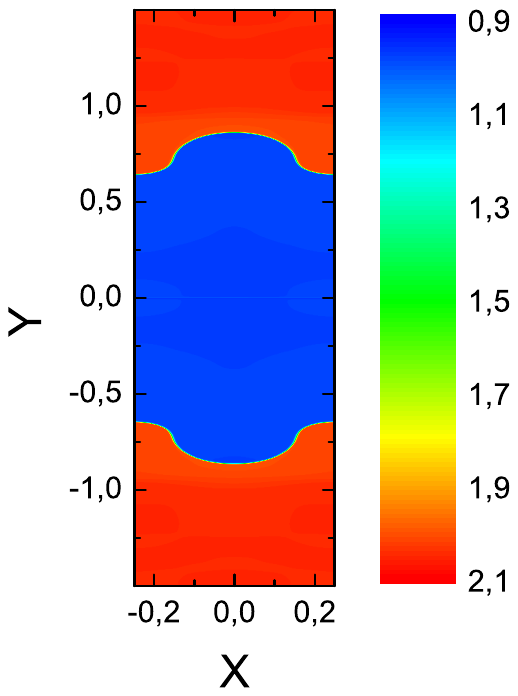} 
\includegraphics[bb = 0 0 165 230, clip, width=0.4\linewidth]{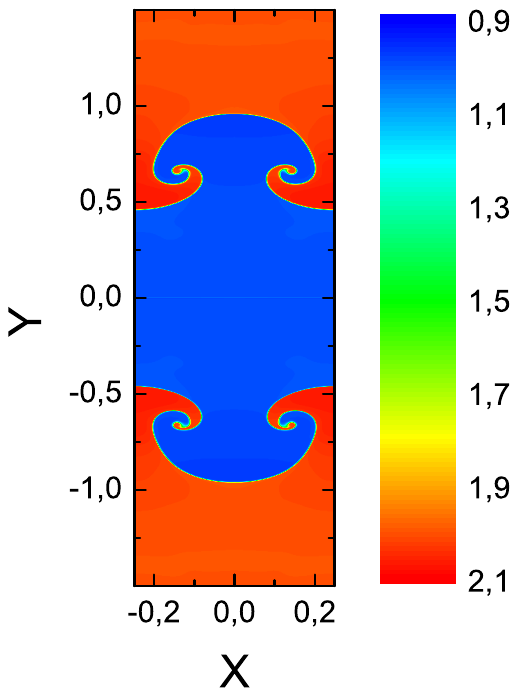}
\includegraphics[bb = 0 0 165 230, clip, width=0.4\linewidth]{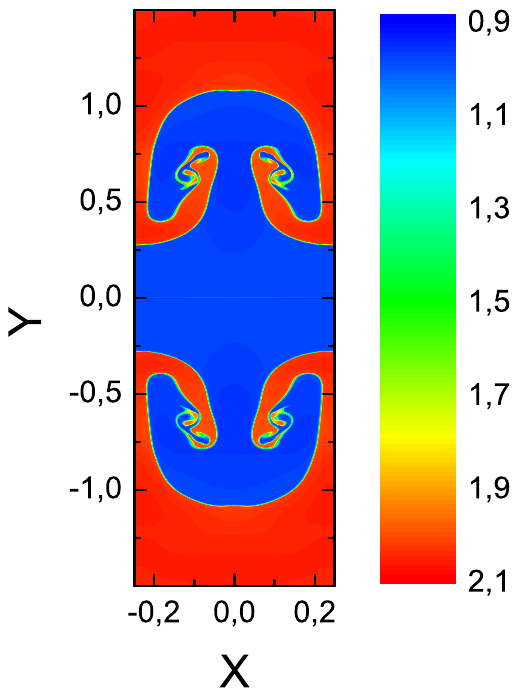}
\caption{Development of the Rayleigh-Taylor instability. The evolution times are 0.0 (top-left), 
7.0 (top-right), 10.0 (bottom-left), and 13.0 (bottom-right).}
\label{RTInstability}
\end{figure}

The initial hydrostatic equilibrium receives a perturbation
of the form: $v_{y,0}(x,y) = A(\vert y \vert - 0.75)[1+cos(2 \pi x)][1+cos(2 \pi y)]$, where
$$A(y) = \left\{
\begin{array}{c}
10^{-2}, \vert y \vert \le 0.01 \\ 
0, \vert y \vert > 0.01
\end{array} 
\right. .
$$
 {For numerical simulations, a computational mesh of $100 \times 600$ grid zones was used.} The development of the Rayleigh-Taylor instability is shown in Figure (\ref{RTInstability}).
To analyse the growth of the Rayleigh-Taylor instability, 
we use the following formula for the perturbation amplitude
\begin{equation}
p(t) = 0.01 exp \left( t \sqrt{A g} \right) \simeq 0.01  exp \left( 0.25 t \right),
\end{equation}
where $A = 2/3$ is the Atwood number. For example, at $t = 13$ the perturbation amplitude is 
$p(t=13) \simeq 0.25$, which is in good agreement with our simulations. Indeed, the initial location
of the interface between two fluids is at $y=\pm0.75$. By the time $t=13$, the density 
perturbation has propagated by approximately $dy=\pm 0.25-0.3$.  

\subsection{The Sedov blast wave}
\begin{figure}
\centering
\includegraphics[bb = 0 0 300 230, width=0.48\linewidth]{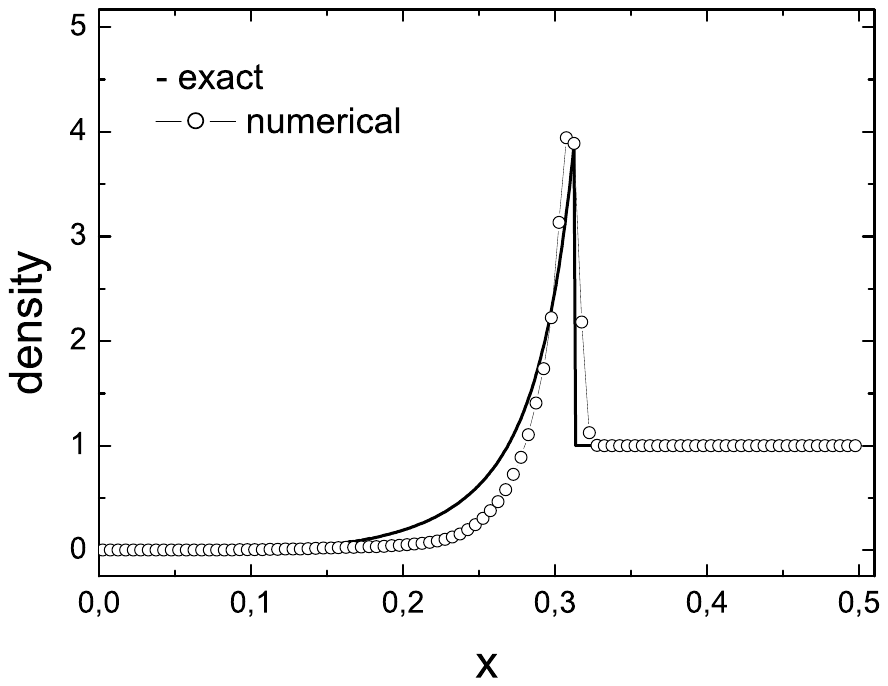}
\includegraphics[bb = 0 0 300 230, width=0.48\linewidth]{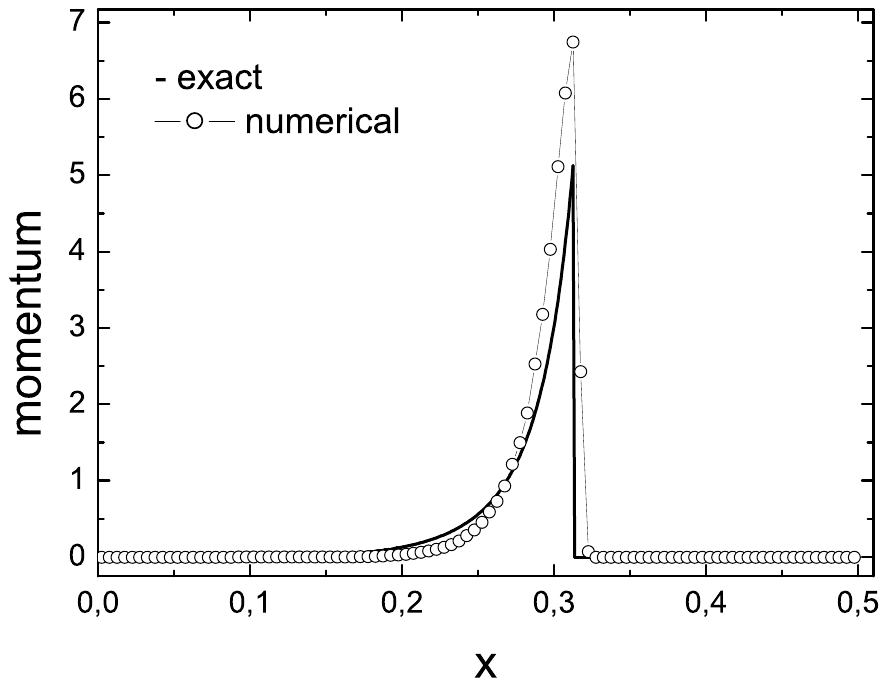}
\caption{Density (left) and momentum (right) in the Sedov blast wave problem. The solid lines 
represent the exact solution.}
\label{SedovSimulation}
\end{figure}

The Sedov blast is a spherical explosion caused by the point injection of energy and in astronomy 
is representative of a supernova explosion. The initial setup for this test is as follows: $[-0.5;0.5]^{3}$ is the computational domain, $\gamma = 5/3$ the adiabatic index, $\rho_{0} = 1$ the initial density, and $p_{0} = 10^{-5}$ the initial pressure. At time $t=0$, the thermal energy 
$E_{0} = 0.6$ is injected. The energy is injected in a central sphere with radius $r_{central} = 0.02$. For numerical simulations, a computational mesh of $100^3$ grid zones was used. The resulting 
radial profiles of the density and momentum at time $t = 0.05$ are shown in Figure (\ref{SedovSimulation}). 

The Sedov blast wave problem is the standard test which verifies the code's ability to
manage strong shocks with high Mach numbers. The sound speed of the background medium is 
negligibly small, so that the Mach number is high $M = 1432$. Our numerical method performs at this difficult problem quite well and reproduces accurately the position of the shock and profile of the shock wave in the wake.

\subsection{The expansion of gas into vacuum}
\begin{figure}
\centering
\includegraphics[bb = 0 0 300 230, width=0.8\linewidth]{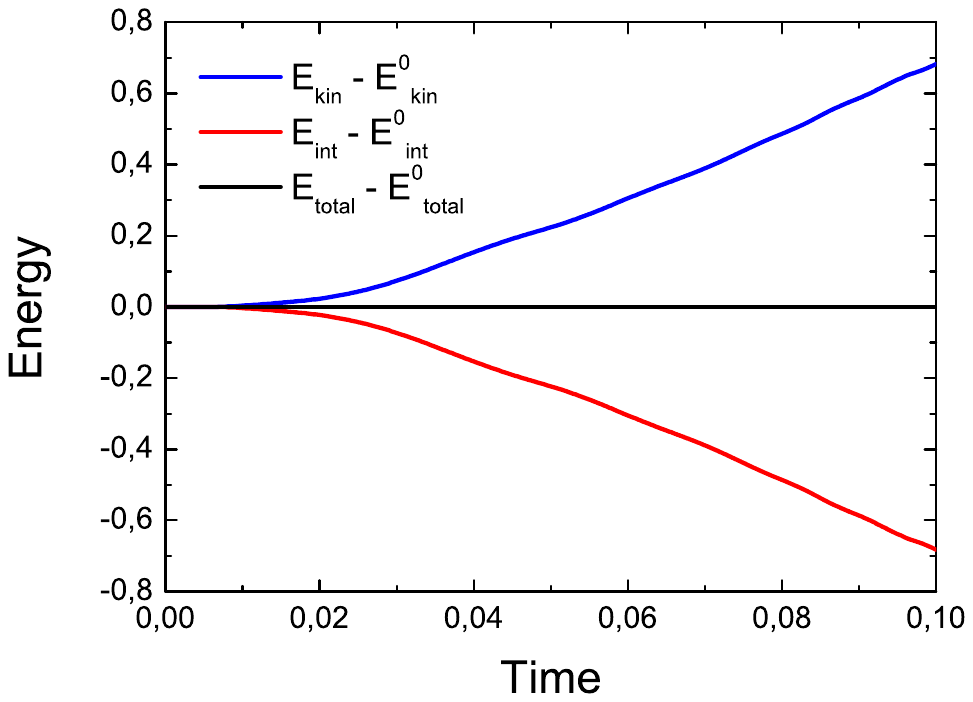}
\caption{Time evolution of the kinetic, internal, and total energies with respect to their initial values in the test problem with expansion of gas into vacuum.}
\label{Vacuum}
\end{figure}

Numerical hydrodynamics methods that use the classic Roe space-averaging scheme for gas density 
behave poorly on the gas-vacuum boundary.
Thanks to the special space-averaging scheme (see Section~1.1), our numerical method can model
the gas-vacuum boundary with conservation of total energy. To demonstrate this, we set up a test problem
describing the expansion of gas into vacuum. More specifically, we set up a gas sphere with unit radius on the $[-2;2]^{3}$ domain: 
$$
p_{0}(r) = \rho_{0}(r) = \left\{
\begin{array}{c}
1, r \le 1 \\ 
0, r > 1
\end{array} 
\right. .
$$ 
The ratio of specific heats $\gamma = 1.4$ and initial velocity  $\vec{v}$ is set to zero.
 {For numerical simulations, a computational mesh of $256^3$ grid zones was used.}
The behaviour of the kinetic, internal, and total energies of gas are shown 
in Figures (\ref{Vacuum}). As expected, the kinetic energy increases owing to expansion of gas into vacuum,
the internal energy decreases owing to adiabatic cooling, but the total energy stays constant 
within the machine precision.

It's the synthetic test for control of total energy behaviour for the expand of gas cloud to vacuum. On realistic astrophysical simulations don't use a vacuum region, but numerical method must be able to simulate such regions with conservation of total energy.

 {
\subsection{Three test problem of Calella \& Woodward (1984)} 
To further verify our numerical hydrodynamics scheme, three test problem from \cite{Woodward_1984} 
were used: two interacting blast waves, a Mach 3 wind tunnel with a step, and double Mach reflection of a strong shock. We note that these test problems do not have analytic solutions or predictions, 
unlike other tests used in our paper. }

\begin{figure}
\centering
\includegraphics[bb = 0 0 300 230, width=0.5\linewidth]{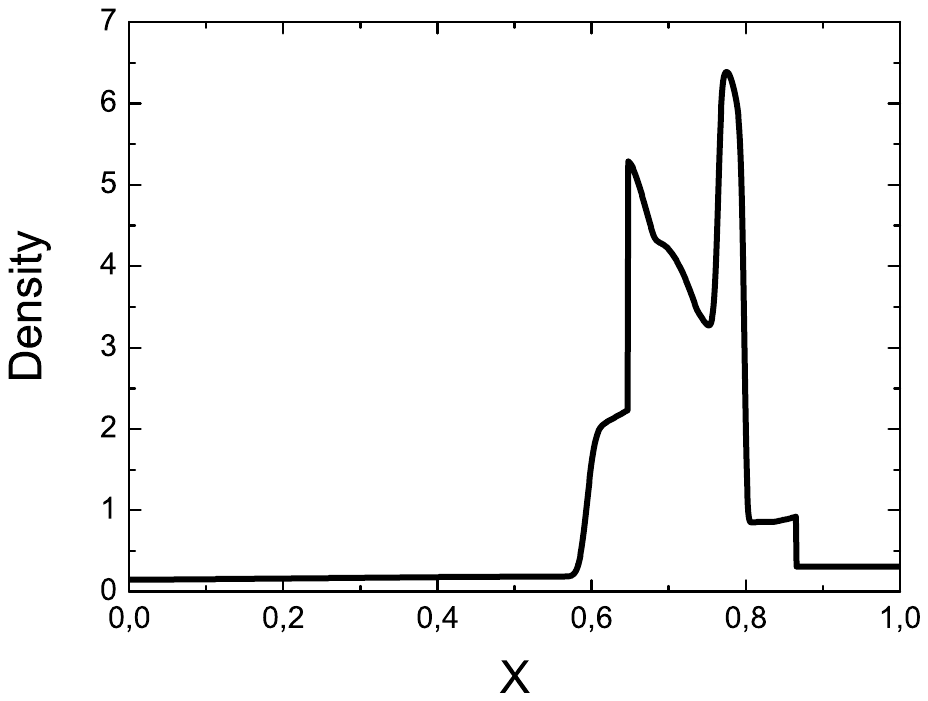}
\includegraphics[bb = 0 0 300 230, width=0.5\linewidth]{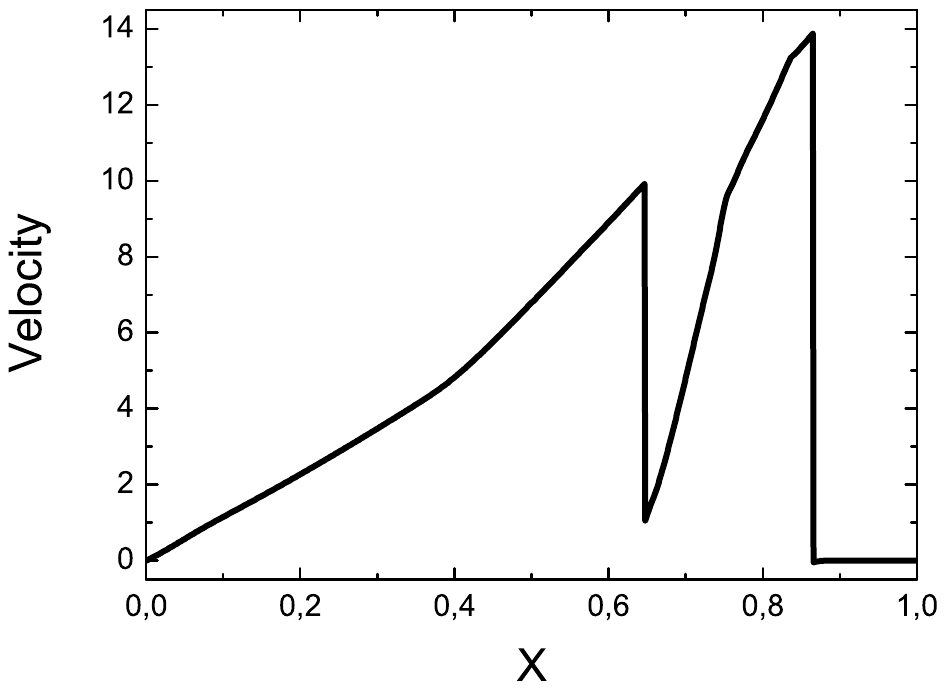} 
\caption{Results of the two interacting blast waves problem.}
\label{ClassicTestTwoWaves}
\end{figure}

 {
\subsubsection{Two Interacting Blast Waves}
This test is designed to check the code's ability to simulate the interaction of strong shock waves in a narrow region. The initial density is set to unity  and the velocity is zero everywhere on the
$[0;1]$ domain. The pressure for $x < 0.1$ is set to $p_L = 1000$, while for $0.1 < x < 0.9$ and
$x > 0.9$it is set to $p_C = 0.01$ and $p_R = 100$, respectively. Reflecting boundary conditions 
are used. For numerical simulations, a computational mesh of $2400$ grid zones was used. 
The result of numerical simulation is presented in Figure (\ref{ClassicTestTwoWaves}). 
Our numerical scheme correctly reproduces the position of shock waves and contact discontinuities,
but similar to many other numerical schemes smears somewhat the contact discontinuity at $x=0.6$.
The shock wave at $x= 0.76$ has a correct amplitude of $\approx 6.5$ and the precursor wave at 
$x=0.86$ has a correct amplitude of $\approx 1.0$. The velocity field having two peak
at $x=0.65$ and $x=0.86$ is also correctly reproduced in our test problem. We note that we used
an equidistant mesh, while Woodward \& Colella used adaptive meshes.
}

\subsubsection{A Mach 3 Wind Tunnel with a Step}
\begin{figure}[ht]
\centering
\includegraphics[bb = 0 0 430 180, width=0.7\linewidth]{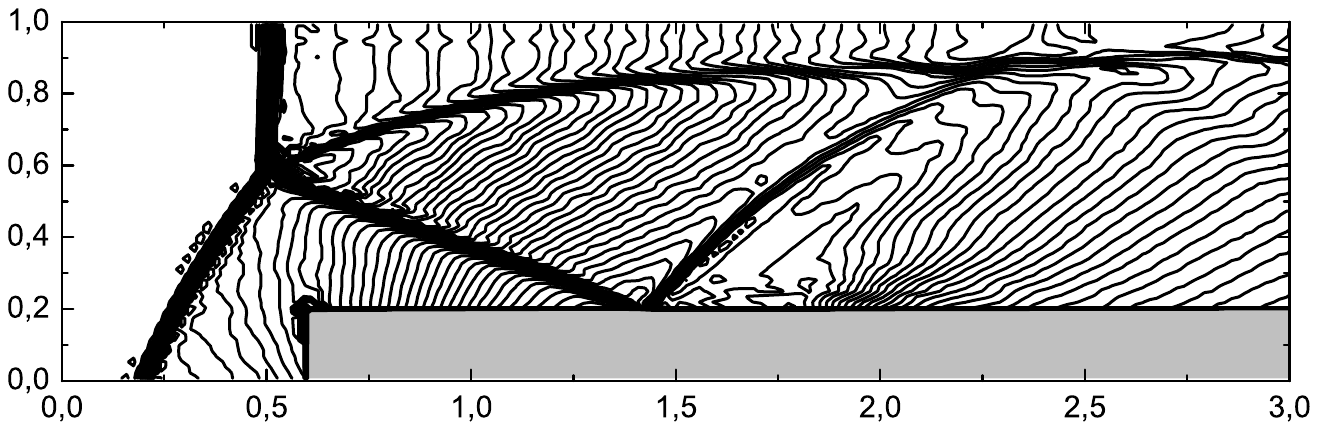}
\caption{The a Mach 3 wind tunnel with a step problem. The isolines of density for 
computational mesh of $720 \times 240$ (bottom) grid zones}
\label{ClassicTestMach3}
\end{figure}

In this test problem, a flow with a Mach number equal to 3 is set in a tunnel with dimensions
$[0;3] \times [0;1]$ containing a step.
More specifically, the initial density and pressure are set to $\rho_0 = 1.4$ and $p_0 = 1$, 
respectively, the $y$-component of velocity is equal to zero, and the $x$-component of velocity 
is equal to 3.0 at $x < 0.5$ and is zero elsewhere. The ratio of specific heats is $\gamma = 1.4$. 
The inflow and outflow boundary conditions are chosen on the left and right boundaries, respectively,
while on the top and bottom walls the reflecting boundary conditions are used.
For numerical simulations, a computational mesh of $720 \times 240$ grid zones was used. 
The results are presented in Figure (\ref{ClassicTestMach3}). Evidently, our numerical scheme performs
well on this difficult test problem. The position of all shock waves is correctly reproduced 
\cite{Yoon_2008}. We note that in our method, we did not use any special adjustments for computing the
singularity point at the corner of the step, in contrast to other numerical schemes \cite{Woodward_1984}.
This is due to a fundamental difference between the PPM and PPML methods, the latter using the local
stencil for calculating the interpolation parabolas.

\begin{figure}[ht]
\centering
\includegraphics[bb = 0 0 430 180, width=0.7\linewidth]{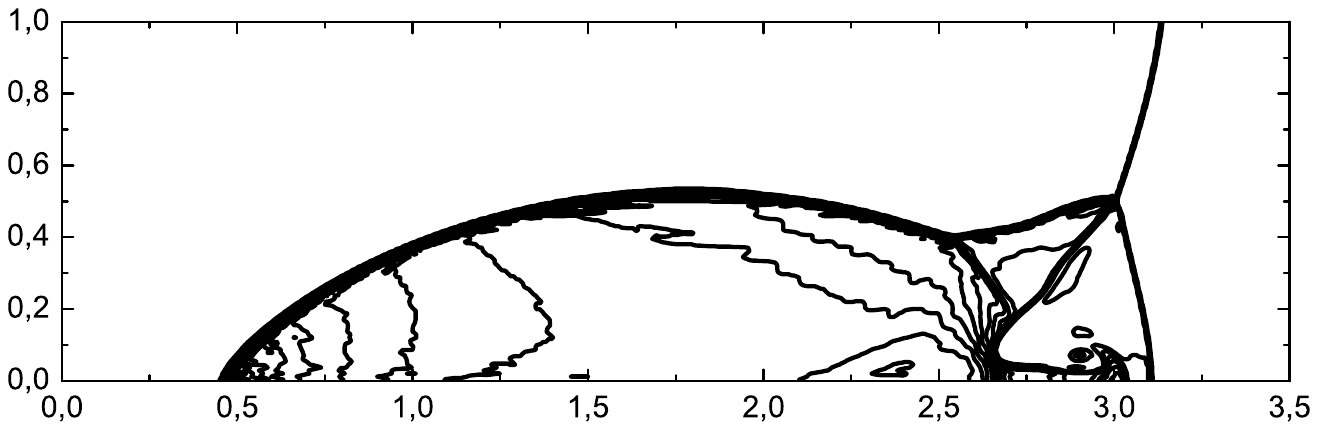}
\caption{The problem double Mach reflection of a strong shock. The isolines of density for 
computational mesh of $840 \times 240$ grid zones}
\label{ClassicTestDoubleMach}
\end{figure}

 {
\subsubsection{Double Mach Reflection of a Strong Shock}
In this test problem, the reflection of a Mach 10 shock from an inclined plane is simulated.  
The initial density, pressure and velocity are set in the $[0;3.5] \times [0;1]$ domain as follows:
$$
\rho_{0}  = \left\{
\begin{array}{c}
8,  x < 1/2 + y / tan(\pi/3) \\ 
1.4, x \geq 1/2 + y / tan(\pi/3)
\end{array} 
\right. 
\qquad
p_{0}  = \left\{
\begin{array}{c}
116.5, x < 1/2 + y / tan(\pi/3) \\ 
1, x \geq 1/2 + y / tan(\pi/3)
\end{array} 
\right.
$$
$$
v_{x,0}  = \left\{
\begin{array}{c}
7.1447, x < 1/2 + y / tan(\pi/3) \\ 
0, x \geq 1/2 + y / tan(\pi/3)
\end{array} 
\right. 
\qquad
v_{y,0}  = \left\{
\begin{array}{c}
-4.125, x < 1/2 + y / tan(\pi/3) \\ 
0, x \geq 1/2 + y / tan(\pi/3)
\end{array} 
\right. ,
$$
the ratio of specific heats $\gamma = 1.4$. On the left wall, the inflow boundary condition is set.
On the right wall, on the top wall and on the bottom wall for $x<0.5$, the outflow boundary condition is used, while at the bottom wall for $x>0.5$ the reflection boundary condition is set.
For numerical simulations, a computational mesh of $840 \times 240$ grid zones was used. The 
results are presented in Figure (\ref{ClassicTestDoubleMach}). Our scheme performs well
on this test problem.  The position of all shock waves and contact discontinuities 
is reproduced correctly, similar to the classic PPM method of Woodward \& Colella (1984).
All shock waves are narrow, implying low dissipation, and exhibit no numerical artifacts near the
boundaries. The reflected wave emerges from the bottom wall at $x=0.5$ and its amplitude 
reaches $\sim 0.5$ at $x=1.8$. At $x=2.6$ and $x=3.0$, the first and second triple Mach
reflections are correctly reproduced.
}

\begin{figure}
\centering
\includegraphics[bb = 0 0 300 230, width=0.5\linewidth]{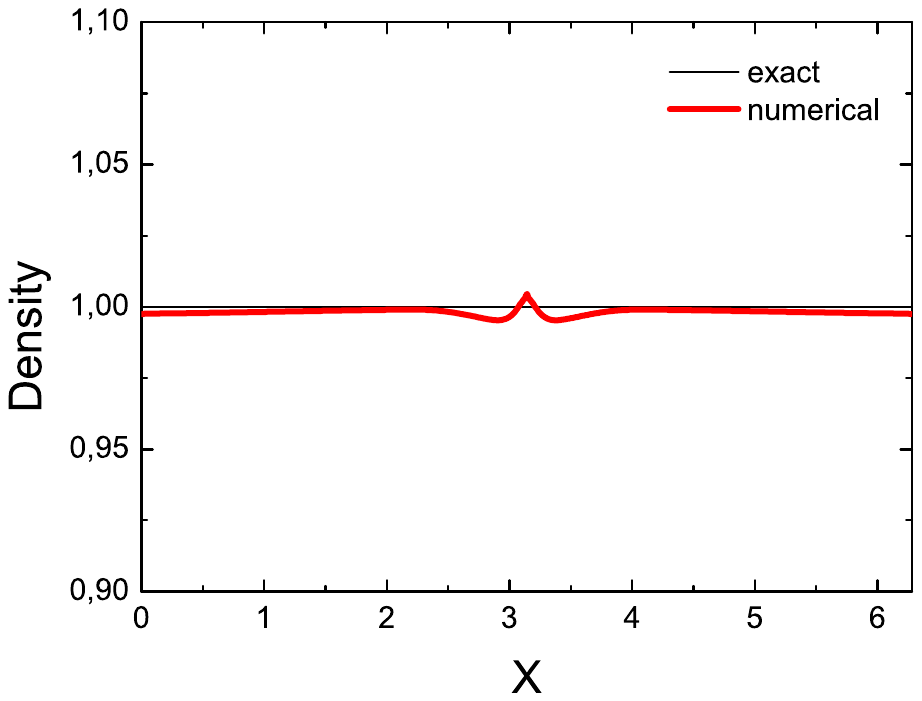}
\includegraphics[bb = 0 0 300 230, width=0.5\linewidth]{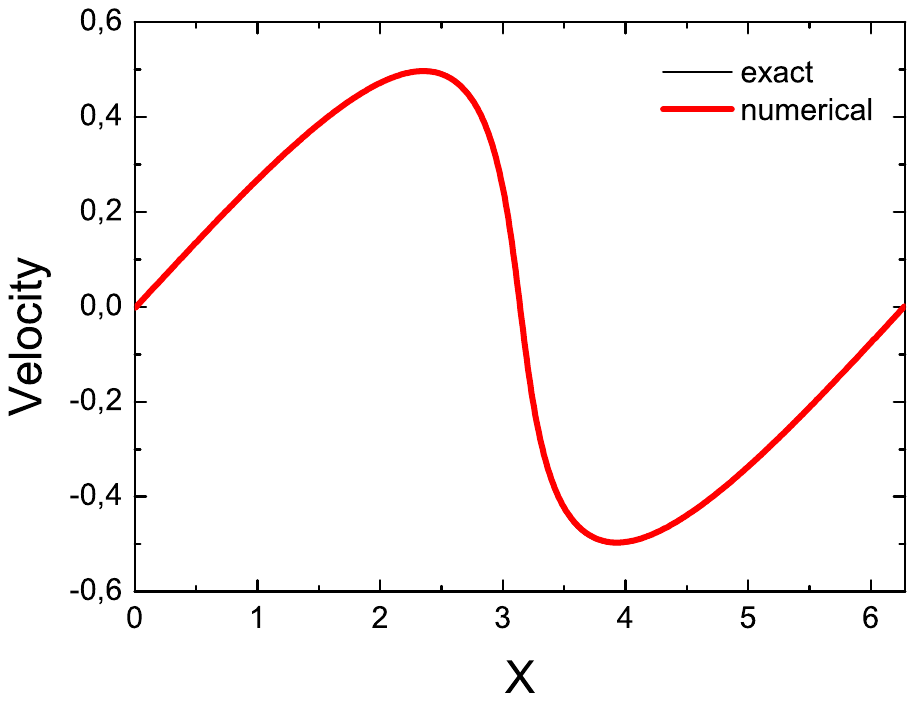} 
\caption{Results of the Aksenov test.}
\label{AksenovSimulation}
\end{figure}

 {
\subsection{The Aksenov Test}
For the accuracy analysis of our numerical method we have chosen the Aksenov test 
\cite{Aksenov_2005}. It is a one-dimensional hydrodynamical test with 
the analytical periodic solution, which belongs to a class of 
infinitely differentiable functions. Let us consider the initial setup consisting of a density
distribution of the following form  
\begin{equation}
\rho = 1 + 0.5 \cos(x)
\end{equation}
on the $x=[0:2\pi]$ domain
with the periodic boundary conditions. The velocity is set to $u=0$ everywhere and
the equation of state has the following form
$p = \rho^{\gamma}$, where $\gamma = 3$. 
Then the analytical solution is described as follows:
\begin{equation}
\rho = 1 + 0.5 \cos(x - u t) \cos(\rho t),
\end{equation}
\begin{equation}
u = 0.5 \sin(x - u t) \sin(\rho t).
\end{equation}
The results of the simulation for time $t = \pi/2$ are given in Figure~(\ref{AksenovSimulation}).
For the analysis of accuracy of our numerical method we use the following equation:
\begin{equation}
T = \log_{2} \left( \frac{E_{N}}{E_{N/2}} \right),
\end{equation}
where $E_{N} = \sum_{1}^N \vert f_{i}^{\rm exact} - f_{i}^{\rm num} \vert$, $f_{i}^{\rm exact}$
is the exact solution of Aksenov test in cell $i$, $f_{i}^{\rm num}$ the numerical solution in 
cell $i$, and $N$ the number of grid zones. For this test problem, a computational mesh of 
$N = 628$ grid zones was used. The resulting accuracy of our numerical scheme using the
density distribution for $f_{i}$ is $T = 1.713$. 
}

\section{NUMERICAL SIMULATION OF THE SPIRAL INSTABILITY IN A GALACTIC DISK}
In this section, we will demonstrate the development of the spiral instability in a galactic disk 
adopting the isothermal approximation with $\gamma = 1$. A thin gaseous disk is unstable to an 
axisymmetric perturbation if the Toomre $Q$-parameter is less than unity \cite{Toomre_1964}:
\begin{equation}
Q = \frac{c_s \kappa}{\pi G \Sigma} < Q_{\rm crit}=1.0,
\end{equation}
where $c_s$ is the sound speed, $\Sigma$ is the column density and $\kappa$ is the epicyclic frequency
defined as:
\begin{equation}
\kappa^2 = \frac{2 \Omega}{r} \frac{d}{dr} \left( r^2 \Omega \right)
\end{equation}
where $\Omega$ is the angular velocity and $r$ is the radial coordinate.
For local non-axisymmetric perturbations and three-dimensional disks with finite thickness,
the critical Toomre parameter for the development of spiral instability can be somewhat greater,
$Q_{\rm crit}\approx 1.5$  \cite{Nelson1998,Polyachenko1997}.

The initial setup consists of a self-gravitating galactic gaseous disk submerged in a fixed 
dark matter (DM) halo of the following form:
\begin{equation}
\rho_{DM} = \frac{\rho_0}{1 + \left( r/r_0 \right)^2}
\end{equation}
where $\rho_0 = 1.97 M_{\odot}$ pc$^{-3}$ is the central DM density, $r_0 = 1.6$ kpc the characteristic scale length of the DM halo and $r_h = 78.55$ kpc the halo radius.  The DM mass is set to 
$5\times10^{10}~M_\odot$. The details of the procedure for generating an equilibrium profile of 
a galactic gaseous disk in the gravitational potential of the DM halo can be found in 
\cite{Vorobyov_2012}. 

\begin{figure}
\centering
\includegraphics[bb = 0 0 300 230, width=0.7\linewidth]{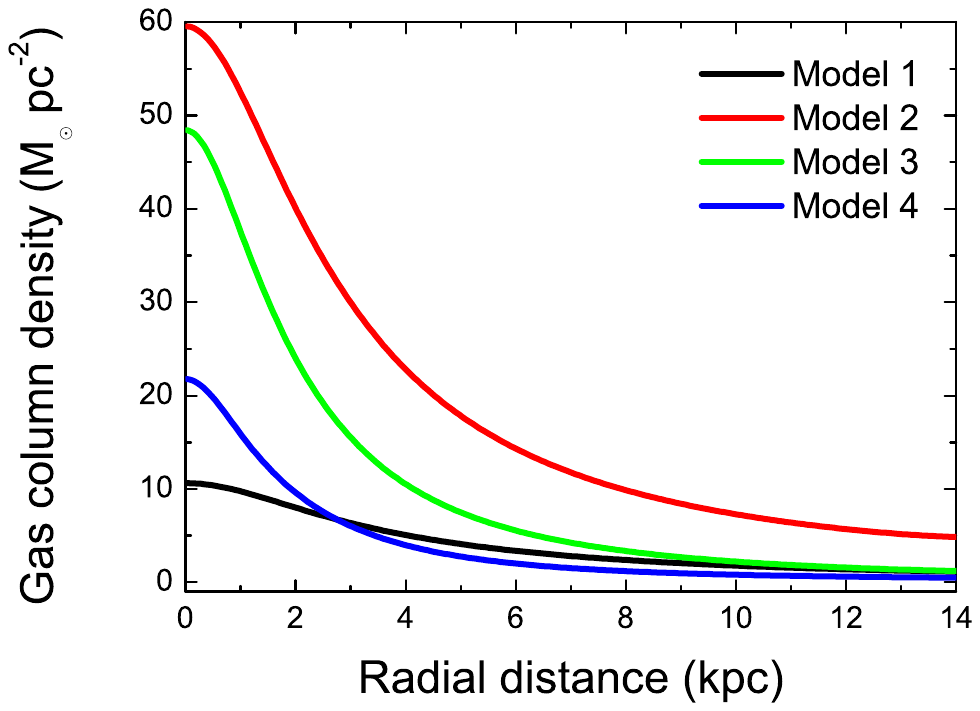}
\includegraphics[bb = 0 0 300 230, width=0.7\linewidth]{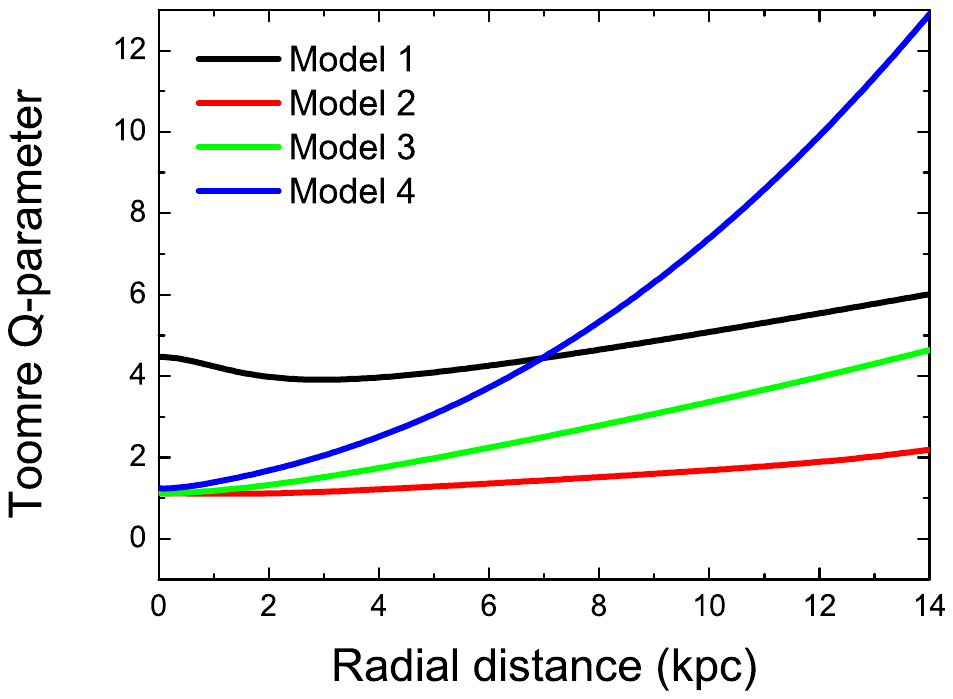}
\caption{Initial profiles of column density (left) and Toomre $Q$-parameter (right) for all models.}
\label{setup}
\end{figure}

We have considered four models, the parameters of which are listed in Table~(\ref{table_models}). The
initial distributions of the gas column density $\Sigma$  and Toomre $Q-$parameter are shown in Figure~(\ref{setup}).
The gas temperature in models~1-3 is set to $T=10^4$~K, while in model~4 it equals to $T=2\times 10^3$~K. At the beginning of simulations, we introduce a small density perturbation characterized by a normal distribution with a zero mean and $10^{-4}$ root-mean-square deviation.
 {For all numerical simulations a CFL number equal 0.2 and a computational mesh of $512^3$ grid zones are used.}

To quantify the strength of spirals modes in numerical simulations of gravitationally 
unstable disks, we calculate the global Fourier amplitudes on the local sub-domain 
$r \leq R_{f}=8$~kpc, where $r$ is the distance from the coordinate center. The total computational box has the size of $[-16; 16]^3$ kpc$^3$:
\begin{equation}
A_m(t) = \frac{\left| \int \limits_{-R_f}^{R_f} \int \limits_{-\sqrt{R_f^2-x^2}}^{\sqrt{R_f^2-x^2}} \Sigma(x,y,t) e^{i m \phi} dxdy \right|}{\int \limits_{-R_f}^{R_f} \int \limits_{-\sqrt{R_f^2-x^2}}^{\sqrt{R_f^2-x^2}} \Sigma(x,y,t) dxdy } 
\end{equation}
where $m = 1,2, \ldots$ is the spiral mode, $\Sigma(x,y,y)$ the gas column density, 
and $\phi = \tan^{-1}(y/x)$ the polar angle.
When the disk is axisymmetric, the amplitudes of all modes
are equal to zero. When, say, $A_m(t) = 0.1$, the amplitude of spiral modes is 10\% that of 
the underlying axisymmetric density distribution.

\begin{figure}
\centering
\begin{minipage}[h]{0.48\linewidth}
\center{ \includegraphics[bb = 5 50 255 255, clip, width=1\linewidth, height=0.9\textwidth]{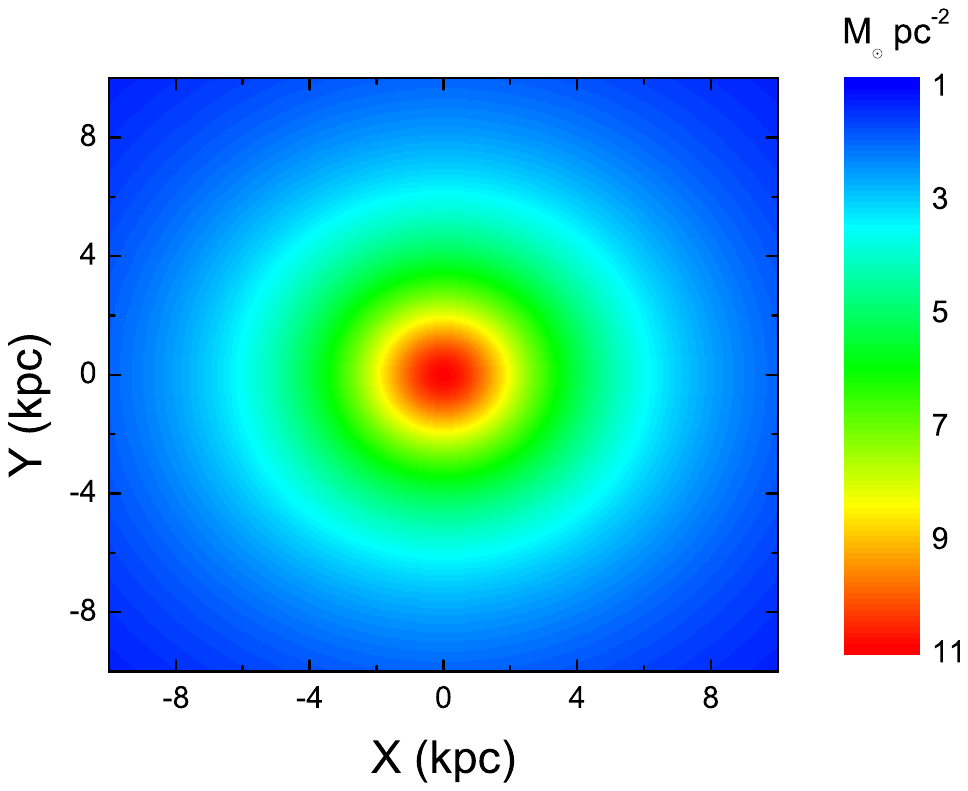} }
\end{minipage}
\hfill
\begin{minipage}[h]{0.48\linewidth}
\center{ \includegraphics[bb = 45 50 295 255, clip, width=1\linewidth, height=0.9\textwidth]{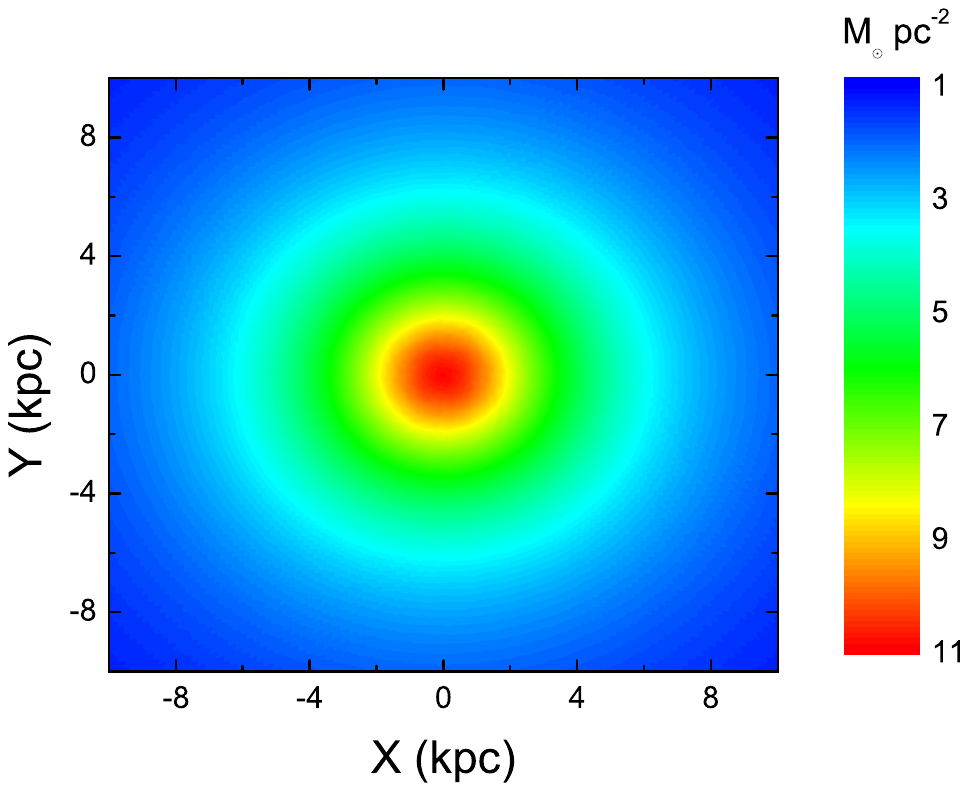} }
\end{minipage}
\hfill
\begin{minipage}[h]{0.48\linewidth}
\center{ \includegraphics[bb = 5 20 255 225, clip, width=1\linewidth, height=0.9\textwidth]{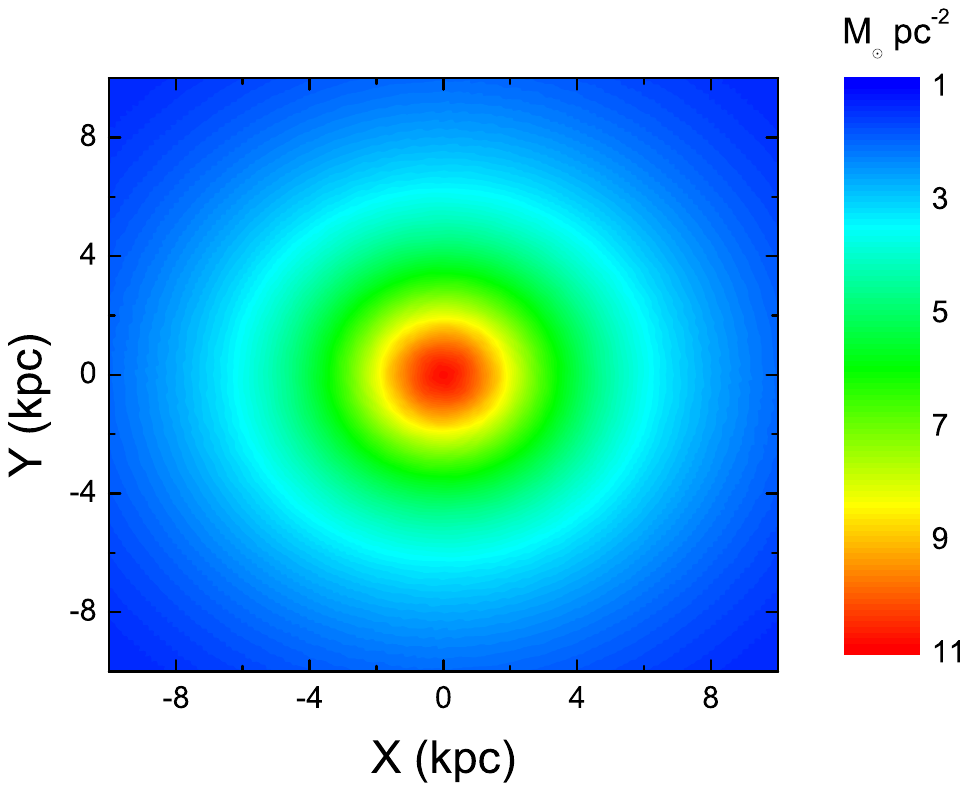} }
\end{minipage}
\hfill
\begin{minipage}[h]{0.48\linewidth}
\center{ \includegraphics[bb = 45 20 295 225, clip, width=1\linewidth, height=0.9\textwidth]{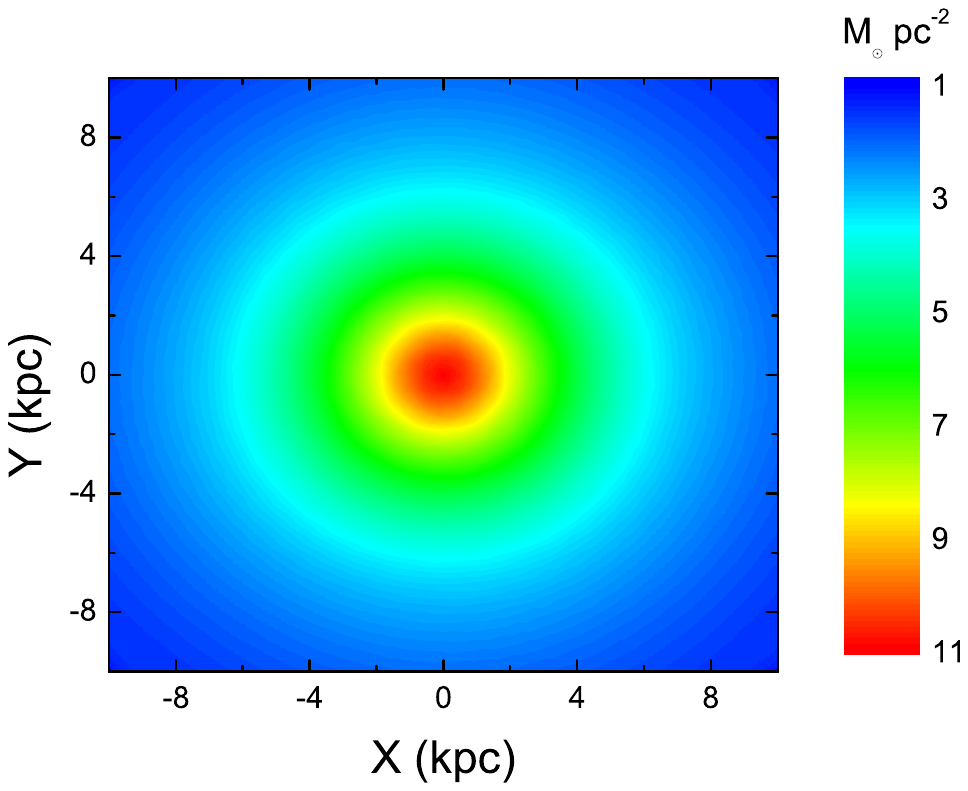} }
\end{minipage}
\caption{Evolution of the column density (in $M_{\odot} pc^{-2}$) in model~1. The evolution times 
are 0~Myr (top-left), 200~Myr (top-right), 300~Myr (bottom-left) and 400~Myr (bottom-right). The model is gravitationally stable.}
\label{simulationarm0}
\end{figure}

\begin{figure}
\centering
\includegraphics[bb = 0 0 300 230, width=0.7\linewidth]{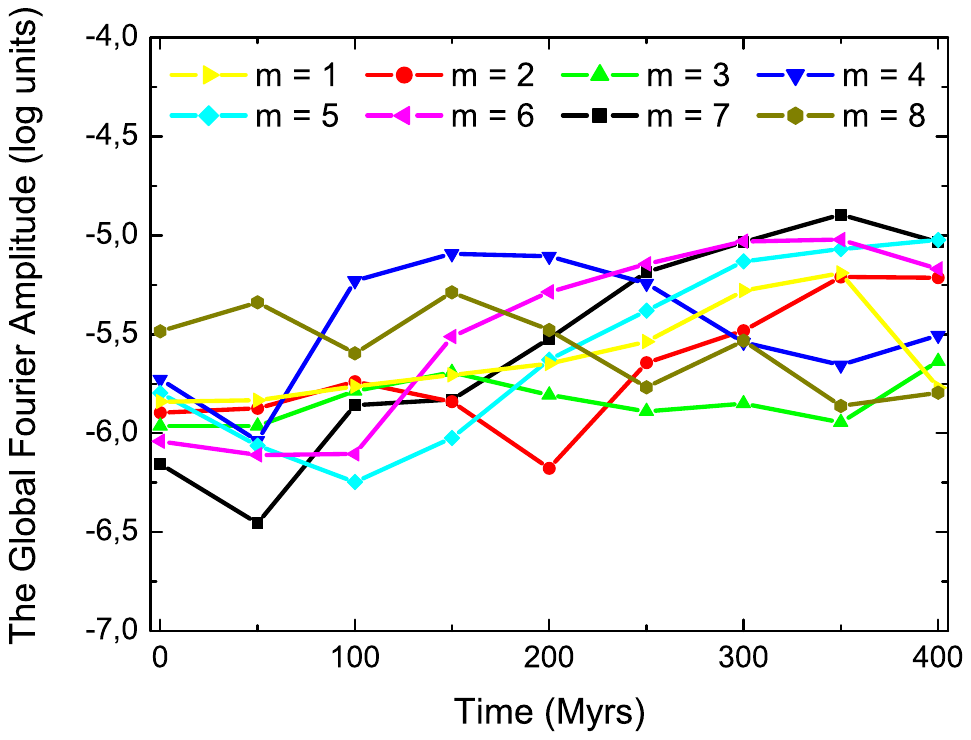}
\caption{Time evolution of the global Fourier amplitudes (in log scale) in model~1.}
\label{analysisarm0}
\end{figure}

Figure~(\ref{simulationarm0}) shows the time evolution of the gas column density in model~1 
during 400~Myr. The $Q$-parameter in this model is everywhere greater than a critical value of 
$Q_{\rm crit}\approx1.5$. As expected, the initial perturbations die out with time and 
this model does not show any significant deviations from the initial near-axisymmetric 
state. The time behaviour of the global Fourier amplitudes shown in Figure~(\ref{analysisarm0})
confirms our findings. The amplitudes of the first eight modes 
stay in the $(-6.0 : -5.0)$ range (in the log scale),
which corresponds to a 0.001\% density perturbation relative to the underlying
axisymmetric disk. The higher-order amplitudes show a similar behaviour. This test
therefore confirms the expected stability of model~1 against small perturbations.

\begin{figure}
\centering
\begin{minipage}[h]{0.48\linewidth}
\center{ \includegraphics[bb = 5 50 255 255, clip, width=1\linewidth, height=0.9\textwidth]{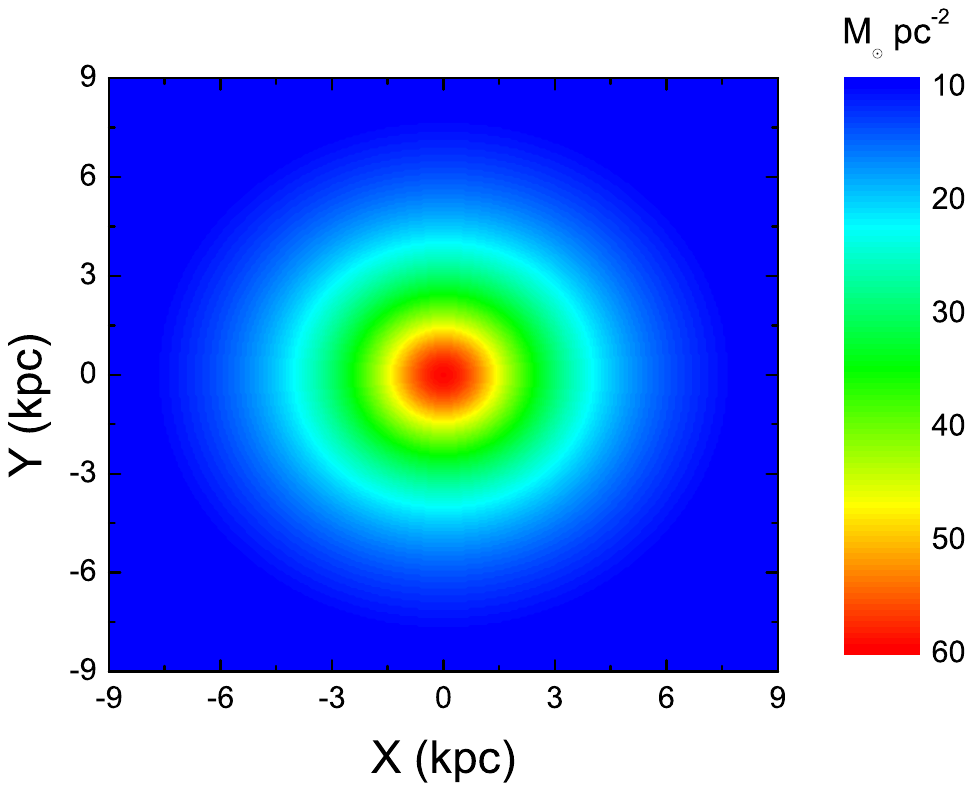} }
\end{minipage}
\hfill
\begin{minipage}[h]{0.48\linewidth}
\center{ \includegraphics[bb = 45 50 295 255, clip, width=1\linewidth, height=0.9\textwidth]{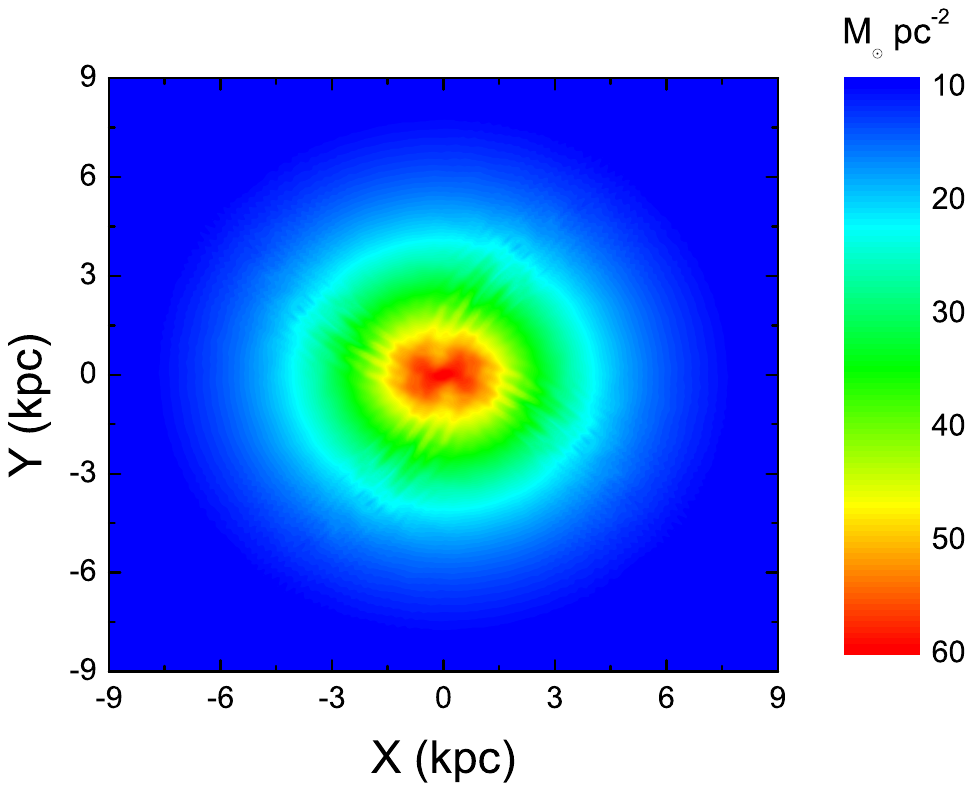} }
\end{minipage}
\hfill
\begin{minipage}[h]{0.48\linewidth}
\center{ \includegraphics[bb = 5 21 255 226, clip, width=1\linewidth, height=0.9\textwidth]{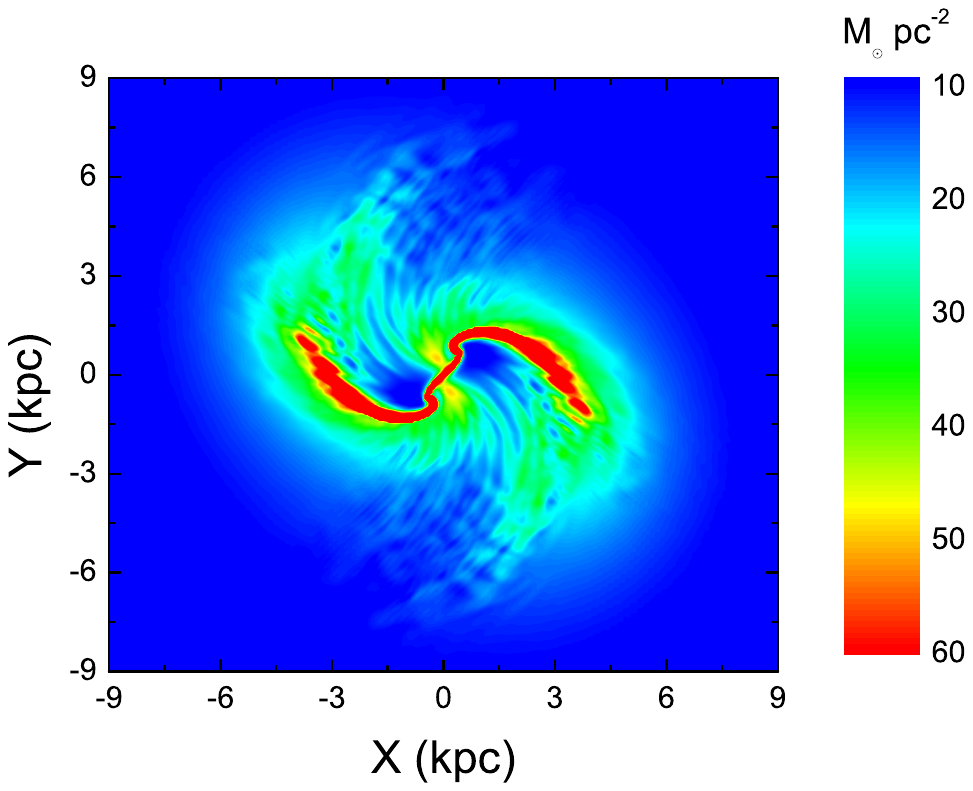} }
\end{minipage}
\hfill
\begin{minipage}[h]{0.48\linewidth}
\center{ \includegraphics[bb = 45 21 295 226, clip, width=1\linewidth, height=0.9\textwidth]{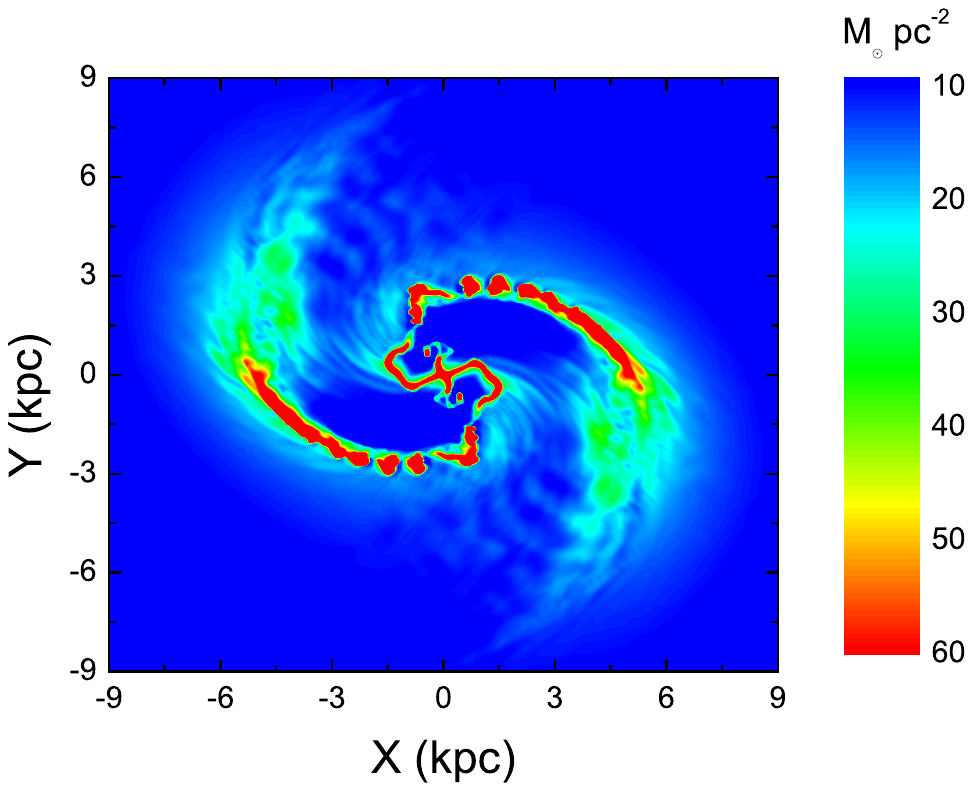} }
\end{minipage}
\caption{Evolution of the column density (in $M_{\odot} pc^{-2}$) in model~2. The evolution times 
are 0~Myr (top-left), 100~Myr (top-right), 200~Myr (bottom-left) and 280~Myr (bottom-right). The model is gravitationally unstable and develops
a two-armed spiral pattern.}
\label{simulationarm2}
\end{figure}

\begin{figure}
\centering
\includegraphics[bb = 0 0 300 230, width=0.7\linewidth]{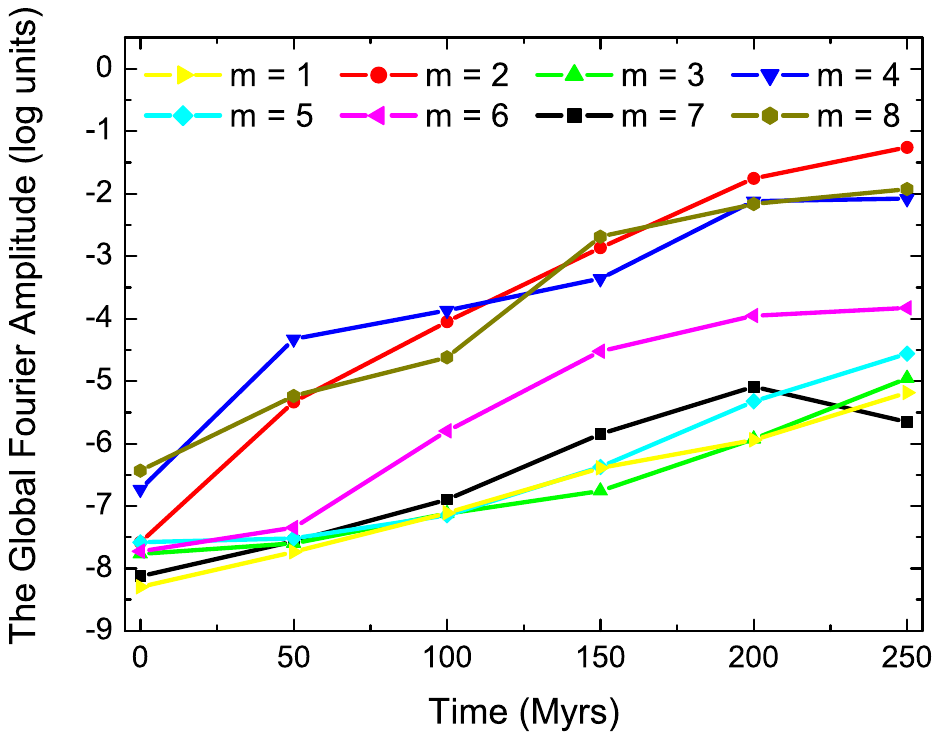}
\includegraphics[bb = 0 0 300 230, width=0.7\linewidth]{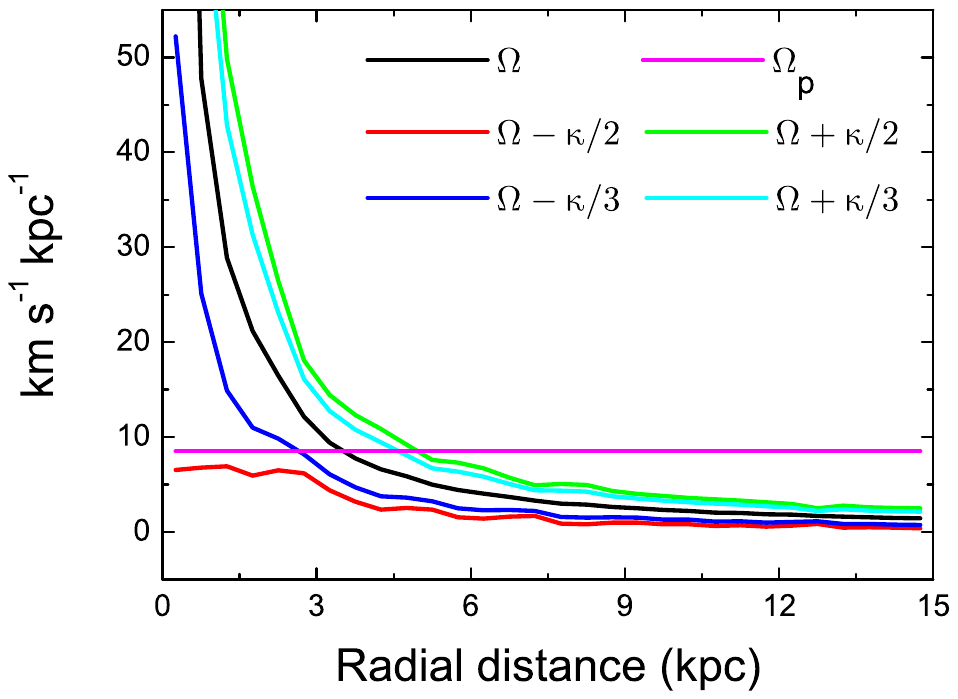}
\caption{{\bf Top:} Time evolution of the global Fourier amplitudes in model~2. {\bf Bottom}: 
radial profiles of $\Omega$, $\Omega_p$, $\Omega \pm \kappa/2$, and $\Omega \pm \kappa/3$. The intersection
of $\Omega_p$ with other profiles gives the position of resonances. In particular, the intersection
of $\Omega_p$ with $\Omega - \kappa/3$ marks the position of the inner Lindblad resonance for the 
$m=3$ mode.}
\label{analysisarm2}
\end{figure}

The time evolution of the gas column density in model~2 is presented in Figure~(\ref{simulationarm2}). This model is characterized by the maximum disk mass and the highest ratio of the disk to DM halo mass, $\xi=0.1708$. The Toomre $Q$-parameter is the lowest among all four models and stays below a critical value of 1.5 in the inner 7~kpc. As a consequence, model~2 is very unstable and shows the development
of a two-armed spiral pattern. The time evolution of the Fourier amplitudes in the upper panel of Figure~(\ref{analysisarm2}) confirms our conclusions, demonstrating that the $m=2$, 4 and 8 modes are  the fastest growing ones. By the end of simulations, the $m=2$ mode dominates and exceeds -1.5 in log units. 

Why does our model disk develop a two-armed global spiral pattern rather than a three-armed 
or any other multi-armed pattern? The reason for the development of the $m=2$ mode can be 
understood when applying the swing amplification theory \cite{Athanas_1984,Goldreich_1965,Toomre_1981}
The initial density perturbation applied to our model galactic disk
creates a spectrum of disturbances. Amplification in a gravitationally unstable disk occurs 
when any leading spiral disturbance unwinds into a trailing one due to differential rotation.
However, feedback loops that turn trailing disturbances into leading ones must be present in 
galactic discs in order for swing amplification to operate continuously \cite{BT_1987}. Reflection of trailing spiral disturbances from a steep, high-$Q$ barrier \cite{Athanas_1984} or propagation of trailing spiral disturbances
through the galactic center naturally lead to the emergence of leading spiral disturbances and 
present positive feedback loops. 

Regions with steeply increasing $Q$-parameters are absent in our model disc 
(see Fig.~\ref{setup}). Therefore, the propagation of  disturbances through the disk centre
presents the only possible feedback loop. This mechanism operates if there is no inner Lindblad  
resonance (ILR) for a specific spiral mode.
The position of Lindblad resonances can be defined as
\begin{equation}
m(\Omega_{\rm p} -\Omega) = \pm \kappa, 
\end{equation}
where $\Omega_{\rm p}$ is the speed of the spiral pattern.

In the bottom panel of Figure~(\ref{analysisarm2}) we 
plot the radial profiles of $\Omega_{\rm p}$, $\Omega\pm \kappa/m$,
and $\Omega$ for the $m=2$ and $m=3$ modes. The intersection of $\Omega_{\rm p}$ 
with $\Omega-\kappa/2$ and $\Omega -\kappa/3$ gives the radial position
of the ILRs for the $m=2$ and $m=3$ modes, respectively. 
Evidently, the ILR for the $m=2$ mode is absent, while it s present for the $m=3$ mode.
Therefore, the $m=2$ trailing disturbances can propagate
through the centre and emerge on the other side as leading ones,
thus providing a feedback for the swing amplifier. As a result, the $m=2$ mode grows while
the $m=3$ (and higher-$m$ modes) die out with time.

\begin{figure}
\centering
\begin{minipage}[h]{0.48\linewidth}
\center{ \includegraphics[bb = 5 50 255 255, clip, width=1\linewidth, height=0.9\textwidth]{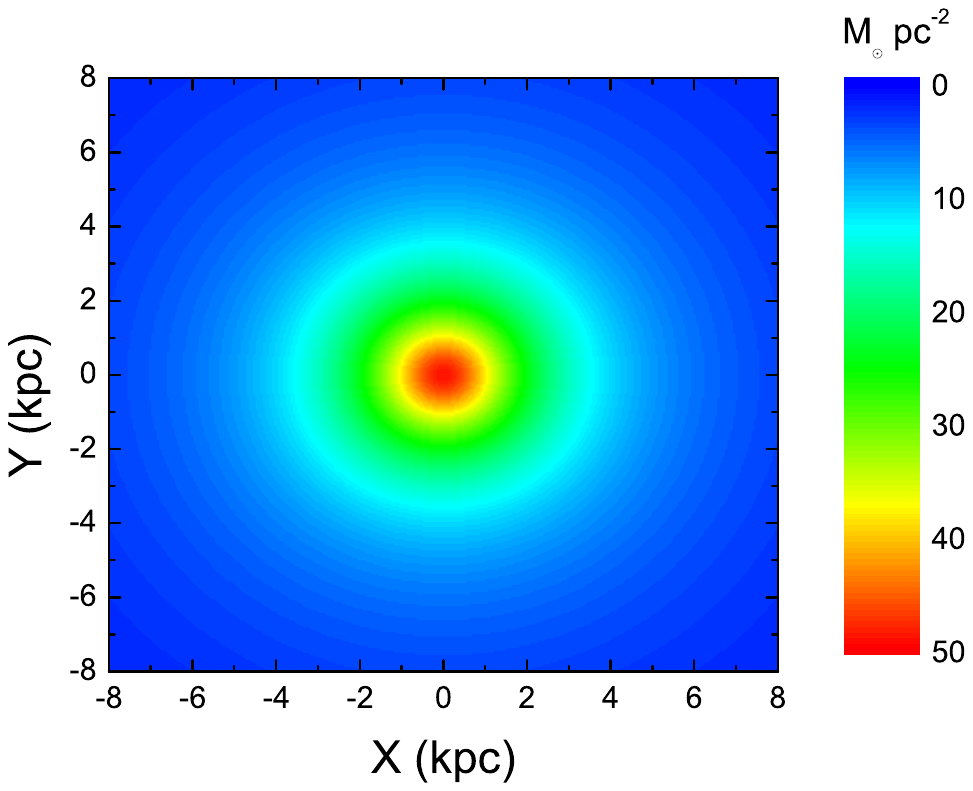} }
\end{minipage}
\hfill
\begin{minipage}[h]{0.48\linewidth}
\center{ \includegraphics[bb = 45 50 295 255, clip, width=1\linewidth, height=0.9\textwidth]{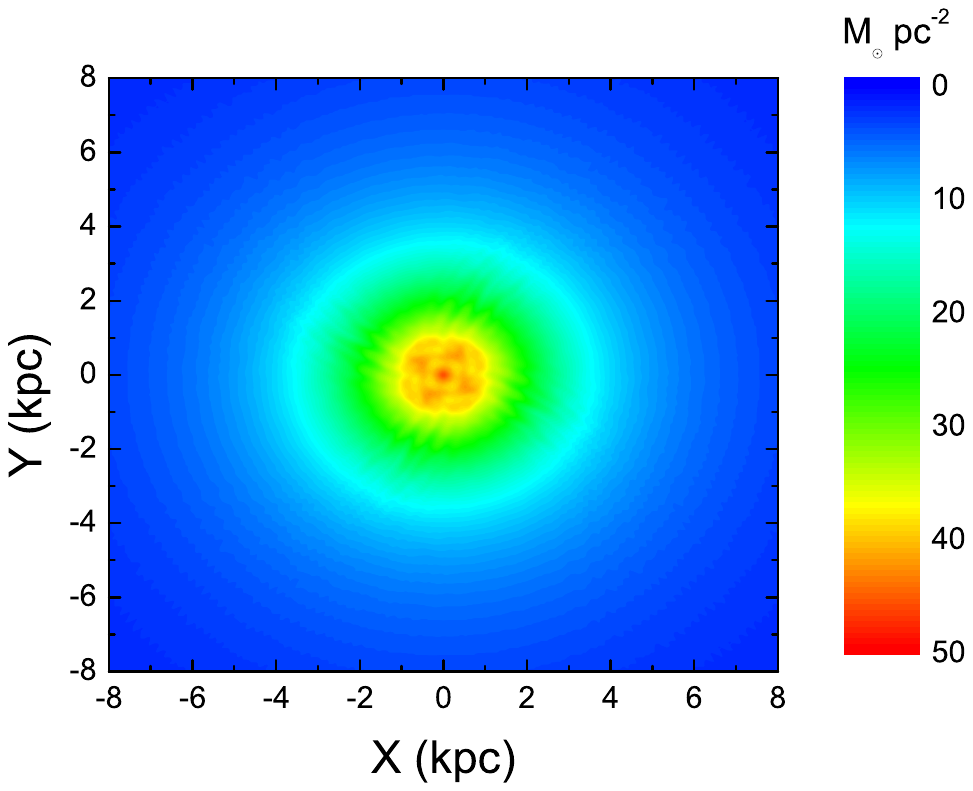} }
\end{minipage}
\hfill
\begin{minipage}[h]{0.48\linewidth}
\center{ \includegraphics[bb = 5 21 255 226, clip, width=1\linewidth, height=0.9\textwidth]{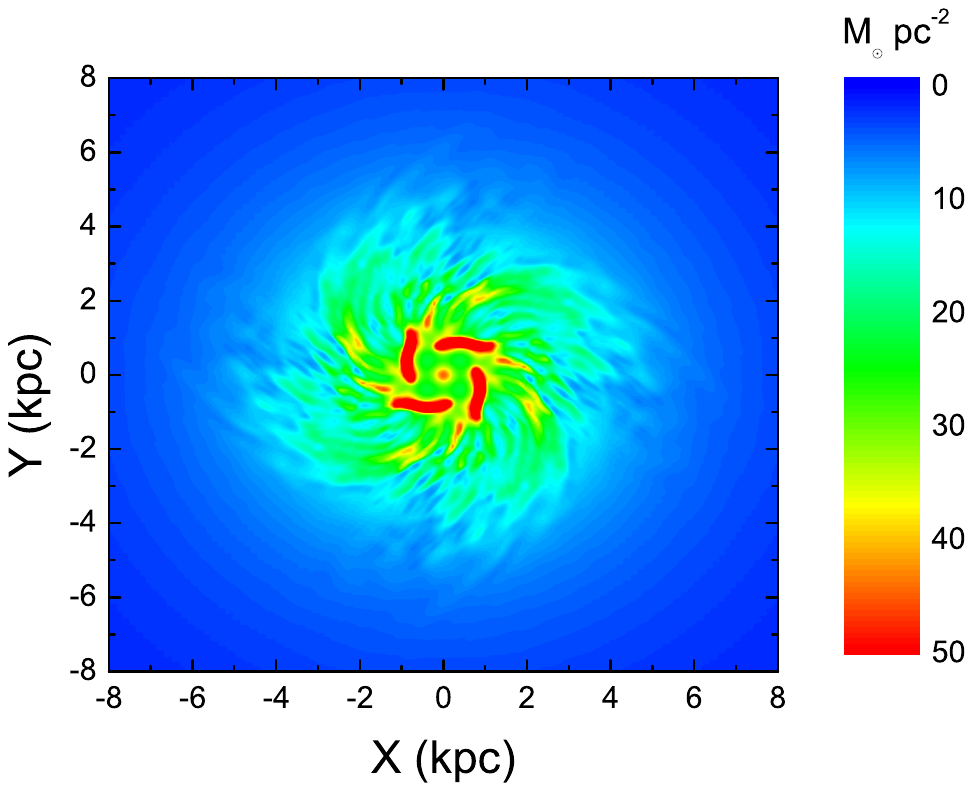} }
\end{minipage}
\hfill
\begin{minipage}[h]{0.48\linewidth}
\center{ \includegraphics[bb = 45 21 295 226, clip, width=1\linewidth, height=0.9\textwidth]{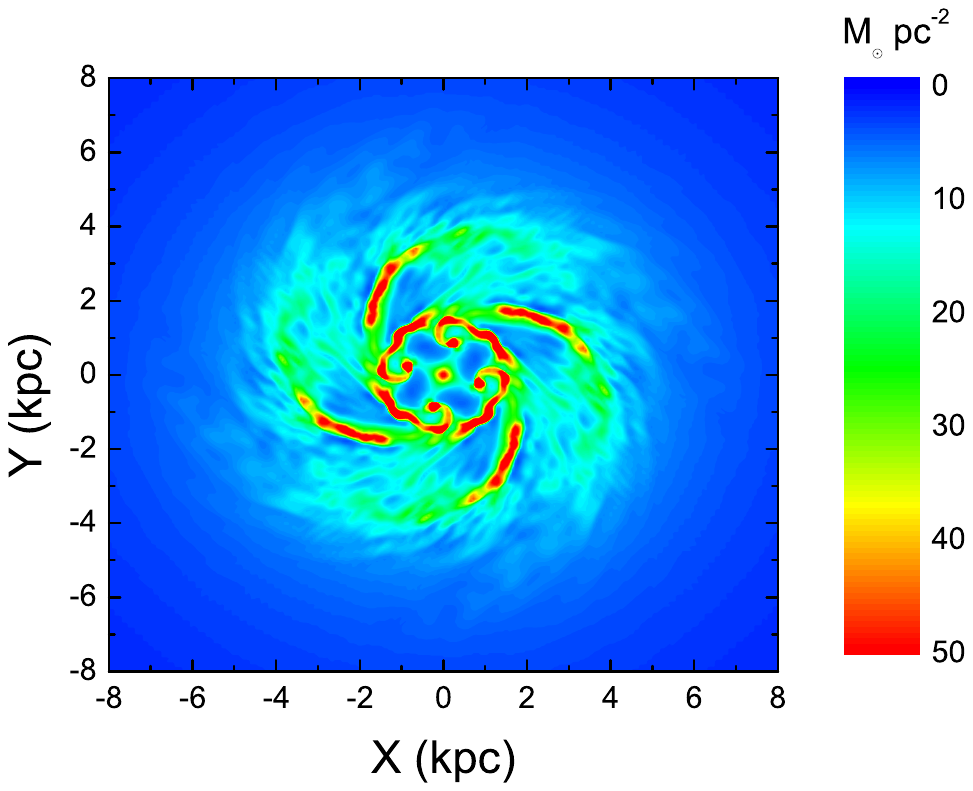} }
\end{minipage}
\caption{Evolution of the column density (in $M_{\odot} pc^{-2}$) in model~3. The evolution times are
0~Myr (top-left), 100~Myr (top-right), 250~Myr (bottom-left) and 360~Myr (bottom-right).}
\label{simulationarm4}
\end{figure}

\begin{figure}
\centering
\includegraphics[bb = 0 0 300 230, width=0.7\linewidth]{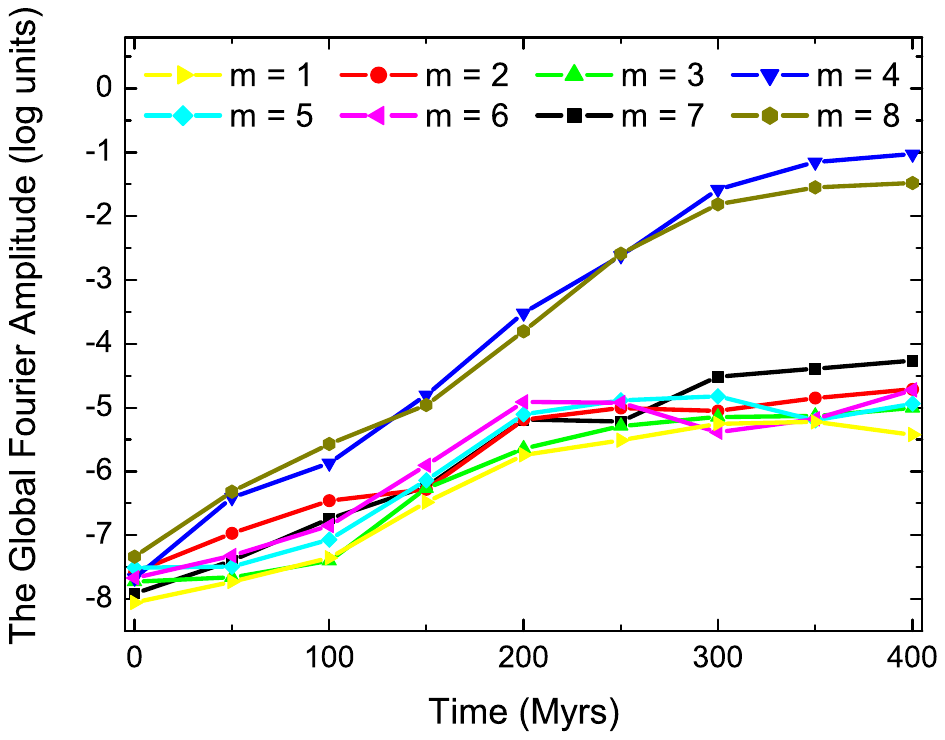}
\includegraphics[bb = 0 0 300 230, width=0.7\linewidth]{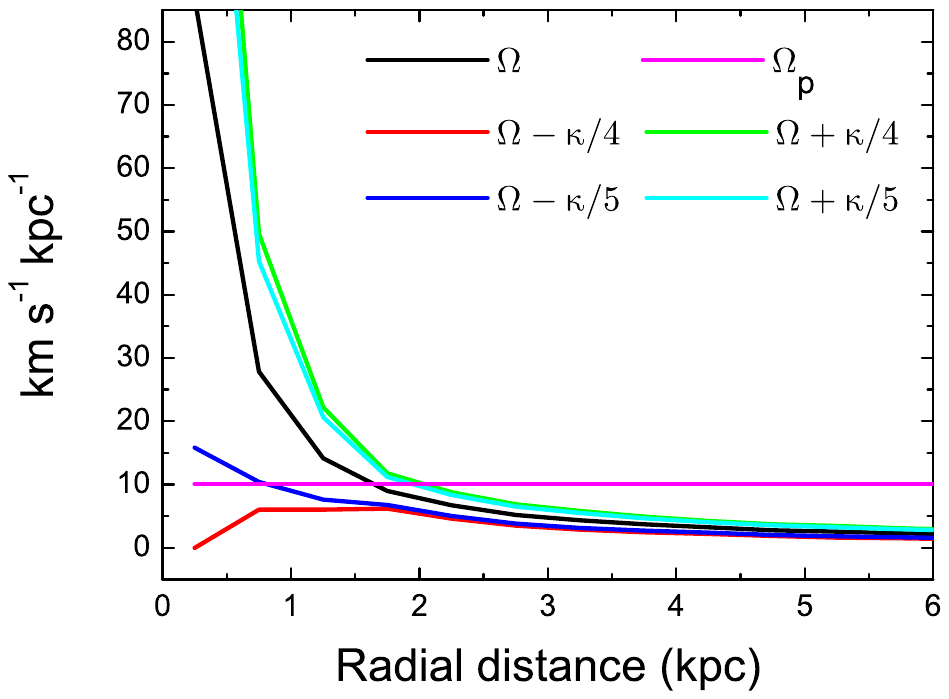}
\caption{{\bf Top}: Time evolution of the global Fourier amplitudes in model~3. {\bf Bottom}: 
radial profiles of $\Omega$, $\Omega_p$, $\Omega \pm \kappa/4$, and $\Omega \pm \kappa/5$. 
The intersection
of $\Omega_p$ with $\Omega - \kappa/5$ marks the position of the inner Lindblad resonance for the 
$m=5$ mode.}
\label{analysisarm4}
\end{figure}

Figure~(\ref{simulationarm4}) demonstrates the development of a four-armed spiral pattern in model~3.
This model is characterized by more than a factor of 2 smaller ratio of the disk to the 
DM halo mass, $\xi=0.0668$, than in model~2. The global Fourier amplitudes shown in the upper
panel of Figure~(\ref{analysisarm4})
confirm that the $m=4$ mode is the dominant one.  The bottom panel in Figure~(\ref{analysisarm4})
shows that the ILR is absent for the $m=4$ mode but is present for the $m=5$ mode (and higher-$m$ modes),
explaining the growth and dominance of the $m=4$ mode in this model. We note, however, that 
the $m=8$ mode is only a factor of 2 smaller in strength than the dominant $m=4$ notwithstanding the
fact that there is the ILR for the $m=8$ mode. The reason for the growth of the $m=8$ mode is not 
clear. It may be caused by the non-linear interaction of the $m=4$ and $m=8$ modes, leading to 
a partial redistribution of energy between the modes.

\begin{figure}
\centering
\begin{minipage}[h]{0.48\linewidth}
\center{ \includegraphics[bb = 5 50 255 255, clip, width=1\linewidth, height=0.9\textwidth]{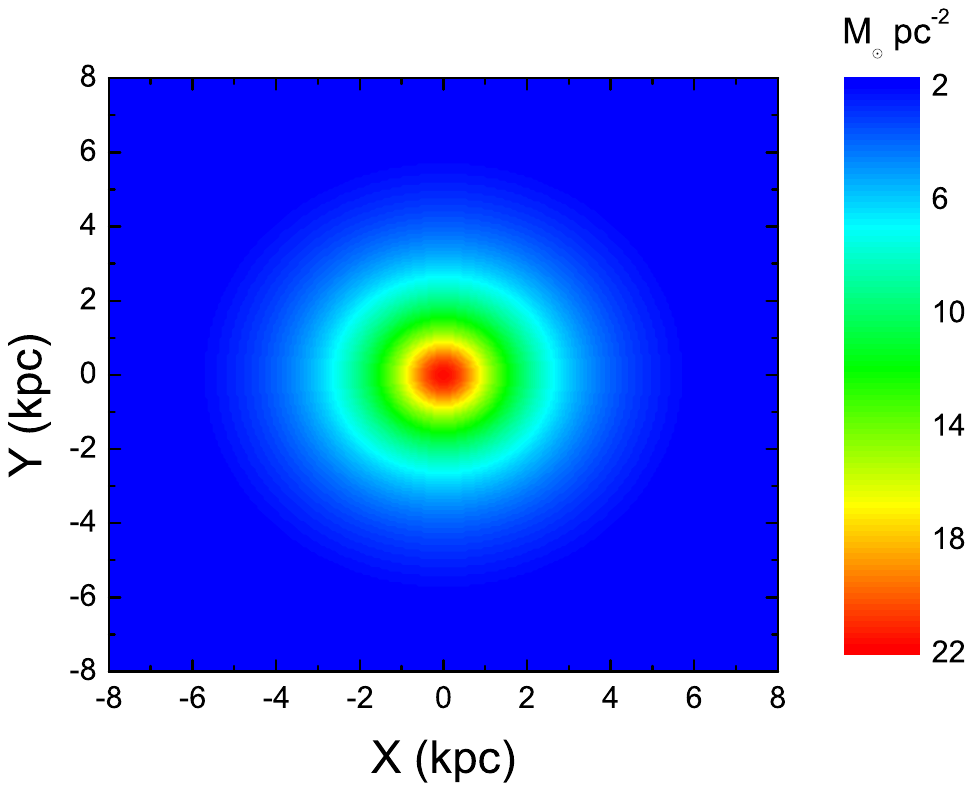} }
\end{minipage}
\hfill
\begin{minipage}[h]{0.48\linewidth}
\center{ \includegraphics[bb = 45 50 295 255, clip, width=1\linewidth, height=0.9\textwidth]{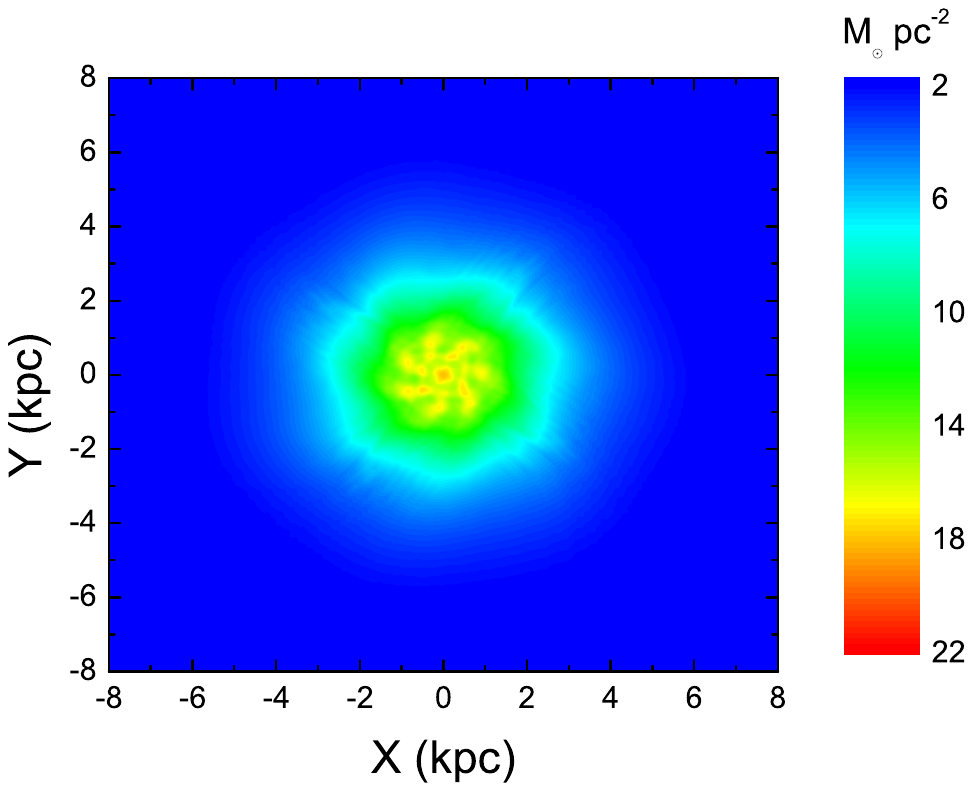} }
\end{minipage}
\hfill
\begin{minipage}[h]{0.48\linewidth}
\center{ \includegraphics[bb = 5 21 255 226, clip, width=1\linewidth, height=0.9\textwidth]{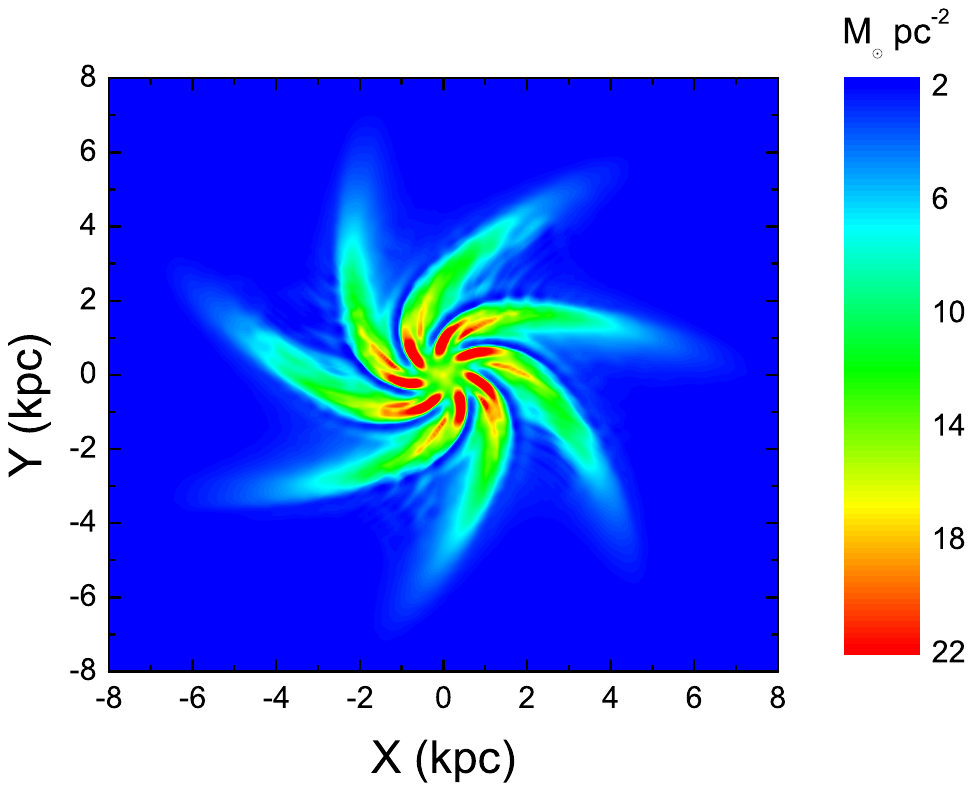} }
\end{minipage}
\hfill
\begin{minipage}[h]{0.48\linewidth}
\center{ \includegraphics[bb = 45 21 295 226, clip, width=1\linewidth, height=0.9\textwidth]{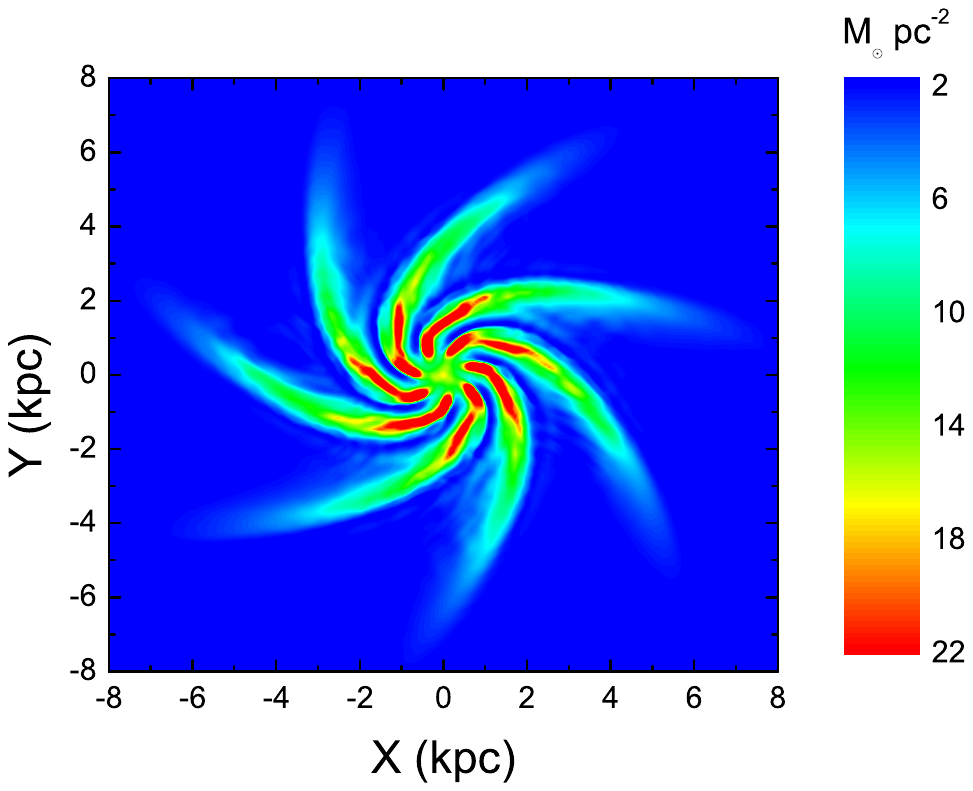} }
\end{minipage}
\caption{Evolution of the column density (in $M_{\odot} pc^{-2}$) in model~4. The evolution times are 0~Myr (top-left), 100~Myr (top-right), 150~Myr (bottom-left) and 200~Myr (bottom-right).}
\label{simulationarm7}
\end{figure}

\begin{figure}
\centering
\includegraphics[bb = 0 0 300 230, width=0.7\linewidth]{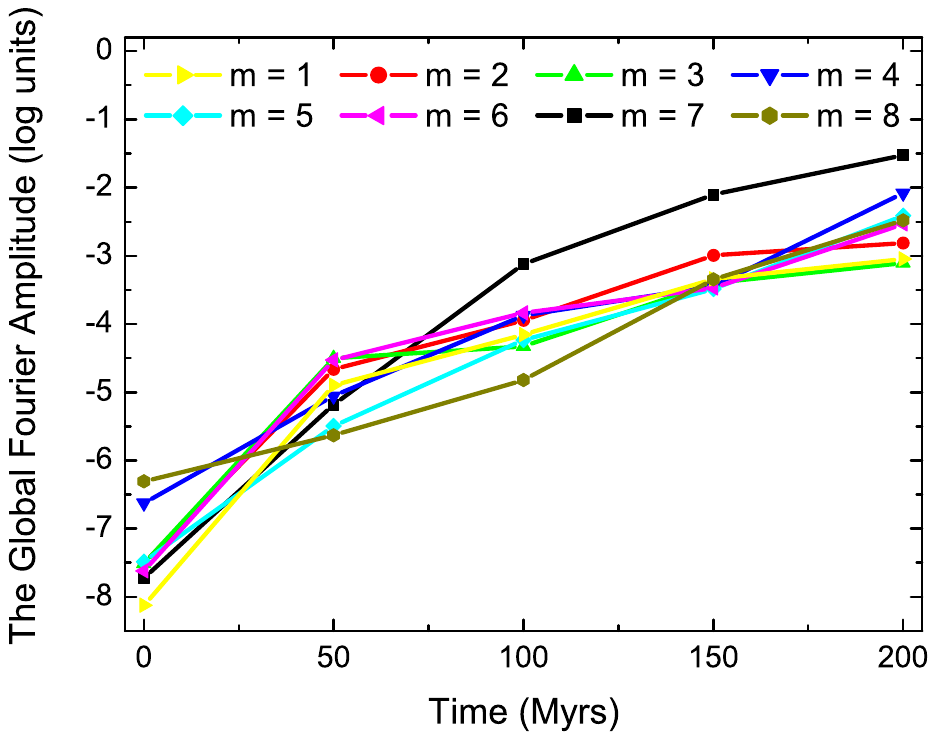}
\includegraphics[bb = 0 0 300 230, width=0.7\linewidth]{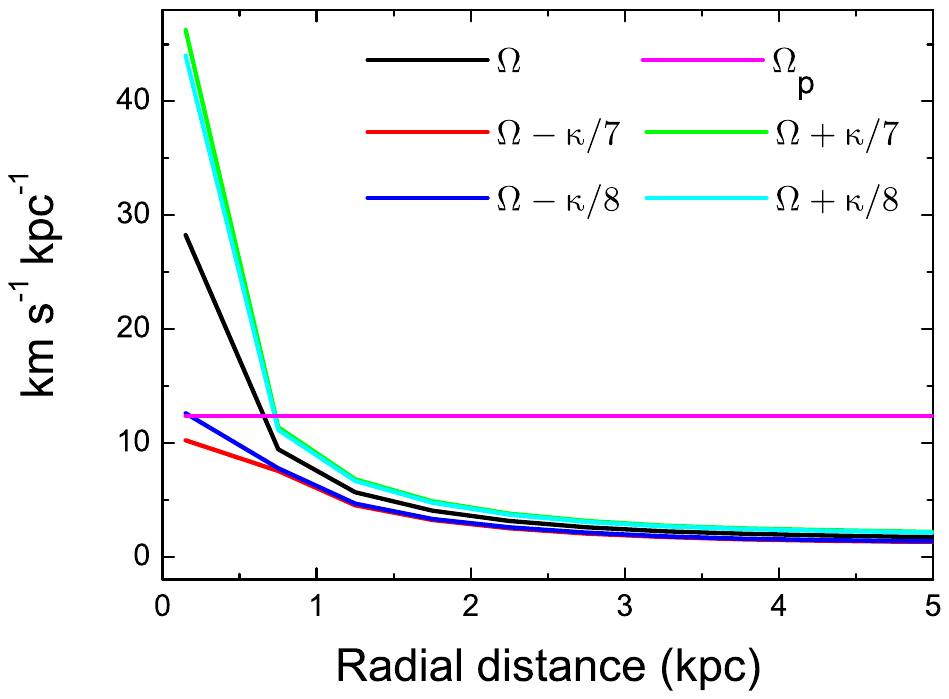}
\caption{{\bf Top}: Time evolution of the global Fourier amplitudes in model~4. {\bf Bottom}: 
radial profiles of $\Omega$, $\Omega_p$, $\Omega \pm \kappa/7$, and $\Omega \pm \kappa/8$. 
The intersection of $\Omega_p$ with $\Omega - \kappa/8$ marks the position of the inner Lindblad resonance for the $m=8$ mode.}
\label{analysisarm7}
\end{figure}

Finally, in Figure~(\ref{simulationarm7}) we show the development a seven-armed spiral pattern in model~4. This model has the smallest ratio of the disk mass to the DM mass among all models, $\xi=0.0236$. The time evolution of the global Fourier amplitudes shown in the upper panel of 
Figure~(\ref{analysisarm7}) confirms that the $m=7$ mode is the fastest growing one. 
The bottom panel in Figure~(\ref{analysisarm7}) demonstrates that 
the ILR for the $m=7$ is absent, while the ILR for the $m=8$ mode is present in model~4.

To conclude, our numerical code behaves in accordance with 
the theoretical expectations on this difficult test problem. We find that the systems with a smaller
disk to DM mass ratio develop patterns with a higher number of spiral arms, which may have interesting
implications for the studies of global non-axisymmetric structures in disk galaxies.

\section*{CONCLUSIONS}
In this paper, we present a new approach for constructing a low-dissipation numerical scheme to simulate
hydrodynamic flows in astrophysics. The method is based on a combination of the operator-splitting 
method, Godunov method, and piecewise-parabolic method on the local stencil. Our numerical method 
was tested on an extensive suite of hydrodynamic test problems. The performance of the method is 
demonstrated on a global astrophysical problem showing the development of a multi-arm spiral structure in a gravitationally unstable gaseous galactic disk in accordance with the swing amplification theory. 

Our numerical method has the following advantages:
1) High-order numerical solution on continuous functions and low-dissipation numerical solution
on discontinuous functions, 2) absence of artificial viscosity and flux limiters, 3) Gallilean-invariance of numerical solutions, 4) guaranteed non-decrease of entropy, 5) extensibility on other numerical hyperbolic systems, 6) simplicity of parallel implementation on hybrid and classic supercomputers, 7) high scalability. 

Main disadvantages of the numerical method is a rather simple finite-difference scheme for the time derivative. In future works, we will focus on extending our numerical method to include the moments of collisionless Boltzmann equations to describe the dynamics of stars and dark matter in galaxies \cite{Mitchell_2013,Kulikov_2014,Vorobyov_2006,Vorobyov_2008,Protasov_2016}. We also plan to include the magnetohydrodynamics effects. This model will be used for numerical modelling of galactic dynamics and evolution of protostellar disks \cite{Vorobyov_2015}. We also plan to implement our numerical scheme on GPU supercomputer (SSCC SB RAS) \cite{Kulikov_2014} and on Intel Xeon Phi supercomputers (JSCC RAS) \cite{Kulikov_2015}. In addition, the numerical method can be used for non astrophysical problem, for example to model explosion welding problems \cite{Godunov_2013}.

\section*{ACKNOWLEDGMENTS}
The research work was supported by the Grant of the President of Russian Federation for the support of young scientists number MK -- 6648.2015.9, RFBR grants 15-31-20150 and 15-01-00508. This project was partly supported by the Russian Ministry of Education and Science Grant 3.961.2014/K.

\section*{Appendix. The construction of piecewise-parabolic functions}
In this section, we construct a piecewise-parabolic function $q(x)$ on a regular mesh 
with step $h$. For simplicity, we choose a spacial domain $[x_{i-1/2},x_{i+1/2}]$.
A parabola on this local domain has the following form \cite{Collela_1984}:
\begin{equation}
q(x) = q_{i}^{L} + \xi \left( \bigtriangleup q_{i} + q_{i}^{(6)} (1 - \xi) \right),
\label{eqA1}
\end{equation}
where $q_{i}$, $q_{i}^L$, and $q_i^R$ are defined at the cell center, at the left-hand side and 
right-hand side interface, respectively,  $\xi = (x - x_{i-1/2})h^{-1}$, $\bigtriangleup q_{i} = q_{i}^{L} - q_{i}^{R}$ and $q_{i}^{(6)} = 6 (q_{i} - 1/2 (q_{i}^{L} + q_{i}^{R}))$. We note that
equation~(\ref{eqA1}) satisfies the conservation law 
$$
q_{i} = h^{-1} \int_{x_{i-1/2}}^{x_{i+1/2}} q(x) dx .
$$
For calculating the  values of $q_{i}^{R} = q_{i+1}^{L} = q_{i+1/2}$,  we used a 4-th order interpolation function of the following form:
$$
q_{i+1/2} = 1/2(q_{i} + q_{i+1}) - 1/6 (\delta q_{i+1} - \delta q_{i}),
$$
where $\delta q_{i}  = 1/2 (q_{i+1} - q_{i-1})$.

Below we present an algorithm for constructing the local parabola $q(x)$. 
The input for this algorithm are the values of $q_{i}$ at the cell centers. The output of 
the algorithm are the local parabolas on each domain $[x_{i-1/2},x_{i+1/2}]$.

 {Step 1.} At the first step, we construct $\delta q_{i}  = 1/2 (q_{i+1} - q_{i-1})$. To do this, we use the values of $q_{i+1}, q_{i-1}$ at the neighbouring cells. To eliminate possible 
extrema on the local parabola, we must modify the formula for $\delta q_{i}$ as follows:
$$
\delta_{m} q_{i} = \left\lbrace \begin{array}{c}
\min(\vert \delta q_{i} \vert, 2 \vert q_{i+1} - q_{i} \vert, 2 \vert q_{i} - q_{i-1} \vert) {\rm sgn} (\delta q_{i}), (q_{i+1} - q_{i})(q_{i} - q_{i-1}) > 0 \\ 
0, (q_{i+1} - q_{i})(q_{i} - q_{i-1}) \leq 0
\end{array}  \right. .
$$
After computing $\delta_{m} q_{i}$, we find the values on the boundary as:
$$
q_{i}^{R} = q_{i+1}^{L} = q_{i+1/2} = 1/2(q_{i} + q_{i+1}) - 1/6 (\delta_{m} q_{i+1} - \delta_{m} q_{i}).
$$

 {Step 2.} At the second step, we can construct  parabolas $q(x)$ on the local domain as follows:
$$
\bigtriangleup q_{i} = q_{i}^{L} - q_{i}^{R},
$$
$$
q_{i}^{(6)} = 6 (q_{i} - 1/2 (q_{i}^{L} + q_{i}^{R})).
$$
In the case of non-monotonic local parabolas, we must reconstruct the values on the boundary $q_{i}^{L}, q_{i}^{R}$ as:
$$
q_{i}^{L} = q_{i}, q_{i}^{R} = q_{i}, (q_{i}^{L} - q_{i})(q_{i} - q_{i}^{R}) \leq 0,
$$ 
and
$$
q_{i}^{L} = 3q_{i} - 2q_{i}^{R},  \bigtriangleup q_{i} q_{i}^{(6)} > (\bigtriangleup q_{i})^{2},
$$
$$
q_{i}^{R} = 3q_{i} - 2q_{i}^{L}, \bigtriangleup q_{i} q_{i}^{(6)} < - (\bigtriangleup q_{i})^{2}.
$$

 {Step 3.} At the third step, we recalculate every parabola on the local domain 
using the following equations:
$$
\bigtriangleup q_{i} = q_{i}^{L} - q_{i}^{R},
$$
$$
q_{i}^{(6)} = 6 (q_{i} - 1/2 (q_{i}^{L} + q_{i}^{R})).
$$
This step completes the construction of local parabolas on each domain $[x_{i-1/2},x_{i+1/2}]$. 
These parabolas may be discontinuous on the cell boundaries. In this case, we must solve the 
Riemann problem. In the classic piecewise-parabolic method parabolas are continuous, but in the PPML
methods local parabolas may be discontinuous, which is an important feature of the latter method. 

To construct the Riemann solver, we must integrate $q(x)$  along the characteristics 
$\pm \lambda t$ to the left and to the right from the cell interface. The result can be written in the following  form:
$$
q_{L}(- \lambda t) = (\lambda t)^{-1} \int_{x_{i+1/2}-\lambda t}^{x_{i+1/2}} q(x) dx =  q_{i}^{R} - \frac{\lambda t}{2h} \left( \bigtriangleup q_{i} - q_{i}^{(6)} \left(1 - \frac{2 \lambda t}{3h} \right) \right),
$$
$$
q_{R}(\lambda t) = (\lambda t)^{-1} \int_{x_{i+1/2}}^{x_{i+1/2}+\lambda t} q(x) dx =  q_{i}^{L} + \frac{\lambda t}{2h} \left( \bigtriangleup q_{i} + q_{i}^{(6)} \left(1 - \frac{2 \lambda t}{3h} \right) \right).
$$
These values can be used in Equations~(\ref{lowriemannvelocity}), ~(\ref{lowriemannpressure}) and (\ref{fluxonlagrangestage}) for $v_L(\pm \lambda t)$, $v_R(\pm \lambda t)$, $p_L(\pm \lambda t)$, $p_R(\pm \lambda t)$, $f_{L}(- \vert \lambda \vert t)$ and $f_{R}( \vert \lambda \vert t)$.

\clearpage

\clearpage

\clearpage

\clearpage

\clearpage

\clearpage

\clearpage

\clearpage

\clearpage

\clearpage

\clearpage

\clearpage

\clearpage

\clearpage

\clearpage

\clearpage

\clearpage

\clearpage

\clearpage

\clearpage

\clearpage

\clearpage

\clearpage

\clearpage
\begin{table}
\centering
\caption{Behavior of Poisson solver on meshes with different resolution}
\vskip 0.4 cm
\begin{tabular}{|c|c|}
 \hline 
  The mesh size & The relative error \\ 
 \hline 
  $16^{3}$ & 4.799777e-005 \\
  $32^{3}$ & 2.978253e-006 \\
  $64^{3}$ & 1.862059e-007 \\
  $128^{3}$ & 1.164110e-008 \\
  $256^{3}$ & 7.278760e-010 \\
  $512^{3}$ & 4.548964e-011 \\
  $1024^{3}$ & 2.843139e-012 \\
 \hline  
\end{tabular}  
\label{PoissonTestSolver}
\end{table} 

\clearpage
\begin{table}
\centering
\caption{Initial state of the shock tube problem}
\vskip 0.4 cm
\begin{tabular}{|c|c|c|c|c|c|c|c|c|}
 \hline 
 N &  $\rho_{L}$ & $v_{L}$ & $p_{L}$ & $\rho_{R}$ & $v_{R}$  &  $p_{R}$ & $x_{0}$ & $t$ \\ 
 \hline 
 1 & 1 & 0 & 1 & 0.125 & 0 & 0.1 & 0.5 & 0.2 \\ 
 \hline 
 2 & 1 & -2 & 0.4 & 1 & 2 & 0.4 & 0.5 & 0.15 \\ 
 \hline 
 3 & 1 & 0 & 1000 & 1 & 0 & 0.01 & 0.5 & 0.012 \\ 
 \hline 
\end{tabular}  
\label{ShockTubeProblem}
\end{table}

\clearpage
\begin{table*}
\renewcommand{\arraystretch}{1.2}
\center
\caption{Model parameters}
\label{table_models}
\begin{tabular}{cccccc}
\hline\hline
Model & $M_{\rm disk}$ & $\xi=M_{\rm disk}/M_{\rm DM}$ & $Q_{\rm min}$ & $T$ & dominant mode  \\
 & ($10^{10} M_\odot$) &  &  & K  \\
\hline
1 & 0.217 & 0.0434 & 3.912 & $10^4$ & --- \\
2 & 0.854 & 0.1708 & 1.109 & $10^4$ & 2 \\
3 & 0.334 & 0.0668 & 1.122 & $10^4$ & 4 \\
4 & 0.118 & 0.0236 & 1.243 & $2\times 10^3$ & 7 \\
\hline
\end{tabular}
\end{table*}


\begin{thebibliography}{100}
\bibitem{Agertz_2007} 
O. Agertz, et al., Fundamental differences between SPH and grid methods, \textit{Monthly Notices of the Royal Astronomical Society.}  {380}, 963-978, (2007).

\bibitem{Tasker_2008} 
E.J. Tasker, et al., A test suite for quantitative comparison of hydrodynamic codes in astrophysics, \textit{Monthly Notices of the Royal Astronomical Society.}  {390}, 1267-1281, (2008).

\bibitem{Vshivkov_2009} 
I. Kulikov, G. Lazareva, A. Snytnikov and V. Vshivkov, Supercomputer Simulation of an Astrophysical Object Collapse by the Fluids-in-Cell Method, \textit{Lectures Notes of Computer Science.}  {5698}, 414-422, (2009).

\bibitem{Attwood_2007} 
R. Attwood, S. Goodwin and A. Whitworth, Adaptive smoothing length in SPH, \textit{Astronomy \& Astrophysics.}  {464}, 447-450, (2007). 

\bibitem{Sijacki_2006} 
D. Sijacki and V. Springel, Physical Viscosity in Smoothed Particle Hydrodynamics Simulations of Galaxy Clusters, \textit{Monthly Notices of the Royal Astronomical Society.}  {371}, 1025-1046, (2006).

\bibitem{Wadsley_2008} 
J. Wadsley, G. Veeravalli and H. Couchman, On the treatment of entropy mixing in numerical cosmology, \textit{Monthly Notices of the Royal Astronomical Society}  {387}, 427-438, (2008).

\bibitem{Mitchell_2013} 
N. Mitchell, E. Vorobyov and G. Hensler, Collisionless Stellar Hydrodynamics as an Efficient Alternative to N-body Methods, \textit{Monthly Notices of the Royal Astronomical Society.}  {428}, 2674-2687, (2013).

\bibitem{Murphy_2008} 
J. Murphy and A. Burrows, BETHE-Hydro: An Arbitrary Lagrangian-Eulerian Multidimensional Hydrodynamics Code for Astrophysical Simulations, \textit{The Astrophysical Journal Supplement Series.}  {179}, 209-241, (2008).

\bibitem{Springel_2010} 
V. Springel, E pur si muove: Galilean-invariant cosmological hydrodynamical simulations on a moving mesh, \textit{Monthly Notices of the Royal Astronomical Society.}  {401}, 791-851, (2010).

\bibitem{Kulikov_2013} 
I. Kulikov, PEGAS: Hydrodynamical code for numerical simulation of the gas components of interacting galaxies, \textit{Book Series of the Argentine Astronomical Society.}  {4}, 91-95, (2013).

\bibitem{Vshivkov_2007} 
V. Vshivkov, G. Lazareva and I. Kulikov, A modified fluids-in-cell method for problems of gravitational gas dynamics, \textit{Optoelectronics, Instrumentation and Data Processing.}  {43}, 530-537, (2007).

\bibitem{Vshivkov_2011_a} 
V. Vshivkov, G. Lazareva, A. Snytnikov, I. Kulikov and A. Tutukov, Hydrodynamical code for numerical simulation of the gas components of colliding galaxies, \textit{The Astrophysical Journal Supplement Series.}  {194}, 47, (2011).

\bibitem{Vshivkov_2011_b} 
Vshivkov, V., Lazareva, G., Snytnikov, A., Kulikov, I., Tutukov, A., Computational methods for ill-posed problems of gravitational gasodynamics, \textit{Journal of Inverse and Ill-posed Problems.}  {19}, 151-166, (2011).

\bibitem{Tutukov_2011} 
A. Tutukov, G. Lazareva and I. Kulikov, Gas Dynamics of a Central Collision of Two Galaxies: Merger, Disruption, Passage, and the Formation of a New Galaxy, \textit{Astronomy Reports.}  {55}, 770-783, (2011).

\bibitem{KurganovTadmor_2000} 
A. Kurganov and E. Tadmor, New High-Resolution Central Schemes for Nonlinear Conservation Laws and Convection-Diffusion Equation, \textit{J. Comput. Phys.}  {160}, 214-282, (2000).

\bibitem{VanLeer_1979} 
B. Van Leer, Towards the Ultimate Conservative Difference Scheme, V. A Second Order Sequel to Godunov's Method, \textit{J. Comput. Phys.}  {32}, 101-136, (1979).

\bibitem{Jin_1995} 
S. Jin and Z. Xin, The Relaxation Schemes for Systems of Conservation Laws in Arbitrary Space Dimensions, \textit{Communications on Pure and Applied Mathematics.}  {48}, 235-276, (1995).

\bibitem{Jiang_1996}
G.-S. Jiang and C.-W. Shu, Efficient Implementation of Weighted ENO Schemes \textit{J. Comput. Phys.}  {126}, 202-228, (1996).

\bibitem{Balsara_2000}
D. Balsara and C.-W. Shu, Monotonicity Preserving Weighted Essentially Non-oscillatory Schemes with Increasingly High Order of Accuracy, \textit{J. Comput. Phys.}  {160}, 405-452, (2000).

\bibitem{Balsara_2009}
D. Balsara, T. Rumpf, M. Dumbser and C.-D. Munz, Efficient, high accuracy ADER-WENO schemes for hydrodynamics and divergence-free magnetohydrodynamics, \textit{J. Comput. Phys.}  {228}, 2480-2516, (2009).

\bibitem{Henrick_2005}
A. Henrick, T. Aslam and J. Powers, Mapped weighted essentially non-oscillatory schemes: Achieving optimal order near critical points, \textit{J. Comput. Phys.}  {207}, 542-567, (2005).

\bibitem{Collela_1984} 
P. Collela and P.R. Woodward, The Piecewise Parabolic Method (PPM) Gas-Dynamical simulations, \textit{J. Comput. Phys.}  {54}, 174-201, (1984).

\bibitem{Watersona_2007} 
N. Watersona and H. Deconinck, Design principles for bounded higher-order convection schemes -- a unified approach, \textit{J. Comput. Phys.}  {224}, 182-207, (2007).

\bibitem{Ustyugov_2007} 
M. Popov and S. Ustyugov, Piecewise parabolic method on local stencil for gasdynamic simulations, \textit{Computational Mathematics and Mathematical Physics.}  {47}, 1970-1989, (2007).

\bibitem{Ustyugov_2008} 
M. Popov and S. Ustyugov, Piecewise parabolic method on a local stencil for ideal magnetohydrodynamics, \textit{Computational Mathematics and Mathematical Physics.}  {48}, 477-499, (2008).

\bibitem{Balsara_1999}
D. Balsara and D. Spicer, Maintaining Pressure Positivity in Magnetohydrodynamic Simulations, \textit{J. Comput. Phys.} 148, 133вЂ“148, (1999).

\bibitem{Ryu_1993}
D. Ryu, J. Ostriker, H. Kang and R. Cen, A cosmological hydrodynamic code based on the total variation diminishing scheme, \textit{The Astrophysical Journal.}  {414}, 1-19, (1993).

\bibitem{Springel_2002}
V. Springel and L. Hernquist, Cosmological smoothed particle hydrodynamics simulations: the entropy equation, \textit{Monthly Notices of the Royal Astronomical Society.}  {333}, 649-664, (2002).

\bibitem{GodunovKulikov_2014} 
S. Godunov and I. Kulikov, Computation of Discontinuous Solutions of Fluid Dynamics Equations with Entropy Nondecrease Guarantee, \textit{Computational Mathematics and Mathematical Physics.}  {54}, 1012-1024, (2014).

\bibitem{Roe_1997} 
P. Roe, Approximate Riemann solvers, parameter vectors, and difference solvers, \textit{J. Comput. Phys.}  {135}, 250-258, (1997).

\bibitem{Balsara_2010}
D. Balsara, Multidimensional HLLE Riemann solver: Application to Euler and magnetohydrodynamic flows, \textit{J. Comput. Phys.}  {229}, 1970-1993, (2010).

\bibitem{Balsara_2012b} 
D. Balsara, A two-dimensional HLLC Riemann solver for conservation laws: Application to Euler and magnetohydrodynamic flows, \textit{J. Comput. Phys.}  {231}, 7476-7503, (2012).

\bibitem{Balsara_2014}  
D. Balsara, M. Dumbser and R. Abgrall, Multidimensional HLLC Riemann solver for unstructured meshes -- With application to Euler and MHD flows, \textit{J. Comput. Phys.}  {261}, 172-208, (2014).

\bibitem{Balsara_2014b}  
D. Balsara, Multidimensional Riemann problem with self-similar internal structure. Part I вЂ“- Application to hyperbolic conservation laws on structured meshes, \textit{J. Comput. Phys.}  {277}, 163-200, (2014).

\bibitem{Balsara_2015}
D. Balsara and M. Dumbser, Multidimensional Riemann problem with self-similar internal structure. Part II -- Application to hyperbolic conservation laws on unstructured meshes, \textit{J. Comput. Phys.}  {287}, 269-292, (2015).

\bibitem{Balsara_2015b}
D. Balsara, Three dimensional HLL Riemann solver for conservation laws on structured meshes: Application to Euler and magnetohydrodynamic flows, \textit{J. Comput. Phys.}  {295}, 1-23, (2015).

\bibitem{Balsara_2015c}
D. Balsara and M. Dumbser, Divergence-free MHD on unstructured meshes using high order finite volume schemes based on multidimensional Riemann solvers, \textit{J. Comput. Phys.}  {299}, 687-715, (2015).

\bibitem{Boscheri_2014} 
W. Boscheri, D. Balsara and M. Dumbser, Lagrangian ADER-WENO finite volume schemes on unstructured triangular meshes based on genuinely multidimensional HLL Riemann solvers, \textit{J. Comput. Phys.}  {267}, 112-138, (2014).

\bibitem{Boscheri_2014b}
W. Boscheri, D. Balsara, and M. Dumbser, High-order ADER-WENO ALE schemes on unstructured triangular meshes-application of several node solvers to hydrodynamics and magnetohydrodynamics. \textit{International journal for numerical methods in fluids.}  {76}, 737-778, (2014).

\bibitem{Bryan_2014} 
G. Bryan, et al., ENZO: An Adaptive Mesh Refinement Code for Astrophysics, \textit{The Astrophysical Journal Supplement Series.}  {211}, 19 (2014).

\bibitem{Clarke_2010} 
D. Clarke, On the Reliability of ZEUS-3D, \textit{The Astrophysical Journal Supplement Series.}  {187}, 119-134, (2010).

\bibitem{Balsara_2012}
D. Balsara, Self-adjusting, positivity preserving high order schemes for hydrodynamics and magnetohydrodynamics, \textit{J. Comput. Phys.}  {231}, 7504-7517, (2012).

\bibitem{Zhang_2010}
X. Zhang and C.-W. Shu, On positivity-preserving high order discontinuous Galerkin schemes for compressible Euler equations on rectangular meshes, \textit{J. Comput. Phys.}  {229}, 8918-8934, (2010).

\bibitem{Frigo_1998} 
M. Frigo and S. Johnson, The Design and Implementation of FFTW3, \textit{Proceedings of the IEEE.}  {93}, 216-231, (2005).

\bibitem{Goloviznin_1998} 
V. Goloviznin and S. Karabasov, Nonlinear correction of Cabaret scheme, \textit{Mathematical Modelling Journal.}  {10}, 107-123, (1998).

\bibitem{Shu_1988} 
C.-W. Shu, Total-Variation-Diminishing time discretizations, \textit{SIAM Journal on Scientific and Statistical Computing.}  {9}, 1073-1084, (1988).

\bibitem{ShuOsher_1988} 
C.-W. Shu and S. Osher, Efficient implementation of essentially non-oscillatory shock-capturing schemes, \textit{J. Comput. Phys.}  {77}, 439-471, (1988).

\bibitem{Kulikov_2014} 
I. Kulikov, GPUPEGAS: A New GPU-accelerated Hydrodynamic Code for Numerical Simulations of Interacting Galaxies, \textit{The Astrophysical Journal Supplements Series.}  {214}, 12, (2014).

\bibitem{Kulikov_2015} 
I. Kulikov, I. Chernykh, A. Snytnikov, B. Glinskiy and A. Tutukov, AstroPhi: A code for complex simulation of dynamics of astrophysical objects using hybrid supercomputers, \textit{Computer Physics Communications.}  {186}, 71-80, (2015).

\bibitem{Godunov_2013} 
S. Godunov, S. Kiselev, I. Kulikov and V. Mali, Numerical and experimental simulation of wave formation during explosion welding, \textit{Proceedings of the Steklov Institute of Mathematics.}  {281}, 12-26, (2013).

\bibitem{Vorobyov_2006} 
E. Vorobyov and Ch. Theis, Boltzmann moment equation approach for the numerical study of anisotropic stellar discs, \textit{Monthly Notices of the Royal Astronomical Society.}  {373}, 197-208, (2006).

\bibitem{Woodward_1984}
P. Woodward and P. Colella, The Numerical Simulation of Two-Dimensional Fluid Flow with Strong Shocks, \textit{J. Comput. Phys.}  {54}, 115-173, (1984).

\bibitem{Yoon_2008}
S.-H. Yoon, C. Kim and K.-H. Kim, Multi-dimensional limiting process for three-dimensional flow physics analyses, \textit{J. Comput. Phys.}  {227}, 6001-6043, (2008).

\bibitem{Aksenov_2005}
A.V. Aksenov, Linear Differential Relations Between Solutions of the Class of Euler-Poisson-Darboux Equations, \textit{Journal of Mathematical Sciences.}  {130}, 4911-4940, (2005).

\bibitem{Toomre_1964} 
A. Toomre, On the gravitational stability of a disk of stars, \textit{The Astrophysical Journal.}  {139}, 1217-1238, (1964).

\bibitem{Nelson1998} 
A. Nelson, W. Benz, F. Adams and D. Arnett, Dynamics of Circumstellar Disks, \textit{The Astrophysical Journal.}  {502}, 342вЂ“371, (2008).

\bibitem{Polyachenko1997} 
V. Polyachenko, E. Polyachenko and A. StrelвЂ™nikov, Stability criteria for gaseous self-gravitating disks, \textit{Astronomy Letters.}  {23}, 483-491, (1997).

\bibitem{Vorobyov_2012} 
E. Vorobyov, S. Recchi and G. Hensler, Self-gravitating equilibrium models of dwarf galaxies and the minimum mass for star formation, \textit{Astronomy \& Astrophysics}  {543}, A129, (2012).

\bibitem{Athanas_1984} 
E. Athanassoula, The spiral structure of galaxies, \textit{Physics Reports.}  {114}, 319-403 (1984).

\bibitem{Goldreich_1965} 
P. Goldreich and D. Lynden-Bell II. Spiral arms as sheared gravitational instabilities, \textit{Monthly Notices of the Royal Astronomical Society.}  {130}, 125-158, (1965).

\bibitem{Toomre_1981} 
A. Toomre, What amplifies the spirals, \textit{In ''The Structure and Evolution of Normal Galaxies'' Fall S. M., Lynden-Bell D., eds, Cambridge University Press, Cambridge.} 283, (1981).

\bibitem{BT_1987} 
J. Binney and S. Tremaine, Galactic Dynamics, \textit{Princeton Univ. Press, Princeton, NJ.} 904 (2008).

\bibitem{Vorobyov_2008} 
E. Vorobyov and Ch. Theis, Shape and orientation of stellar velocity ellipsoids in spiral galaxies, \textit{Monthly Notices of the Royal Astronomical Society.}  {383}, 817-830, (2008).

\bibitem{Protasov_2016}
V. Protasov, A. Serenko, V. Nenashev, I. Kulikov and I. Chernykh, High-Performance Computing in Astrophysical Simulations, \textit{Journal of Physics: Conference Series.}  {681}, 012022, (2016).

\bibitem{Vorobyov_2015} 
E. Vorobyov, D. Lin and M. Guedel, The effect of external environment on the evolution of protostellar disks, \textit{Astronomy \& Astrophysics.}  {573}, A5, (2015).
\end{thebibliography}
\end{document}